\documentclass[journal=jpccck,manuscript=article]{achemso}
\setkeys{acs}{etalmode=truncate,maxauthors=0,articletitle=true}
\usepackage{amsmath,amssymb,bm}
\usepackage{array, longtable, booktabs, makecell}
\usepackage{lscape}
\newcommand{\etal}{\emph{et al.}}

\def\sAs{\rm s_A^2}
\def\sA {\rm s_A}

\def\zAs{\rm z_A^2}
\def\zA {\rm z_A}
\def\zAb{\rm \overline{z}_A}
\def\xAs{\rm x_A^2}
\def\xA {\rm x_A}
\def\xAb{\rm \overline{x}_A}
\def\yAs{\rm y_A^2}
\def\yA {\rm y_A}
\def\yAb{\rm \overline{y}_A}
\def\sBs{\rm s_B^2}
\def\sB {\rm s_B}
\def\sBb{\rm \overline{s}_B}
\def\zBs{\rm z_B^2}
\def\zB {\rm z_B}
\def\zBb{\rm \overline{z}_B}
\def\xBs{\rm x_B^2}
\def\xB {\rm x_B}
\def\xBb{\rm \overline{x}_B}
\def\yBs{\rm y_B^2}
\def\yB {\rm y_B}
\def\yBb{\rm \overline{y}_B}

\def\gss{\sigma_s^2}
\def\uss{\sigma_s^{*2}}
\def\gps{\sigma_p^2}
\def\ups{\sigma_p^{*2}}
\def\pxs{\pi_x^2}
\def\uxs{\pi_x^{*2}}
\def\pys{\pi_y^2}
\def\uys{\pi_y^{*2}}

\def\gs{\sigma_s}
\def\us{\sigma_s^{*}}
\def\gp{\sigma_p}
\def\up{\sigma_p^{*}}
\def\px{\pi_x}
\def\ux{\pi_x^{*}}
\def\py{\pi_y}
\def\uy{\pi_y^{*}}

\def\gsb{\overline{\sigma_s}}
\def\usb{\overline{\sigma_s^{*}}}
\def\gpb{\overline{\sigma_p}}
\def\upb{\overline{\sigma_p^{*}}}
\def\pxb{\overline{\pi_x}}
\def\uxb{\overline{\pi_x^{*}}}
\def\pyb{\overline{\pi_y}}
\def\uyb{\overline{\pi_y^{*}}}

\def\S{$\Sigma_g^+$}
\def\D{$\Delta_g$}

\author{Alexander F. Sax}
\email{alexander.sax@uni-graz.at}
\affiliation{Department of Chemistry, University of Graz,  Graz,
Austria}

\title[ChemBond]
{Chemical Bonding in the C$_2$ Molecule}

\begin{document}

\section{Abstract}
Bonding in the C$_2$ molecule is investigated with CAS(8,8) wave functions using canonical MOs. In a subsequent step, orthogonal atomic orbitals are constructed by localizing the CASSCF MOs  on the two carbon atoms with  an orthogonal transformation. This orbital transformation causes an orthogonal transformation of the configuration state functions (CSF) spanning the function space of the singlet ground state of C$_2$. Instead of CSFs built from canonical MOs one gets CSFs of orthogonal deformed atomic orbitals (AO). This approach resembles the orthogonal valence bond methods (OVB) CSFs which are very different from conventional VB, based on non-orthogonal AOs. To get used to the different argumentation, the bonding situation in ethane (single bond), ethene (double bond), and the nitrogen molecule (triple bond)  are also studied. The complex bonding situation in C$_2$ is caused by the possibility to excite an electron with spin flip from the doubly occupied 2s AO into the 2p subshell, the resulting high-spin $^5S_u$ state of the carbon atom allows for  a better reduction of the Pauli repulsion. But the electron structure around the equilibrium distance does not allow to say that C$_2$ in its ground state has a double, or triple, or even a quadruple bond.

\section{Introduction and Basics}
Covalent bonding is a central concept in chemistry but its semantic is not unique. In physical parlance, bonding means the energetic stabilization of unspecified size of a system composed of interacting subsystems by any kind of interactions\cite{Sax1} or any kind of attraction\cite{Vo2022}; the result of bonding is a bonded system and often it is said that there is a bond in the stabilized system.\cite{Vo2022} Depending on the amount of energy released during bonding, one can distinguish between weak (secondary) and strong (primary) bonding. In chemistry, chemical bonding is the thermodynamic stabilization of a molecular system at ambient conditions composed of atoms, free radicals, ions  or molecules. Frequently Coulomb interaction between negatively charged particles like electrons or atomic anions and positively charged nuclei or atomic cations are said to be responsible for this type of chemical bonding. This view can be justified in case of, e.g.,  ionic solids, but bonding of non-charged subsystems needs a different description of system stabilization, called covalent bonding. The high reactivity of radicals with odd numbers of electrons and the observation that most stable molecules have an even number of electrons led Lewis to the assumption that linking radicals with one unpaired electron each yields a stabilized  molecular system with an even number of electrons, this is the rule of two.\cite{Lewis1916} According to this view, the stabilization is caused by the formation of an electron pair that is shared by the atoms where the unpaired electrons are located.  In the bonded system, these two atoms form a group with a characteristic short distance between them. This atom group with short distance, is evidence for the formation of a covalent bond between the atoms. If more than one unpaired electron is located at each of the interacting atoms, multiple bonds can be formed. The number of bonds each atom can form is its valence. However, this captivating Lewis model gives no convincing physical explanation for what causes the energetic stabilization.

The purely electrostatic model was first proposed by Slater\cite{Slater1933} and is still most often used - not only in introductory textbooks - to explain covalent bonding. The line of reasoning is as follows: Electrons between the bonded atoms are attracted by both nuclei; if the electron density in the midbond region increases due to electron sharing, the (negative) potential energy and therefore also the total energy is lowered.  However, this model does not agree with Earnshaws theorem\cite{Earnshaw1842}, which says that electrostatic interactions alone can never hold a system of charged particles in a stable, stationary state, the charges must be moving, and, therefore, the kinetic energy must play a central role in the stabilization of the system. According to Hellmann,\cite{Hellmann1933} the increase of the region of space where the shared electrons can be found  causes a decrease of the kinetic energy and thus of the total energy, and this is the main reason for the stabilization of the system.   The role of kinetic energy was neglected for decades in both the physical and the chemical community. 1962 showed Ruedenberg\cite{Ruedenberg1962} in a seminal paper that the kinetic energy is responsible for energetic stabilization, but it took another 40 years until again Ruedenberg and coworkers\cite{Ruedenberg2007,Bitter2007,Ruedenberg2009,Bitter2010,Schmidt2014} in a series of high level calculations could substantiate the claim very convincingly. And, at the same time, they could also demonstrate that the energetic stabilization during covalent bonding is indeed a 1-electron effect, not a 2-electron effect. That the first molecule treated with quantum theory (Heitler and London, 1927)\cite{Heitler1927} was the hydrogen molecule is in accordance with Lewis' view that covalent bonding is caused by shared electron pairs. The bonding electron pair was represented by the Heitler-London wave function $\Phi_{\rm HL}$, which is the product of a two electron spatial wave function and the singlet spin function $\alpha(1)\beta(2)+\beta(1)\alpha(2)$, which is an eigenfunction of the square of the spin operator. The spatial part is the product of a linear combination of products of atomic orbitals (AO), $1s_A(1)1s_B(2)+ 1s_B(1)1s_A(2)$,  each of the two hydrogen atoms A and B contributes one 1s AO. (Normalization factors are omitted.) All wave functions that are eigenfunctions of the square of the spin operator are called  configuration state functions (CSF), therefore, because of its product form, $\Phi_{\rm HL}$ is a CSF. Because the Heitler-London CSF $\Phi_{\rm HL}$ described qualitatively correct  bonding of two univalent hydrogen atoms by an electron pair, it was called the covalent wave function; the valence bond (VB) method uses wave functions that are generalizations of $\Phi_{\rm HL}$. Around the same time, Hund and Mulliken supposed the existence of molecular orbitals (MO) in molecules as in many-electron atoms.\cite{Hund1926,Hund1927,Hund1927a,Hund1927b,Hund1928,Hund1930,Mulliken1928,Mulliken1928a,Mulliken1929} For diatomic molecules, correlation diagrams correlating the orbital energies of the molecule to the orbital energies of the separated atoms  and the united atoms allowed to guess the energetic ordering of the MOs and to classify them as bonding and antibonding. Starting from the dissociated molecules, Lennard-Jones was the first to introduce positive and negative linear combinations of AOs to approximate MOs and to make the first quantitative calculation using MOs in the LCAO approximation.\cite{Lennard1929} The acronym LCAO (linear combination of atomic orbitals) was coined by Mulliken in 1932,\cite{Mulliken1932} the acronym is still used although in actual quantum chemical calculations not orbitals of free atoms but AO-like basis functions are used. The first quantitative SCF (self consistent field) calculation of the H$_2$ molecule with MOs was done by Coulson\cite{Coulson1938} using MOs in elliptic coordinates. But it took another 20 years until SCF calculations with the LCAO approximation could be made.\cite{Roothaan1951,Hall1951} The MOs were calculated as eigenfunctions of the Hermitian Hamiltonian and, thus, known to be orthogonal to each other, which was not only a great computational advantage over the VB method but had also  great conceptual importance, e.g.,  an electron occupying a bonding MO can never occupy also an antibonding MO.   Using a single Slater determinant $|\sigma \overline{\sigma}|$, which is also a CSF, with the doubly occupied bonding $\sigma$  MO, the bond energy of H$_2$ calculated by Goodisman\cite{Goodisman1963} was about 0.5\,eV lower than the bond energy calculated with $\Phi_{\rm HL}$. On the other hand, it was also known that the single CSF $|\sigma \overline{\sigma}|$ gives reasonable results only  for molecular structures close to the equilibrium geometry,\cite{Mulliken1935} but not when bonds are highly stretched; in contrast to the VB method, the dissociated H$_2$ system has an energy that is much too high.  So both wave functions have deficiencies that must be corrected.

Weinbaum\cite{Weinbaum1933} showed that a linear combination $\Psi_{\rm W} \propto \Phi_{\rm HL} + \mu \Phi_{\rm ion}$ of $\Phi_{\rm HL}$ and the ionic  CSF $\Phi_{\rm ion} \propto 1s_A(1)1s_A(2)+ 1s_B(1)1s_B(2)$ improves the bond energy considerably, if the parameter $\mu$ is variationally optimized. If the  LCAO approximation is used for the bonding MO $\sigma\propto 1s_A + 1s_B$, and
if the CSF $|\sigma \overline{\sigma}|$ is expanded, one gets a linear combination $|\sigma \overline{\sigma}| \propto \Phi_{\rm ion} + \Phi_{\rm HL}$ with equal coefficients of the linear combination for all bond lengths. It is this equal contribution of Heitler-Londont and  ionic VB CSFs to the MO CSF $|\sigma \overline{\sigma}|$, which is the reason for the inability of the SCF wave function to describe the dissociation. Analogously, expansion of  $|\sigma^* \overline{\sigma^*}|$ with the doubly occupied antibonding MO $\sigma^*\propto 1s_A - 1s_B$ gives again a linear combination of $\Phi_{\rm HL}$ and $\Phi_{\rm ion}$ with coefficients of equal modulus but noe with a different relative phase, $|\sigma^* \overline{\sigma^*}| \propto \Phi_{\rm ion} -\Phi_{\rm cov}$. Consequently, by using a linear combination of these two MO CSFs, $\Psi_{\rm MO} \propto |\sigma \overline{\sigma}| - \lambda |\sigma^* \overline{\sigma^*}|$, $\lambda >0$ the ionic contribution can be reduced and will disappear for long intermolecular distances. Linear combinations of Slater determinants or CSFs are called CI (configuration interaction) wave functions; the variationally optimized CI wave function $\Psi_{\rm MO}$ is equivalent with the Weinbaum function.

This shows, that CI wave functions both with VB and MO CSFs give a qualitative correct and quantitative satisfying description of the H$_2$ molecule and the dissociation reaction. The reason for the failure of the Slater determinant $|\sigma \overline{\sigma}|$ to describe correctly the dissociations is that two electrons occupying the bonding MO can never separate completely. During the dissociation, each electron should locate at different atoms but the form of the bonding MO allows and forces the electrons to come close. The antibonding MO, with the node between the atoms, describes an electron distribution where the electrons can never meet in the mid-bond region. $\Psi_{\rm HL}$ describes an electron distribution with one electron preferentially at atom A if the other electron is at atom B. This spatial correlation of the electrons is called the left-right correlation. Electrons that tend to stay on different sides of a plane in a molecule show angular correlation,  To describe these two correlation types only  AOs of the valence shell are needed, a third correlation type, in-out correlation, needs AOs with an additional radial nodal surface. Left-right and angular correlation  contribute essentially to what is called non-dynamic correlation.

The most important reason for electron correlation is not charge redistribution caused by Coulomb interaction but the fermionic character of the electrons. The Pauli exclusion principle says that identical electrons, which are electrons that agree also in the spin projection, avoid to come spatially close. So if the total spin of the electrons in an atom changes from a low-spin state to a high-spin state the electrons must locate in different spatial regions. And this spatial correlation of identical electrons is much more effective than the correlation due to the Coulomb repulsion. Non-dynamic correlation covers both spin redistribution and charge redistribution.

Analysis of the Weinbaum functions shows that, for long interatomic distances, $\Phi_{\rm HL}$ describes two neutral atoms and $\Phi_{\rm ion}$ an anion-cation pair. When the interatomic distance is zero and atom A and atom B coalesce, not only the two different AOs but also the CSFs $\Phi_{\rm HL}$  and $\Phi_{\rm ion}$  become identical. That means, with decreasing interatomic distance $\Phi_{\rm HL}$ acquires ionic character and $\Phi_{\rm ion}$ acquires neutral character.  If orthogonal AOs $1s_A$ and $1s_B$ are used, the CSFs $\Phi^o_{\rm HL}$ and $\Phi^o_{\rm ion}$ are also orthogonal to each other for all interatomic distances, and the electronic character of the wave functions never changes: $\Phi^o_{\rm HL}$ is always neutral and $\Phi^o_{\rm ion}$ is always ionic. Consequently, $\Phi^o_{\rm HL}$ alone cannot describe a bonded molecule because the ionic contribution is completely missing. A correct description of the ground state needs always a linear combination of  $\Phi^o_{\rm HL}$  and $\Phi^o_{\rm ion}$.  This was shown by McWeeny\cite{McWeeny1954} and discussed by Pilar.\cite{Pilar1968} Mathematically, the situation is clear: the description of the electronic ground state of H$_2$ needs  CI wave functions, either a linear combination of MO CSFs, or a linear combination of VB CSFs made with either orthogonal or non-orthogonal AOs. The three sets of CSFs are different bases for the same two-dimensional state space, two bases are orthogonal and the third one is non-orthogonal. The advantage of the orthogonal VB basis is that the squared CI coefficients have indeed the properties of probabilities and allow to measure the ionic character of the state; when non-orthogonal AOs are used, weights of the VB CSFs can only approximately calculated, for example with the Chirgwin-Coulson formula.\cite{Chirgwin1950}

Use of orthogonal AOs in VB calculations is not a trivial task, after all,  the orthogonality of  two basis functions or AOs depends on the molecular geometry. Atomic basis functions or AOs located at the position of atoms in a molecule are in general not orthogonal, but they can be orthogonalized with, e.g., L\"owdin's  symmetric orthogonalization method, and then used in a VB calculation. Alternatively, one can calculate the electronic state of a molecule with a conventional CI wave function using MO CSFs and then localize the MOs by an orthogonal transformation. I developed a method where delocalized MOs obtained with CASSCF are localized at predefined fragments with the help of an orthogonal transformation giving orthogonal fragment MOs (FMO). The advantage of this procedure is that the orthogonal transformation in the MO space causes an orthogonal transformation in the CSF space leaving the CASSCF wave function invariant. Most transformed MOs will be delocalized FMOs but some FMOs resemble atom centered AOs or hybrid orbitals (HO), these FMOs are called orthogonal AOs (OAO). OAOs include the deformation of the atomic electron distribution due to polarization caused by the molecular environment; in this respect they are very similar to orthogonalized quasi AOs introduced by Ruedenberg \etal\cite{West2013,West2015} The CASSCF wave function constructed from  orthogonal FMOs instead of orthogonal MOs is a linear combination of OVB CSFs with doubly occupied non-active FMOs and active OAOs. The VB like character of the transformed CSFSCF wave functions is due to the active OAOs. This OVB method\cite{Sax2012,Sax2015} was used to study symmetry allowed and forbidden reactions.\cite{Sax2017}
In this paper, the method is used to analyse the ground state of the C$_2$ molecule, which has  $^1\Sigma_g^+$  symmetry. From two carbon atoms in their $^3P_g$ ground states one can derive three molecular states with \emph{gerade} parity, two are $^1\Sigma_g^+$ states one is a $^1\Delta_g$ state; from the carbon atoms in the $^5S_u$ state one can derive another $^1\Sigma_g^+$ state. In $D_{\infty h}$ symmetry, $\Sigma$ and $\Delta$ states are automatically orthogonal to each other, but because actual calculations can only be  done in the largest Abelian subgroup $D_{2h}$, in which $\Sigma_g^+$ and one component of the $\Delta_g$ state are in the same irreducible representation $A_g$,   they can mix. The lowest three of the four $^1A_g$ states of C$_2$ are studied in this paper. To see how CASSCF wave functions are composed that describe the dissociation of single, double and triple bonds, the molecules ethane, ethene, and  N$_2$ are also studied.

\section{\'Etude: The description of single, double and triple bonds}
A wave function derived from a closed shell  electron configuration is in most cases a single Slater determinant, e.g., the electron configuration $\sigma^2$ with a bonding $\sigma$ MO leads immediately to the MO CSF $|\overline{\sigma}\sigma|$; similarly when doubly degenerate bonding $\pi$ MOs are fully occupied, the MO CSF $|\overline{\pi}_x\pi_x\overline{\pi}_y\pi_y|$ corresponds to the electron configuration $\pi^4$.
If the $\pi$ MOs are not fully occupied, e.g., when the electron configuration  is $\sigma^2\pi^2$, two MO CSFs are possible,
$|\overline{\sigma}\sigma\overline{\pi}_x\pi_x|$ and $|\overline{\sigma}\sigma\overline{\pi}_y\pi_y|$ and a wave function that has rotational symmetry must be a linear combination of them, either $|\overline{\sigma}\sigma\overline{\pi}_x\pi_x|+|\overline{\sigma}\sigma\overline{\pi}_y\pi_y|$ or $|\overline{\sigma}\sigma\overline{\pi}_x\pi_x|-|\overline{\sigma}\sigma\overline{\pi}_y\pi_y|$.

These wave functions are not able to describe dissociation because all MOs are bonding MOs; for a correct description of dissociation all bonding and the corresponding antibonding MOs must be included into the set of active MOs. The CAS problem is defined by giving the number of active MOs and active electrons, and the order of the active orbitals. Non-active MOs are not explicitly mentioned.

The 2s AO and the 2p AOs located at atom A will be labelled $\rm s_A$ and  $\rm x_A, y_A, z_A$, respectively; analogously the AOs on atom B. For HOs no separate symbol is used, they are labelled by their respective dominant AO.
For all molecules discussed, the molecular axis will be the z-axis, $\sigma$ MOs will be made by HOs, if the HO has dominantly s-character the $\sigma$ MO will be labelled $\sigma_s$, a $\sigma_p$ is made with HOs dominated by the $\rm 2p_z$ AO.

\subsection{The dissociation of the C-C single bond in ethane}
\begin{figure}[ht]
\caption{ \label{fig:CAS22Poto}The potential energy curves calculated with MO CASSCF and the OVB method.}
\includegraphics[width=0.48\textwidth]{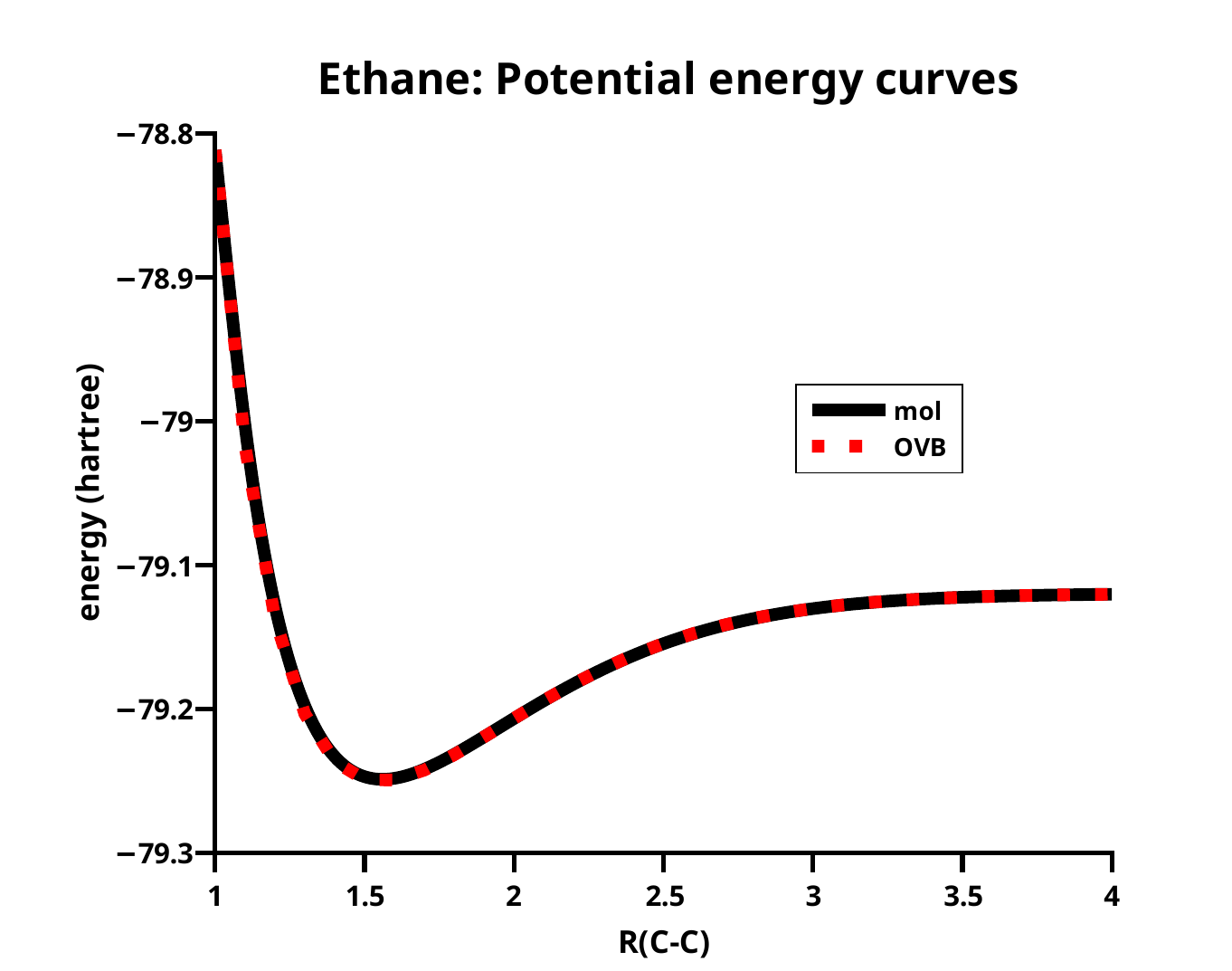}
\end{figure}

The equilibrium C-C distance of ethane is about 1.55\,\AA; Figure \ref{fig:CAS22Poto} shows also that the orthogonal transformation of the MOs leaves the total energy indeed invariant.

The lowest level wave function that correctly describes the dissociation of the $\sigma$ single bond in ethane is a CAS(2,2) wave function, with $\sigma_p$ and the antibonding $\sigma^*_p$ MO as active MOs and 2 active electrons. Two frozen core MOs and six MOs describing the CH bonds are doubly occupied, they are not mentioned in the following.

\begin{figure}[ht]
\caption{ \label{fig:CAS22MO}Energies (left) and weights (right) of the two MO CSFs of ethane.}
\includegraphics[width=0.48\textwidth]{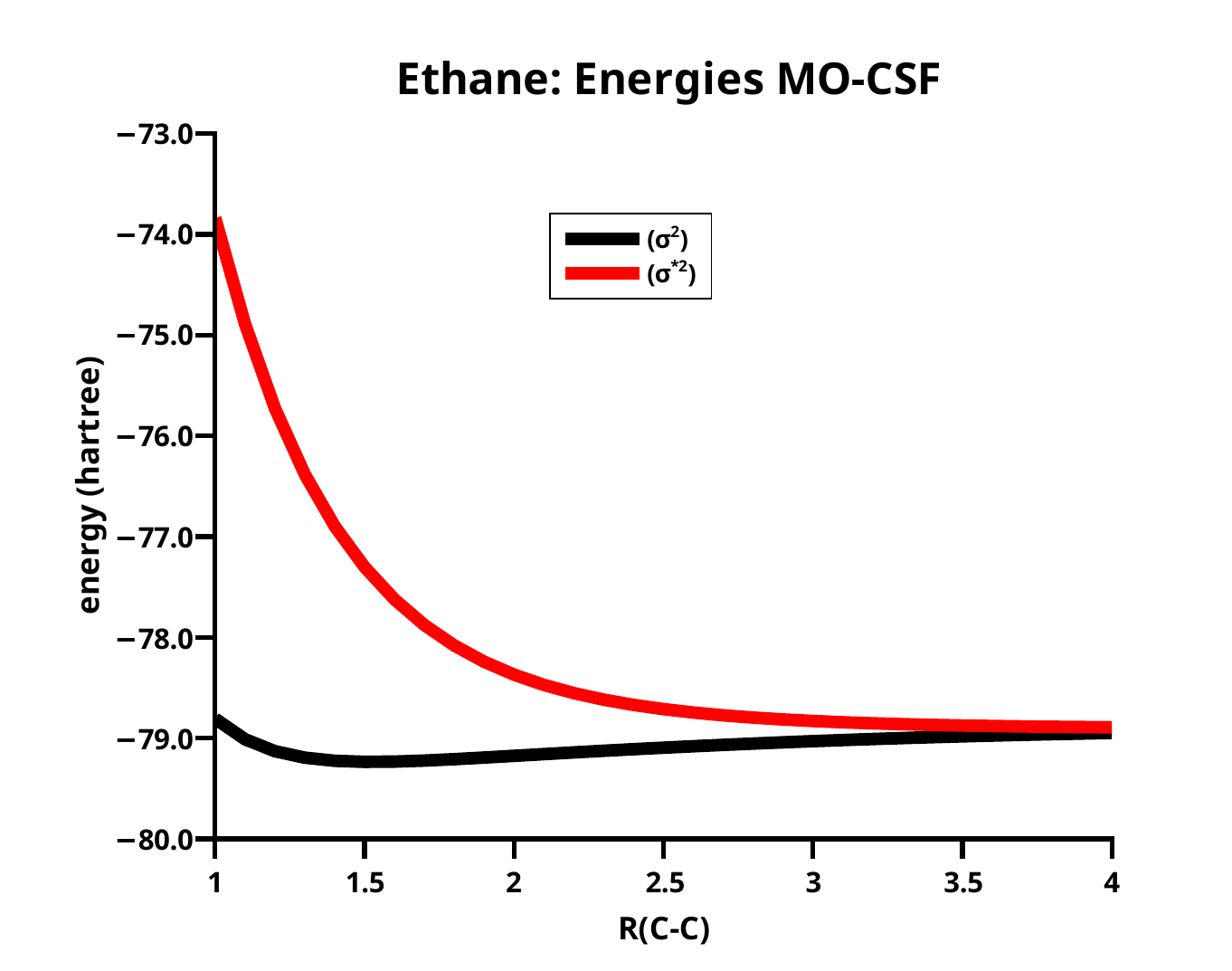}\hfill
\includegraphics[width=0.48\textwidth]{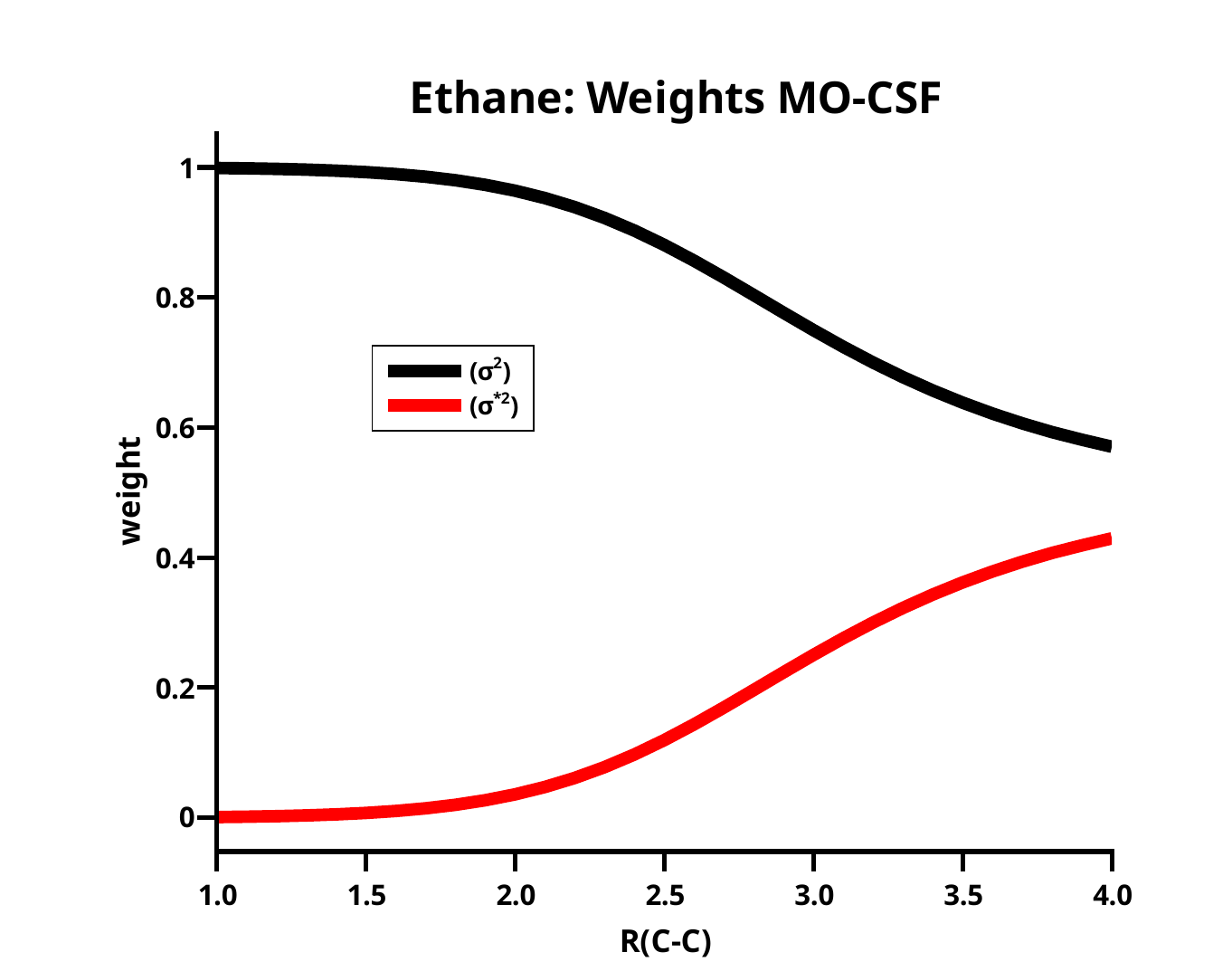}
\end{figure}

The weight curves of the MO CSFs, see Figure \ref{fig:CAS22MO},  show that the $|\sigma_p^2|$  CSF is at short distances a good description of the ethane ground state, but at long distances only a linear combination of $|\sigma_p^2|$ and $|{\sigma_p^*}^2|$ can describe the dissociation into two methyl fragments.

Localization of the 10 MOs onto the two methyl fragments yields two equivalent sets of FMOs, each having one frozen core MO describing the 1s AO, three delocalized FMOs describing the three CH bonds, and one localized FMO having the character of an sp-type HO. The eight non-active FMOs, denoted $1_A,2_A,3_A,4_A$ and $1_B,2_B,3_B,4_B$,  will not be mentioned. With the two $\rm 2p_z$  dominated hybride OAOs, $z_A$ and $z_B$,  one can make CSFs, which  have the form of the Heitler-London VB wave function, $\Phi^o_{\rm HL} = |(z_A \overline{z}_B -\overline{z}_A z_B)|$, and the ionic CSF  $\Phi^o_{\rm ion} = |(z_A \overline{z}_A + z_B\overline{z}_B)|$.

$\Phi_{\rm HL}$ describes the singlet coupling of the doublet states of the methyl groups, each methyl group has one unpaired electron that is ready for bonding and, thus, conforms to the beliefs in chemistry that unpaired electrons are necessary for creating the Lewis electron pair representing a covalent single bond.

\begin{figure}[ht]
\caption{ \label{fig:CAS22VB}Energies (left) and weights (right) of the two OVB CSFs.}
\includegraphics[width=0.48\textwidth]{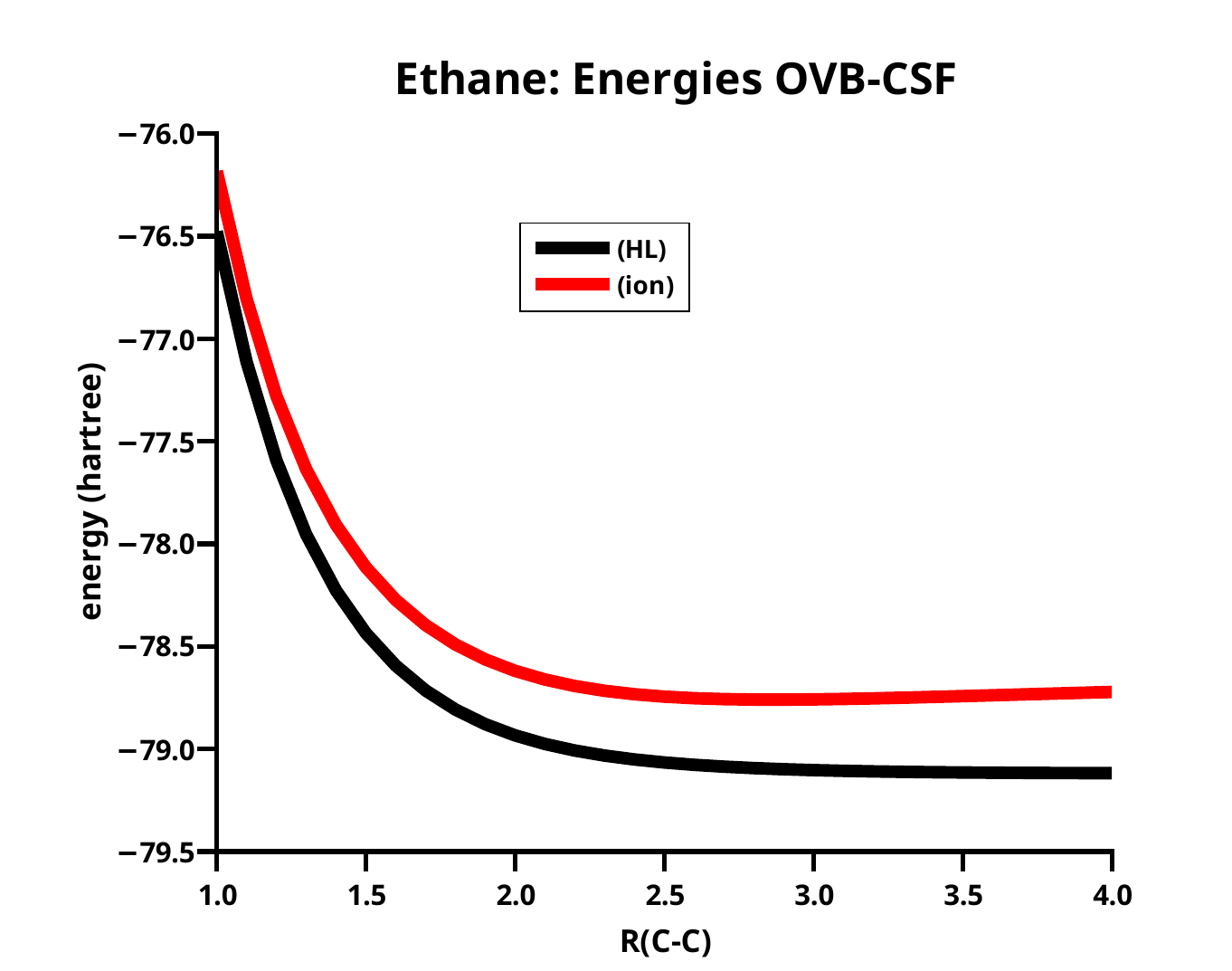}\hfill
\includegraphics[width=0.48\textwidth]{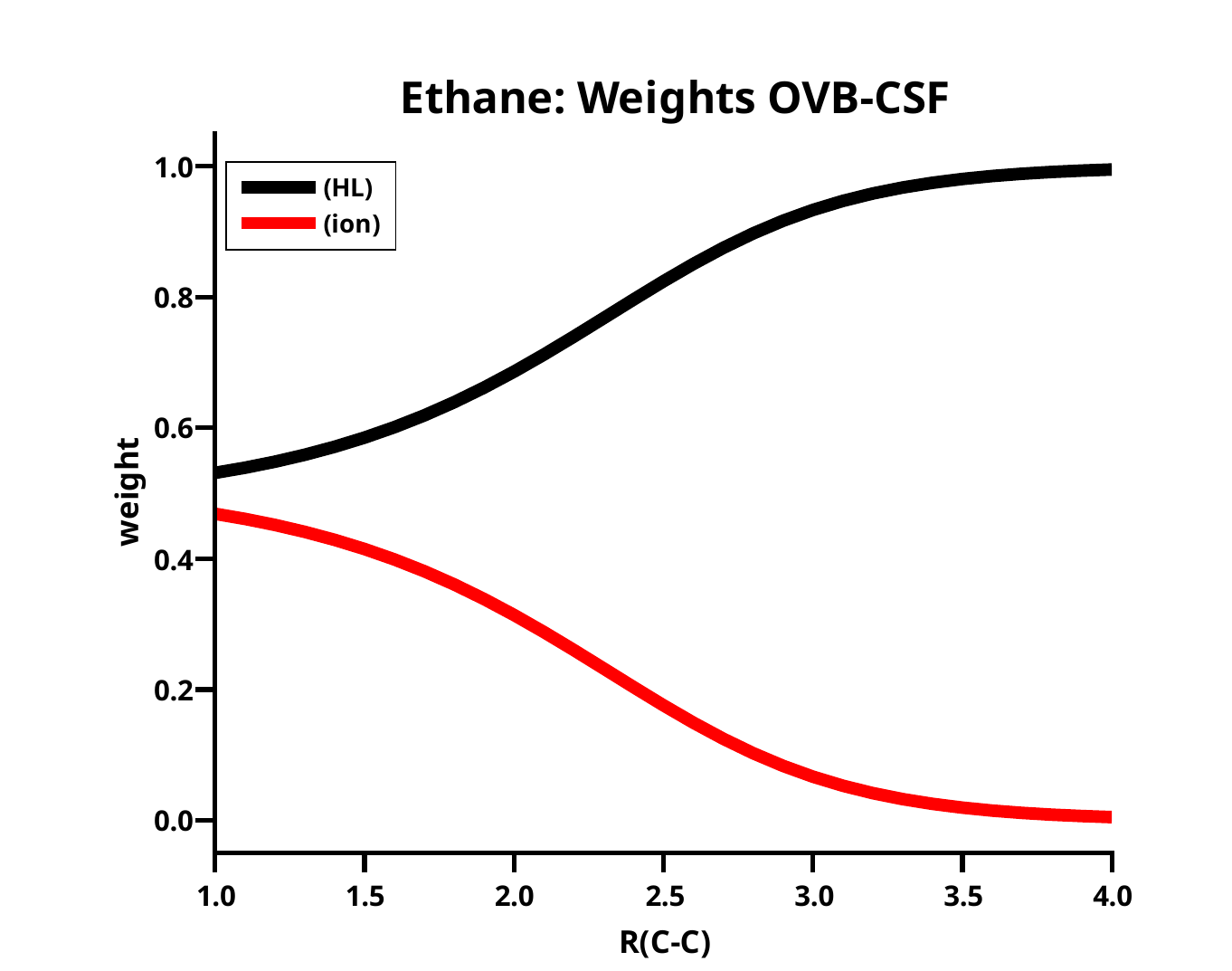}
\end{figure}

The energy curves of $\Phi^o_{\rm HL}$ and $\Phi^o_{\rm ion}$ are completely repulsive, see Figure \ref{fig:CAS22VB}; since McWeeny's early OVB calculations on H$_2$ this is a well known feature of OVB CSFs. The weight curves show that the neutral CSF $\Phi^o_{\rm HL}$ dominates the geometries at long C-C distances but the ionic CSF $\Phi^o_{\rm ion}$ becomes important at shorter distances when polarization and interference cause deviation from the electron distribution of the unperturbed fragments. The ionic CSF $\Phi^o_{\rm ion}$ describes thus a shift of the charge distribution in the covalent bond. When the C-C distance goes to zero the weights of both OVB CSFs become equal. Around the equilibrium geometry, the weight of the neutral CSF $\Phi^o_{\rm HL}$ is larger than that of the ionic CSF $\Phi^o_{\rm ion}$. Comparison of the weight curves of MO CSFs and OVB CSFs show antagonistic behaviour: at long distances, where a single OVB CSF correctly describes the dissociated system, a linear combination of $|\sigma^2|$ and $|{\sigma^*}^2|$ is necessary to do this; at short distances, where $|\sigma^2|$ is a good approximation to the ground state wave function, a linear combination of  $\Phi^o_{\rm HL}$ and $\Phi^o_{\rm ion}$ is necessary to get a qualitative correct ground state wave function. This behaviour is found for all dissociation reactions.

\subsection{The dissociation of the C-C double bond in ethene}
\begin{figure}[ht]
\caption{ \label{fig:CAS44Poto}The potential energy curves calculated with MO CASSCF and the OVB method.}
\includegraphics[width=0.48\textwidth]{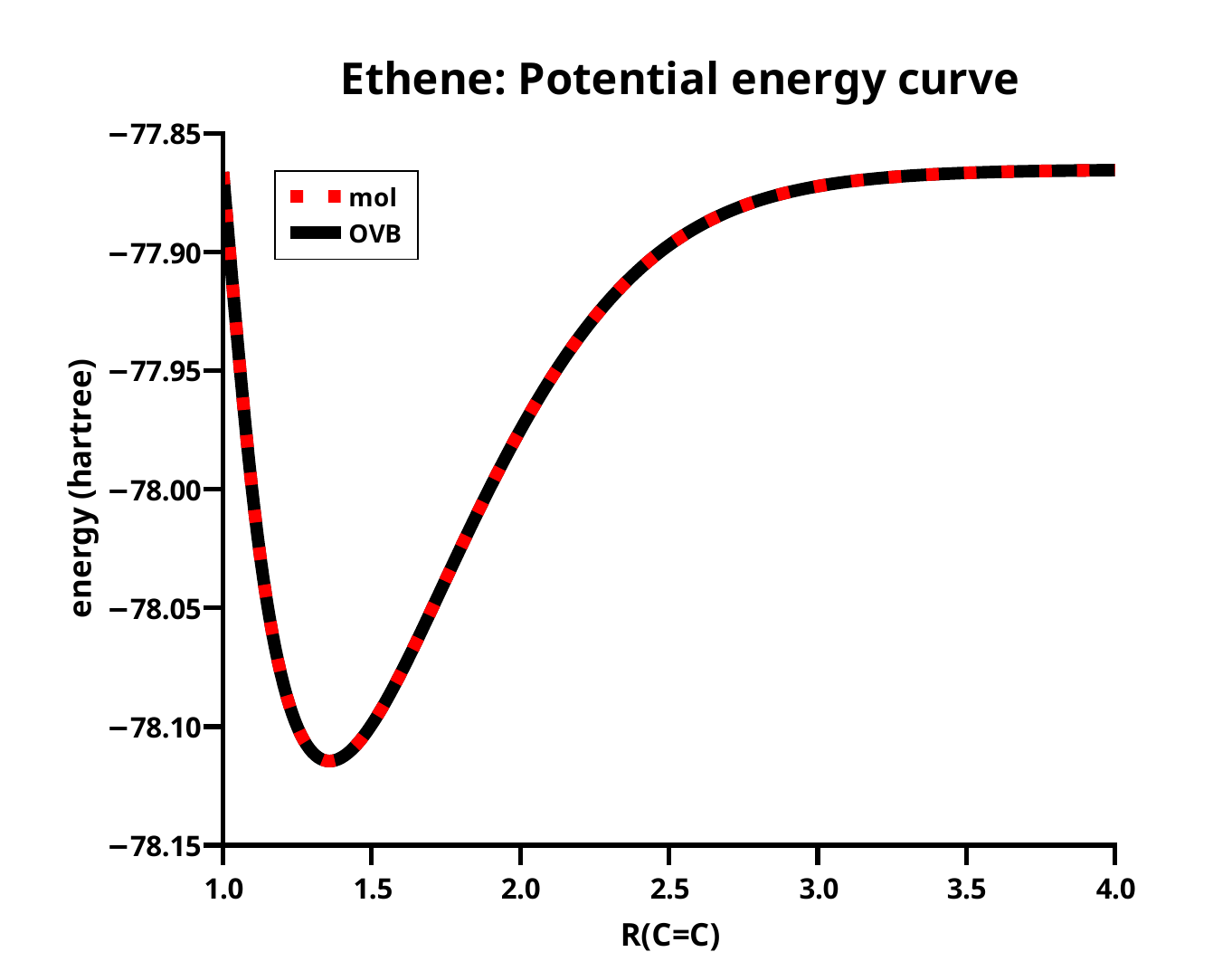}
\end{figure}

The smallest possible wave function that can describe dissociation of the double bond is the CAS(4,4) wave function with the four active MOs $\sigma$, $\pi$, $\sigma^*$, and $\pi^*$, the corresponding electron configuration is $\sigma^2\pi^2$. The HOs used to make the $\sigma$ MO have dominantly z-character, so they are labelled $\rm z_A$ and $\rm z_B$, the $\pi$ MOs are made with  x-OAOs.
Figure \ref{fig:CAS44Poto} shows again that the orthogonal transformation of MOs to FMOs leaves the total energy invariant; the equilibrium C=C distance is about 1.35\,\AA.

The CI space for CAS(4,4) singlet wave functions comprises 20 CSFs, in $D_{2h}$ symmetry only 8 CSFs are totally symmetric.  The order of the four active OAOs in all CSFs is $\rm z_A x_A z_B x_B$.  The following notation is used: $aabb$ means that one $\alpha$ electron occupies the z OAO and one the x OAO on atom A, and $\beta$ electrons occupy the OAOs on  atom B. $a2b0$ means: the z OAO of atom A is singly occupied by an $\alpha$ and the x OAO is doubly occupied, the z OAO of atom B is singly occupied by a $\beta$. The OVB CSF $a2b0$ does not have $D_{2h}$ symmetry, but the positive linear combination with $a0b2$ does. All linear combinations of OVB CSFs are  normalized. In the following, linear combinations of OVB CSFs having the full symmetry of the molecule are labelled LC.

The following 8 linear combinations of OVB CSFs are totally symmetric in $D_{2h}$:
\begin{table}
\caption{\label{tab:CAS44}The totally symmetric linear combinations of singlet OVB LCs for ethene.}
\begin{tabular}{ll}
LC1 & = $\rm |aabb|$\\
LC2 & = $\rm |abba|$\\
LC3 & = $\rm |2020|$\\
LC4 & = $\rm |0202|$\\
LC5 & = $\rm |2002| + |0220|$\\
LC6 & = $\rm |2a0b| + |0a2b|$\\
LC7 & = $\rm |a2b0| + |a0b2|$\\
LC8 & = $\rm |2200| + |0022|$
\end{tabular}
\end{table}

Carbene is a diradical, two electrons occupy two carbon centered, ``nearly-degenerate'' lone pair HOs giving rise to three singlet and one triplet state.\cite{Salem1972} Salem and Rowland classified the two states with singly occupied lone pair orbitals  as diradical, and the two states with doubly occupied lone pair orbitals as zwitterionic. According to our notation, the z OAO represents the sp HO of carbene, and the x OAO the p HO. Using this notation, the four lowest carbene states at the equilibrium geometry, with increasing energy, are $\rm ^3(zx)$, $\rm z^2$, $\rm ^1(zx)$, and $\rm x^2$.

LC1 to LC5 are neutral, LC6 and LC7 are singly ionic, and LC8 is a doubly ionic LC. LC1 describes the two carbenes in their  triplet ground states, coupled to a singlet. The two electrons in the z OAOs form the $\sigma$ bond, the two electrons in the x OAOs form the $\pi$ bond, LC1 is nothing but the Heitler-London portion of the $\sigma$ and the $\pi$ bonds. Since the carbenes are in high spin states, the unpaired electrons are ``ready for bonding''. In the dissociated molecular system, LC1 describes two noninteracting carbenes  in their respective electronic ground states; at all other geometries the carbenes  are no longer in a carbene eigenstates, because interacting subsystems of a system are never in pure states but always in mixed states.\cite{Cohen1977,Sax2022} In these cases, LC1 describes two ``local triplet states'', that is,  ``local high-spin states'', coupled to a singlet. This is, what the Heitler-London CSF represents. That ``local low-spins states'' are rather unimportant for bonding shows LC2, where each carbene is in the singlet diradical state, which is  considerably higher than the triplet diradical state. Moreover, the spins are not unpaired and therefore not ``ready for bonding'' although the same AOs are singly occupied as in case of LC1. LC3 describes two carbenes both with doubly occupied z OAOs, LC4 describes two carbenes with doubly occupied x OAO; in both CSFs the active electrons are singlet coupled and therefore not ``ready for bonding'', the contributions of these CSFs to the ground state wave function are accordingly very small.
The singly ionic LC6 describes the shift of one electron in the $\sigma$ bond, the $\pi$ MO is doubly occupied; the singly ionic LC7 describes the shift in the $\pi$ bond with doubly occupied $\sigma$ MO. These two singly ionic LCs are necessary to describe polarization in the $\sigma$ and the $\pi$ orbital, respectively. Without them covalent bonding cannot be correctly described. The neutral LC5  describes local angular correlation: If atom A is in the low lying zwitterionic carbene state, atom B is in the high lying zwitterionic state. This is the fourth LC that contributes significantly to the ground state of ethene. LC8 describes dianion/dication pairs.

\begin{figure}[ht]
\caption{ \label{fig:CAS44VB}Energies (left) and weights (right) of the large LCs for ethene.}
\includegraphics[width=0.48\textwidth]{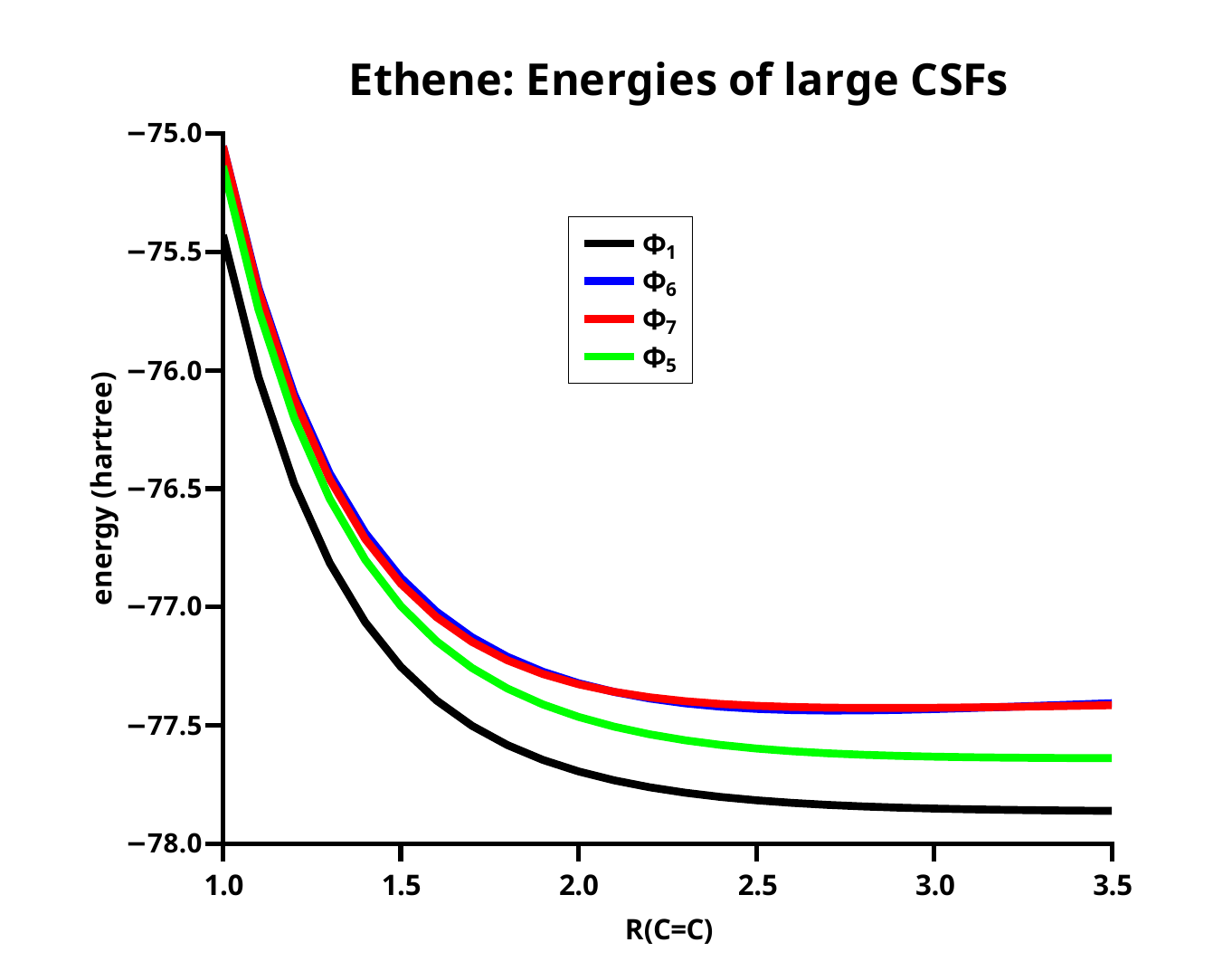}\hfill
\includegraphics[width=0.48\textwidth]{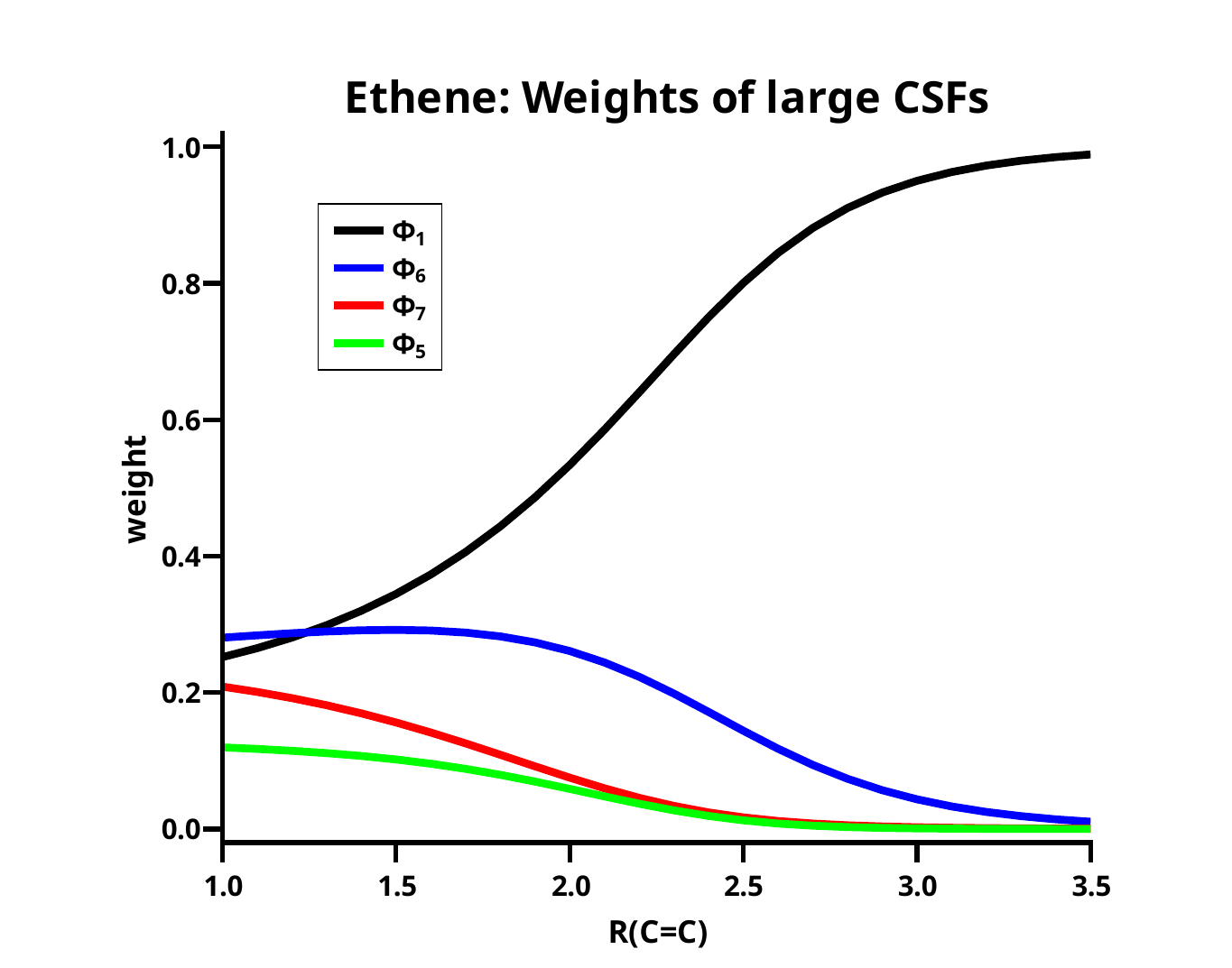}
\end{figure}

\begin{figure}[ht]
\caption{ \label{fig:CAS44sumw}The sum of the weights of the large  LCs.}
\includegraphics[width=0.48\textwidth]{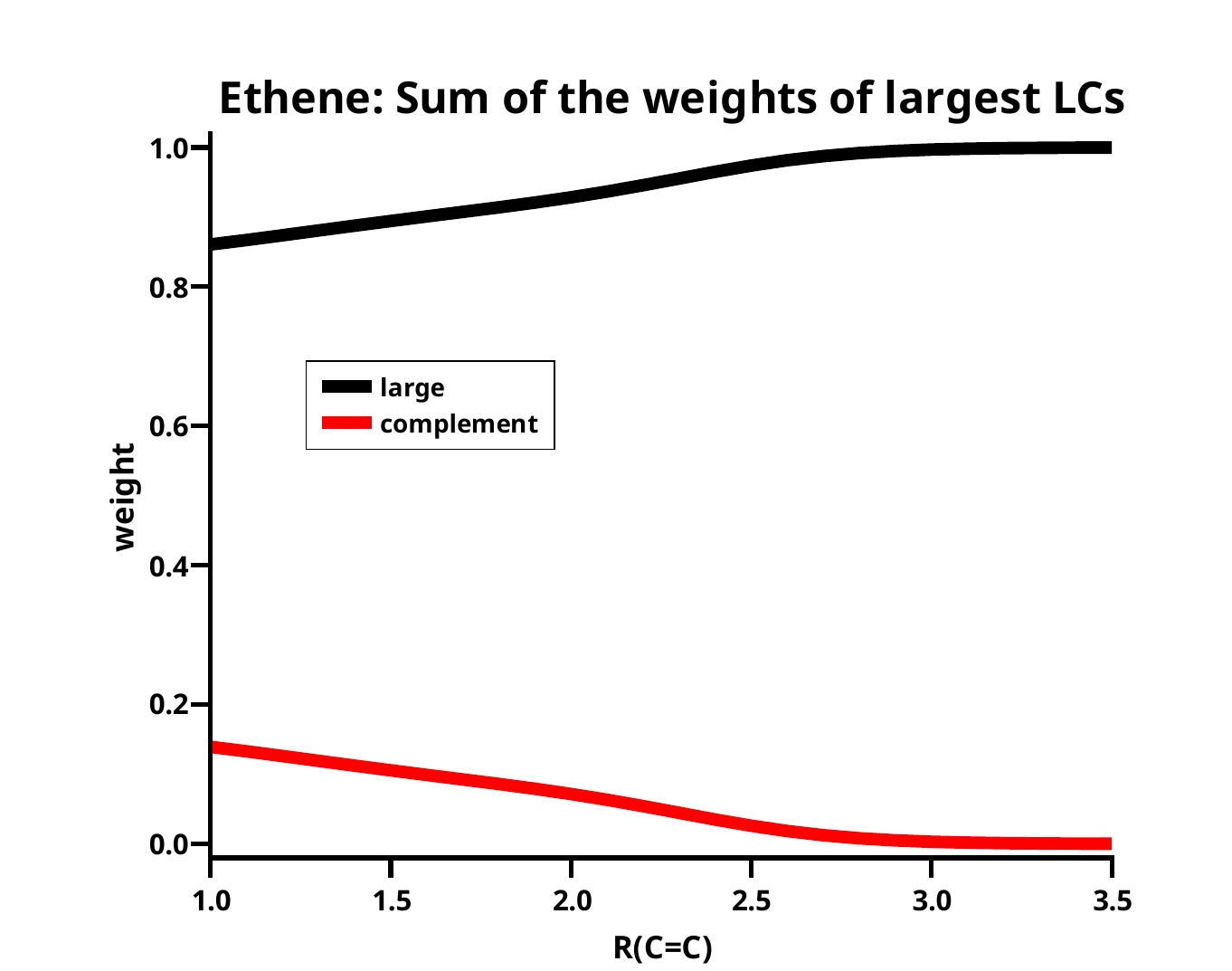}
\end{figure}

Figure \ref{fig:CAS44VB} shows energies and weights of the four large LCs, which are LCs  having a weight larger than 0.1 somewhere along the reaction coordinate. All other LCs are small LCs. LC1 has the lowest energy along the whole reaction coordinate, this demonstrates the importance of the coupling local high-spin states to a global low spin state. Although the ionic LCs LC6 and LC7 have nearly identical energies their weights are very different. The weight of LC6 reaches a value of 0.1 already at a C-C distance of 2.7\,\AA, but that of LC7  only at 1.8\,\AA. LC5 becomes important only when the two carbon atoms are rather close,  the weight is larger than 0.1 only at C-C distances shorter than the equilibrium distance. Note, that LC6 and LC7 have higher energies than LC5 but their weights are much larger. LC6 and LC7 demonstrate the importance of the charge shift for covalent bonding and LC5 that angular correlation becomes important in multiple bonds as soon as the interacting atoms come close. It is noteworthy that, at the equilibrium distance, the weight of the neutral LC1 is 0.32, only slightly larger than the weight of the ionic LC6 (0.29), whereas LC7 has a weight of only 0.17, nevertheless, the sum of the weights of the ionic LCs is much larger than the sum of the two neutral ones.
At C-C distances longer than 3.5\,\AA, the weight of LC1 is 1.; between 3.5 and 2.7\,\AA\, the weight of LC1 decreases and that of LC6 increases, but the sum of both LCs is still close to 1. Then LC7 and LC5 become gradually more important, but the sum of all four large LCs nevertheless decreases down to 0.85 at a C-C distance of 1.0\,\AA. At the same time the weight of the four small LCs increases to 0.15, see Figure \ref{fig:CAS44sumw}, nevertheless,  at the equilibrium distance the sum of their weights is only  0.11.

\subsection{The dissociation of the N-N triple bond in the nitrogen molecule}
The electron configuration of the nitrogen molecule is $\sigma_p^2\pi^4$, where $\sigma_p$ is the bonding linear combination of $\rm z_A$ and $\rm z_B$; $\pi^4$ means $\pi_x^2\pi_y^2$, $\pi_x$ and $\pi_y$ are bonding linear combinations of the respective OAOs.
\begin{figure}[ht]
\caption{ \label{fig:CAS66Poto}The potential energy curves calculated with MO CASSCF and the OVB method.}
\includegraphics[width=0.48\textwidth]{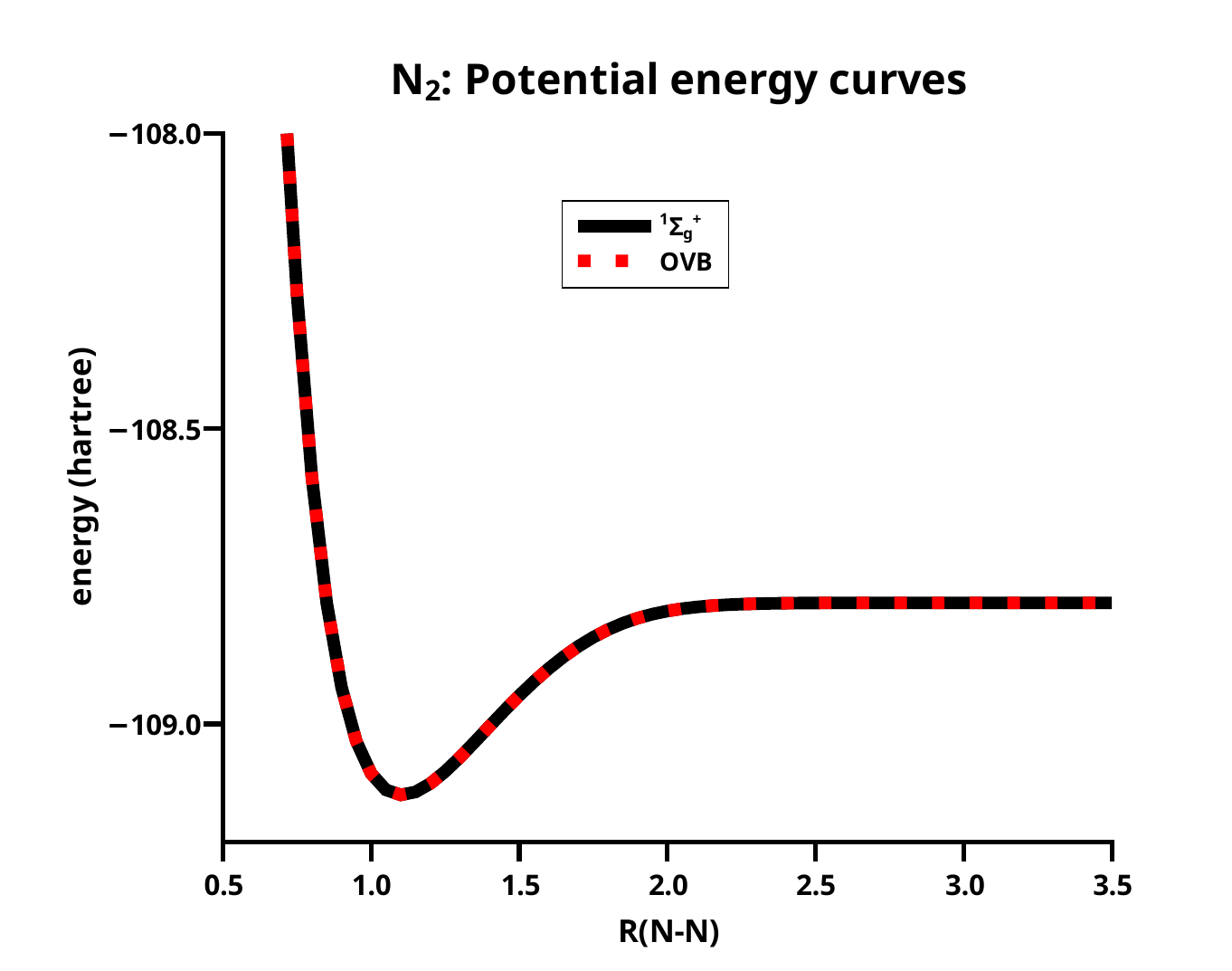}
\end{figure}
Figure \ref{fig:CAS66Poto} shows the potential energy curves obtained with MO CSFs and OVB CSFs, the equilibrium distance of N$_2$ is 1.1\,\AA.

\begin{figure}[ht]
\caption{\label{fig:CAS66VB}Energies (left) and weights (right) of the five large LCs.}
\includegraphics[width=0.48\textwidth]{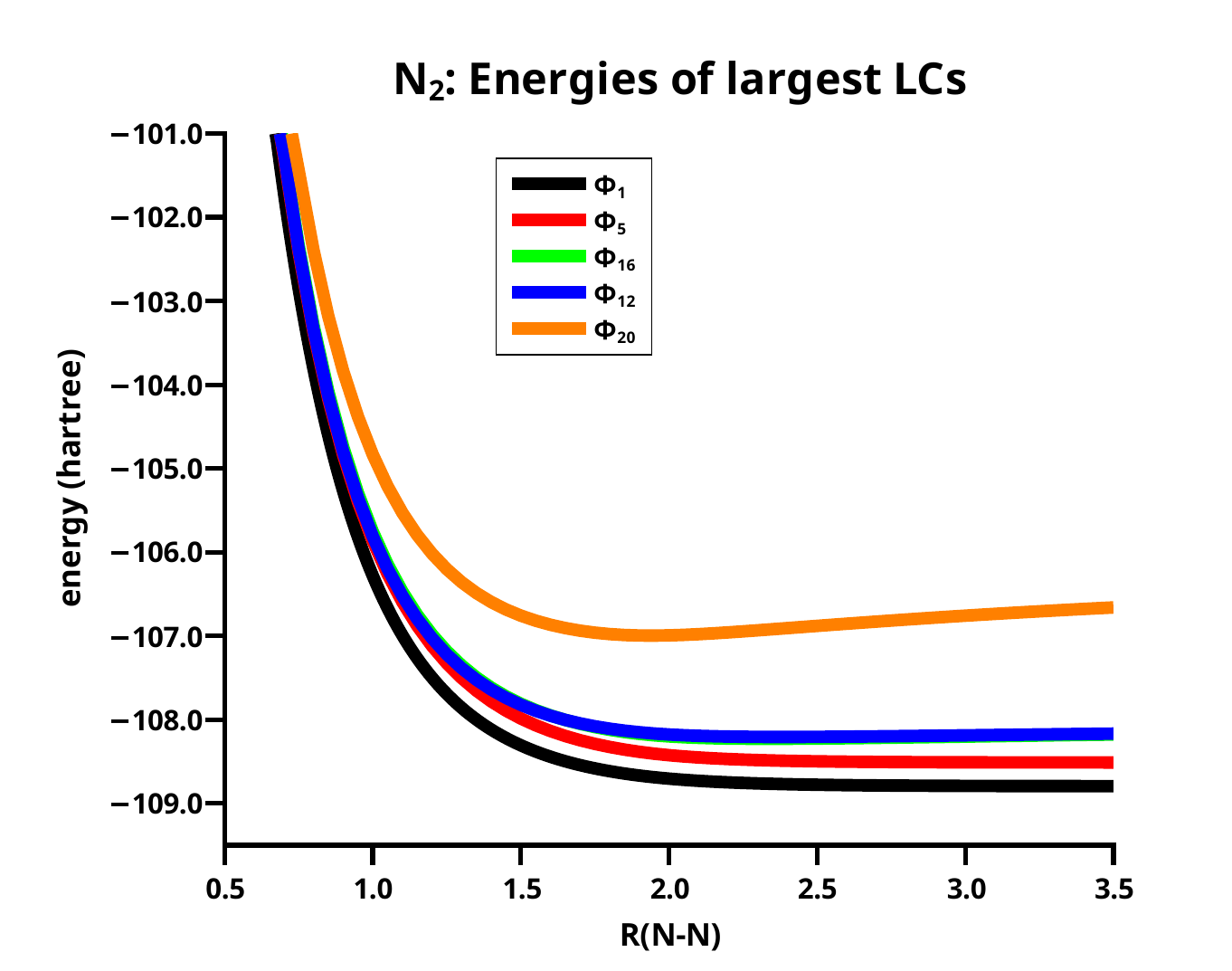}\hfill
\includegraphics[width=0.48\textwidth]{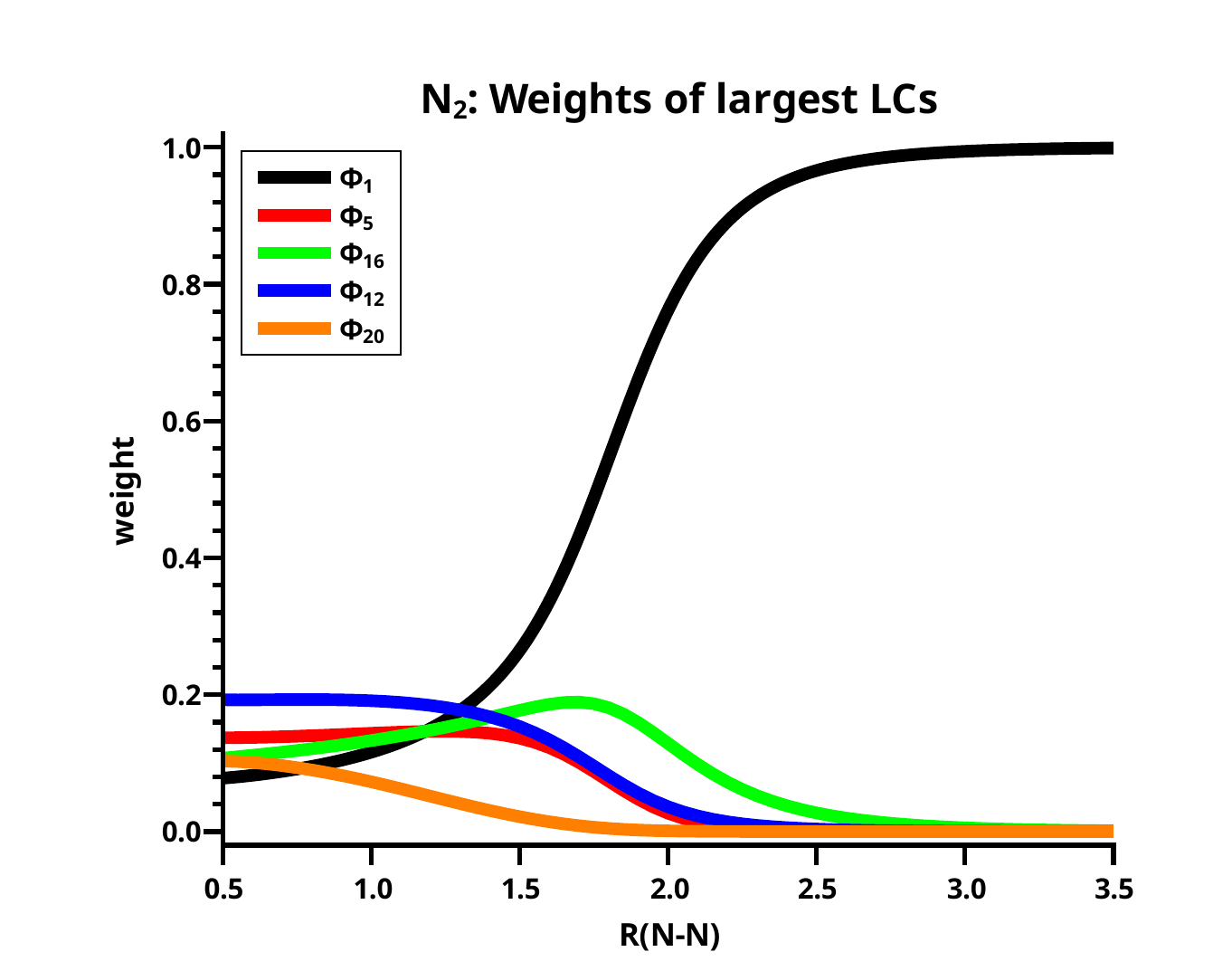}
\end{figure}

The dissociation reaction in the lowest $^1\Sigma_g^+$ state is correctly described by a CAS(6,6) wave function with the six valence MOs $\sigma_p$, $\pi_x$ and $\pi_y$ and the corresponding antibonding MOs; the 2s AOs are always doubly occupied. In $D_{2h}$, there are 32 totally symmetric singlet CSFs, 20 of them are not zero. The OVB calculation is done in $C_{2v}$, from the 55 OVB CSFs only 21 linear combinations are not zero, they are shown in Table \ref{tab:CAS66}. The order of the six OAOs in the CSFs is $y_A,x_A,z_A,y_B,x_B,z_B$. From the 21 linear combinations of OVB CSFs, LC1 to LC9 are neutral, LC10 to LC18 are singly ionic, LC19 and LC20 are doubly ionic, and LC21 is triply ionic.
\begin{table}
\caption{\label{tab:CAS66}The totally symmetric linear combinations of singlet LCs for N$_2$.  }
\begin{tabular}{ll}
LC1  &= $\rm |aaabbb|$\\
LC2  &= $\rm |ababab|$\\
LC3  &= $\rm |baaabb|$\\
LC4  &= $\rm |baabab| + |abaabb|$\\
LC5  &= $\rm |a02b20| - |0a22b0| - |2a00b2| + |a20b02|$\\
LC6  &= $\rm |0a20b2| + |a02b02|$\\
LC7  &= $\rm |02a02b| + |20a20b|$\\
LC8  &= $\rm |20a02b| + |02a20b|$\\
LC9  &= $\rm |a20b20| + |2a02b0|$\\
LC10 &= $\rm |202002| + |022002| + |002022| + |002202|$\\
LC11 &= $\rm |2ba0ab| - |b2aa0b| - |b0aa2b| + |0ba2ab|$\\
LC12 &= $\rm |2aa0bb| - |a2ab0b| - |a0ab2b| + |0aa2bb|$\\
LC13 &= $\rm |220020| + |220200| + |200220| + |020220|$\\
LC14 &= $\rm |202020| + |022200| + |200022| + |020202|$\\
LC15 &= $\rm |022020| + |202200| + |020022| + |200202|$\\
LC16 &= $\rm |aa2bb0| + |aa0bb2|$\\
LC17 &= $\rm |ab2ba0| + |ab0ba2|$\\
LC18 &= $\rm |002220| + |220002|$\\
LC19 &= $\rm |22a00b| + |00a22b|$\\
LC20 &= $\rm |2a20b0| - |a22b00| - |a00b22| + |0a02b2|$\\
LC21 &= $\rm |222000| + |000222|$
\end{tabular}
\end{table}
From these LCs  only five are large: LC1,  LC5, LC12, LC16, and LC20. See Figure \ref{fig:CAS66VB}.

The neutral LC1 describes the singlet coupling of the high spin quartet states, it is the Heitler-London contribution to the triple bond in N$_2$; the neutral LC5 describes  angular correlation and a spin flip from the local quartet state into the low-spin doublet state; the singly ionic LCs LC{16} and LC{20} describe the shift of electrons in the $\sigma$ and in the $\pi$ MOs, respectively. The doubly ionic LC20 describes the simultaneous shift of an electron in the $\sigma$ and an electron in the $\pi$ MOs. LC1, LC16 and LC20 are of major importance for the description of the triple bond in N$_2$, the two ionic LCs have, as found for the ethene molecule, nearly identical energies, but the weights are rather different. LC{16} becomes important at an N-N distance of about 2.7\,\AA, and LC{20}  at a distance of about 2.2\,\AA. The reason for this are the different spatial extensions of the involved AOs: the z AOs, which are aligned along the molecular axis, interfere earlier than the perpendicular x and y AOs, and therefore  charge shift in the $\sigma$ bond starts earlier than in the $\pi$ bonds.

Ten of the remaining 16 LCs are small and six are effectively zero.
\begin{figure}[ht]
\caption{ \label{fig:CAS66sumw}Left: Weights of the small LCs. Right:  Sum of the weights of the large and of the small LCs.}
\includegraphics[width=0.48\textwidth]{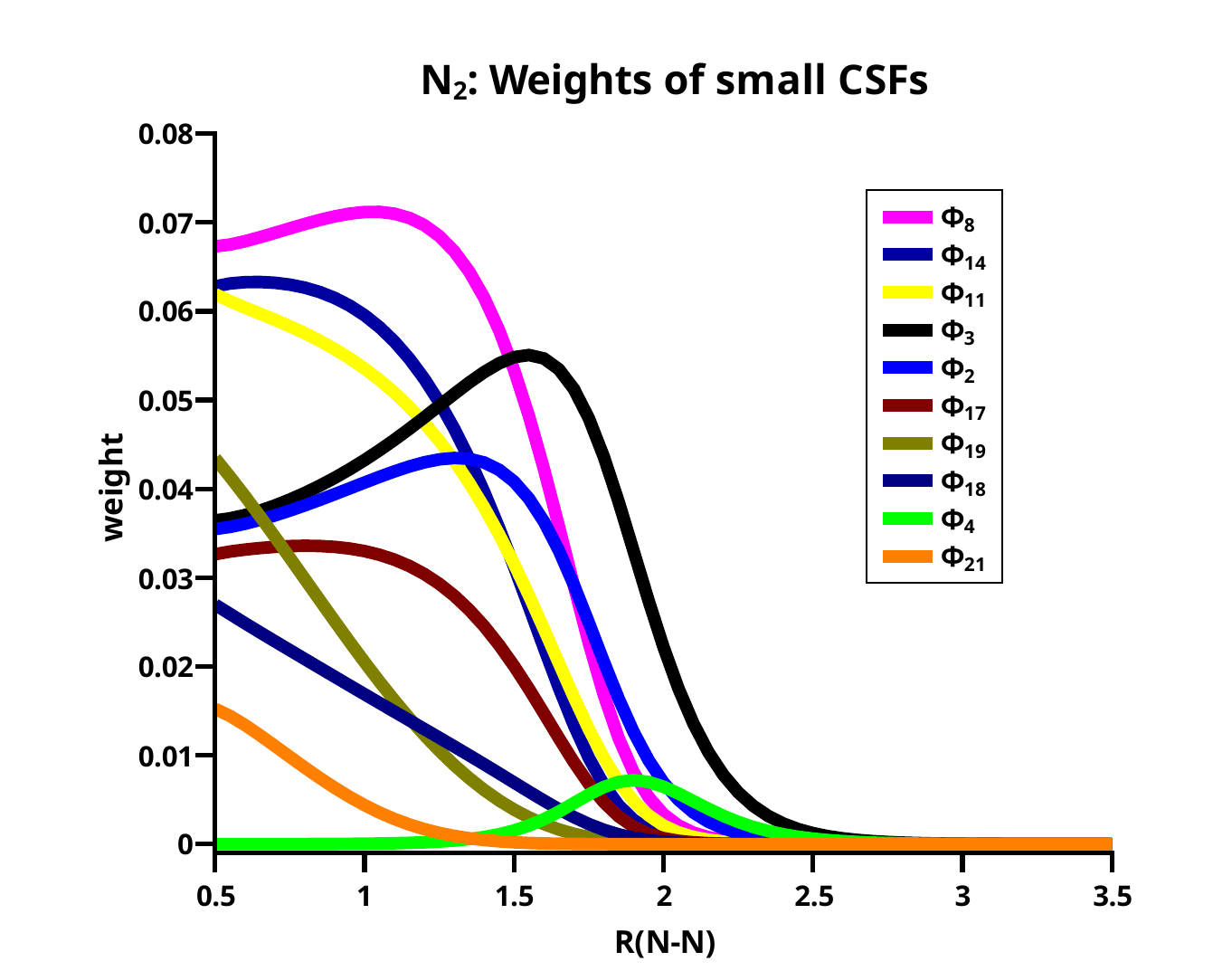}
\includegraphics[width=0.48\textwidth]{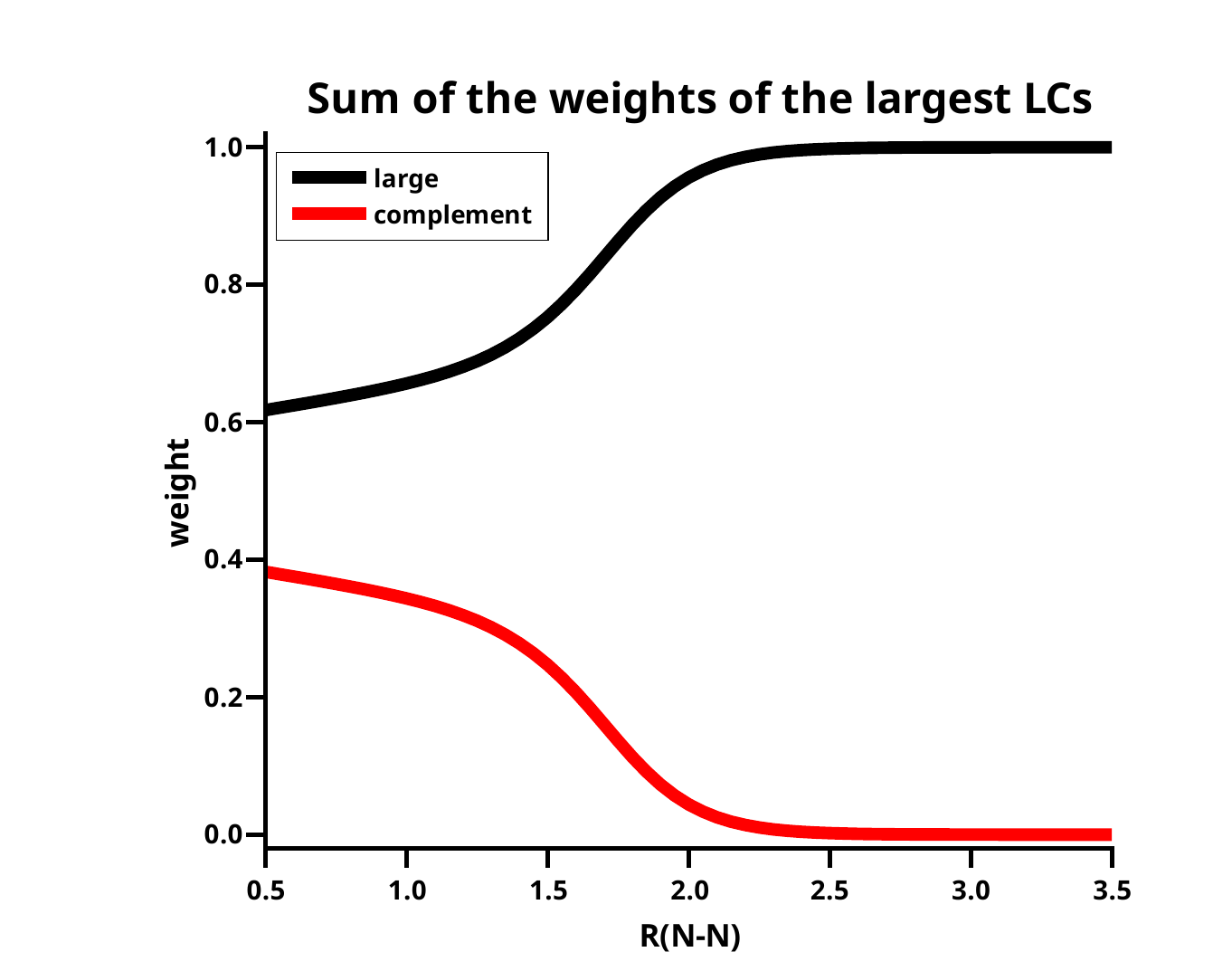}
\end{figure}

The weights of all small LCs are effectively zero for N-N distances longer than 2.5\,\AA, but, as for the large ionic LCs, most of them contribute significantly only at N-N distances smaller than 2.0\,\AA. LC2, LC3, and LC8 are neutral, LC{19} and LC{21} are doubly and triply ionic, respectively, all others are singly ionic. All small LCs are important in describing the deviation of the  atomic charges from  spherical symmetry  during bonding.

\section{Bonding in C$_2$}
\begin{figure}[ht]
\caption{ \label{fig:CAS88PEC}Left: Potential energy curves of the lowest $^1A_g$  states calculated with CAS(4,4), CAS(4,6), and CAS(8,8) wave functions. Right: The CAS(8,8) potential energy curves for the three lowest $^1A_g$   states.}
\includegraphics[width=0.48\textwidth]{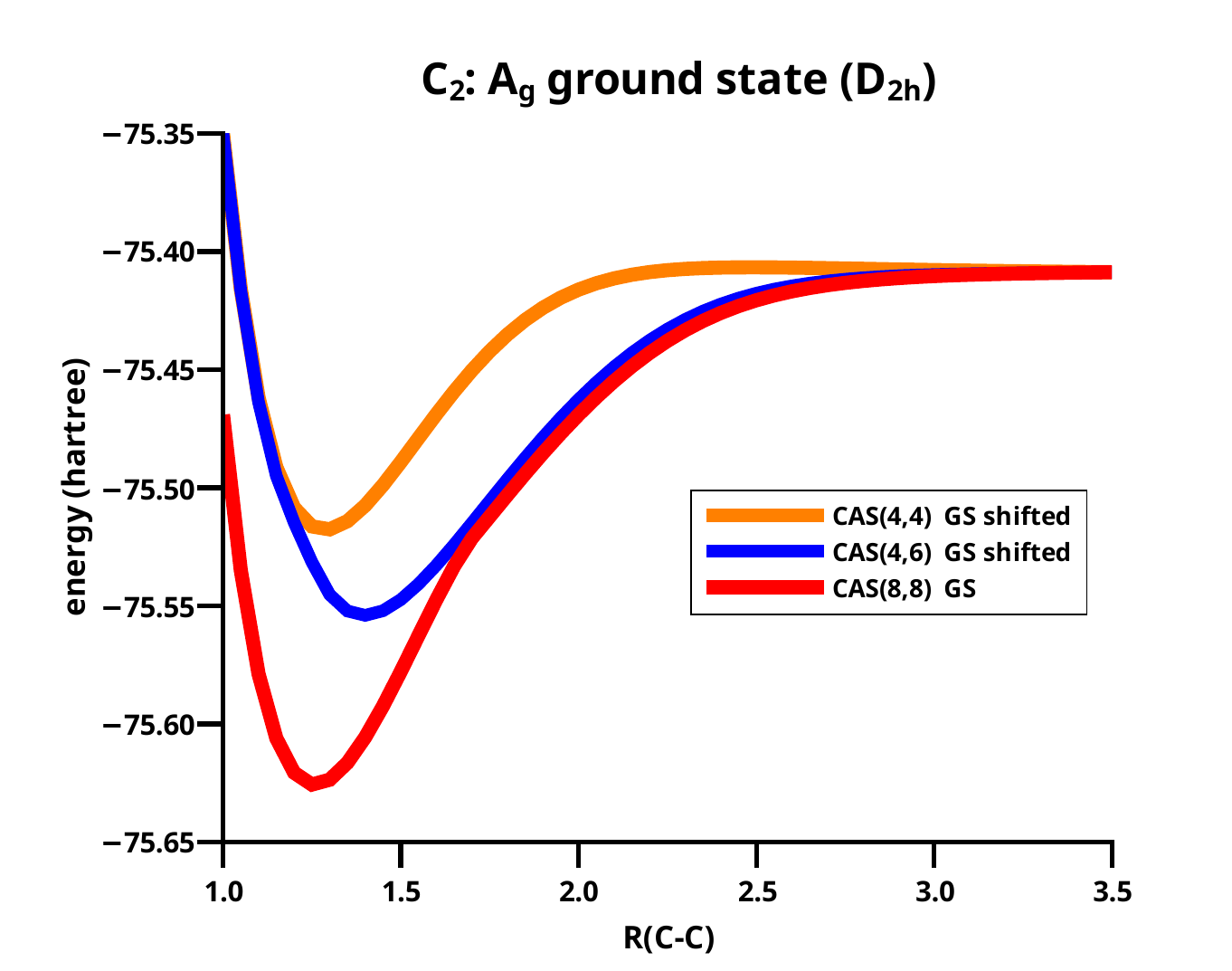}
\includegraphics[width=0.48\textwidth]{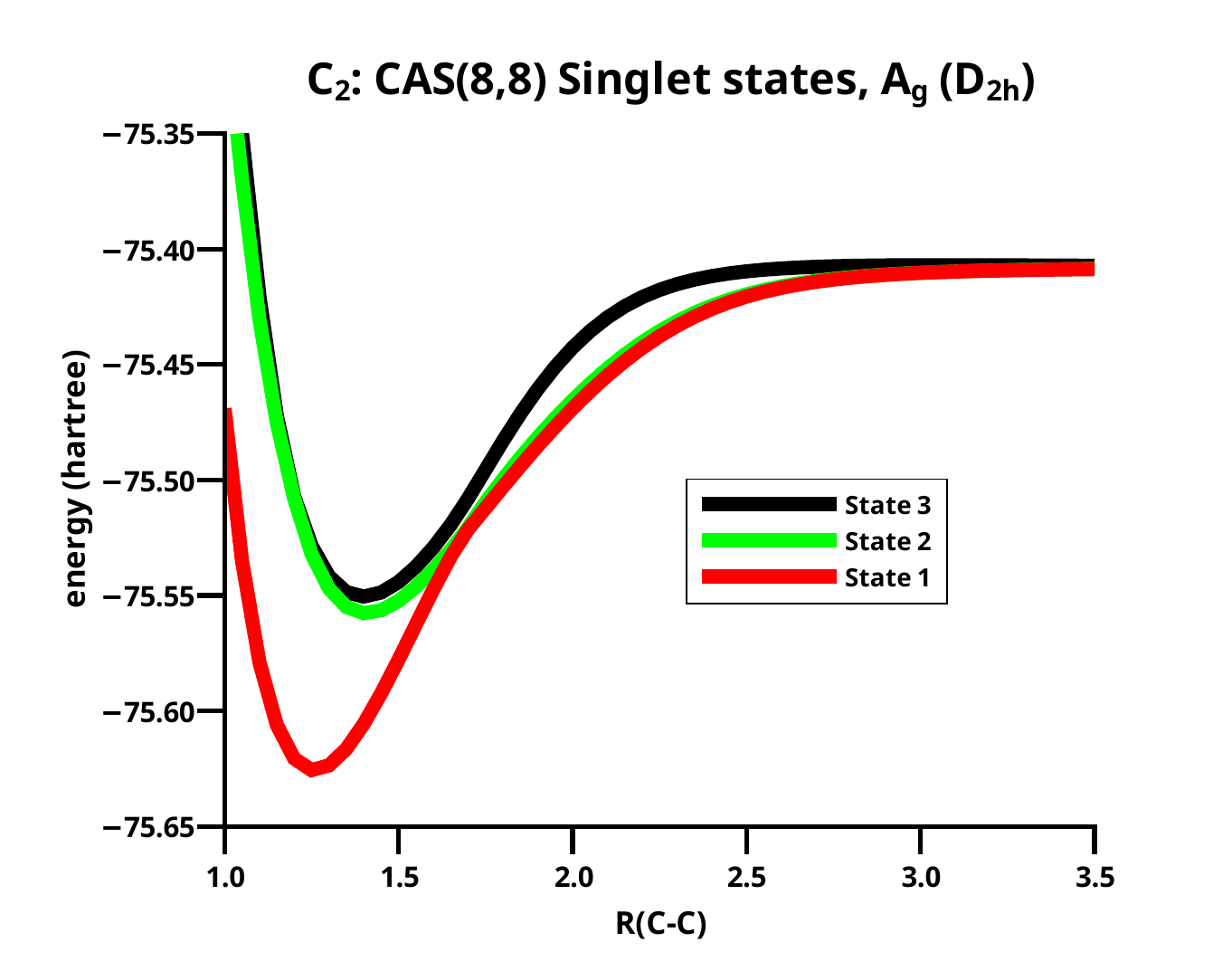}
\end{figure}

The equilibrium distance in the singlet ground state is about 1.25\,\AA.
If one assumes that the electron configuration of C$_2$ is similar to that in  N$_2$, it must be $\sigma_s^2 \sigma_s^{*2} \pi^4$, the bonding $\sigma_s$ and the antibonding $\sigma_s^*$ MOs spanned by the 2s AOs are doubly occupied and the remaining four valence electrons occupy the two bonding $\pi$ MOs. All four MOs are non-active and the wave function is a single Slater determinant. This wave function is not able to describe dissociation, the most simple wave function that can do this is a CAS(4,4) wave function with four active MOs, two bonding and two antibonding $\pi$ MOs, and four active $\pi$ electrons. If one considers that the bonding $\sigma_p$ MO has a lower energy than the bonding $\pi$ MOs, one gets a second possible electron configuration:  $\sigma_s^2 \sigma_s^{*2} \sigma_p^2 \pi^2$. The $\sigma_p$ MO is non-active but the doubly degenerate $\pi$ MOs are occupied by only two electrons, and thus active MOs. The wave function corresponding to this electron configuration are CAS(2,2) wave functions, which cannot describe dissociation, because they contain no antibonding MOs, but CAS(4,6) wave functions with all six MOs spanned by 2p AOs as active MOs can do it. In Figure \ref{fig:CAS88PEC} one can see that the stabilization of the ground state as calculated with both CAS(4,4) and CAS(4,6) are far too low, the equilibrium distance obtained with CAS(4,6) is considerably longer than that obtained with CAS(4,4). The long equilibrium distance stems from the fact that with CAS(4,6) the ground state is a $^1\Delta_g$ whereas with CAS(4,4) it is a $^1\Sigma_g^+$ state, but the poor stabilization indicates that  wave functions without active $\sigma_s$ and $\sigma_s^*$ MOs cannot describe the ground state correctly, the two $\sigma_s$ MOs and the four electrons must become active. Then one has eight active MOs and eight active electrons and with such a CAS(8,8) wave function the lowest singlet state is indeed the ground state of C$_2$ and the stabilization energy is reliable.

On the other hand, the shape of the ground state PEC as calculated with s CAS(8,8) wave function indicates avoided crossings that are not found with the two smaller CAS wave functions. See left side of Figure \ref{fig:CAS88PEC}.

As mentioned above, the avoided crossings are the result of making all CASSCF calculations in $D_{2h}$ symmetry. The totally symmetric CAS(8,8) singlet wave function of $A_g$ symmetry is a linear combination of 264 CSFs. The ground state wave function is indeed a mixture of $^1\Sigma_g^+$ and $^1\Delta_g$ states, which results in avoided crossings, as can be seen for the three lowest singlet states investigated. (Details of the calculations with MO CSFs can be found in the Supporting Informations.) The PECs of the three $A_g$ states agree very well with those reported by Boschen \etal.\cite{Boschen2014} That means that the wave functions for the states have different character in certain regions, e.g., the ground state has $\Sigma_g^+$ character around the equilibrium geometry, the second $A_g$ state has there $\Delta_g$ character, and the third $A_g$ state has again $\Sigma_g^+$ character; the second and the third state are energetically very similar and also the local minima have similar geometries. See right side of Figure\ref{fig:CAS88PEC}.

There is an ongoing debate on the bond order in C$_2$, see for example Ref.\cite{zhao2019}. The concept of bond order in MO theory is based on the system's electron configuration at the equilibrium geometry, an MO can be occupied by two, one or zero electrons. The half difference of the number of valence electrons occupying bonding MOs and the number of electrons occupying antibonding MOs is the bond order, in general, it is an integer but fractions of integer are also possible if degenerate MOs are not fully occupied or the system is ionized. According to this recipe, the bond order corresponding to electron configuration $\sigma_s^2 \sigma_s^{*2} \pi^4$ is 2, for $\sigma_s^2 \sigma_s^{*2} \sigma_p^2 \pi^2$ it is also 2,  and for $\sigma_s^2 \sigma_p^2 \pi^4$ it is 4. But, if the MOs are energetically close the order of the MOs may change when  the  geometry changes giving different electron configurations and bond orders. In such situations multiconfigurational wave functions must be used, for which no bond order is defined. The ground state  has $\Sigma_g^+$ character for C-C distances shorter than 1.70\,\AA, there the CSF derived from the electron configuration $\sigma_s^2 \sigma_s^* \pi_x^2 \pi_y^2$ contributes about 70 percent to the wave function, and the  CSF derived from $\sigma_s^2 \sigma_p^2 \pi_x^2 \pi_y^2$ contributes about 10 percent. Ignoring all  CSFs that contribute in total 20 percent to the wave function, what bond order has a system with 70 percent bond order 2 and 10 percent bond order 4? In my opinion, more important than to answer this question is to find out how spins and charges rearrange on the way from two isolated atoms to the molecule, that is, how the interaction of the atoms disturbs their electron distributions during the recombination reaction, or how the electron and spin arrangement in the molecule readjust to that in the free atoms during dissociation.

\subsection{OVB analysis of bonding in C$_2$}
Every atom in a homonuclear diatomic molecule has $C_{\infty v}$ symmetry, but all actual calculations are made in the Abelian subgroup $C_{2v}$ of $D_{2h}$. With the eight OAOs $\rm s_A, z_A, x_A, y_A,$ $\rm  s_B, z_B, x_B, y_B$ 492 singlet OVB CSFs can be made and only very few OVB CSFs have already $g$ parity or have rotational symmetry, more often than not only LCs have it. This means, a large number of OVB CSFs have zero weight because of symmetry reason. But LCs can also gain zero weight when the molecule's geometry changes, the number of LCs with non-zero weight depends strongly on the geometry,  it is nevertheless rather large. As a consequence, many small LCs can make considerable contributions, and if only large LCs are considered, the description of the wave functions is not satisfying because the weights of many large LCs can be rather small at certain geometries. To consider also small LCs that make, nevertheless, large contributions to the wave functions, significant LCs were defined as those having weights larger than 0.01 somewhere along the reaction coordinate. 128 significant LCs are found to describe the lowest three $A_g$ singlet states along the whole reaction coordinate, in detail, 51 significant LCs are found to contribute to the first $^1A_1$ state (the ground state), 63 LCs to the second state, and 80 LCs to the third singlet state; only 15 LCs of the significant LCs are large. From these LCs, ten contribute to the description of the first $^1A_1$ state, and eight LCs contribute to the second and to the third state, respectively. Details may be found in the Supporting Information.

Table \ref{tbl:table1} lists the ten large LCs found for state 1 along the reaction coordinate. Starting at long C-C distances one can see that only two LCs are important, LC05 ($\approx$ 75 percent) and LC06 ($\approx$ 20 percent). Both LCs describe the singlet coupling of two carbon atoms in their respective $^3P_g$ ground states, in LC06 the electrons are located in the x and the y OAO, therefore LC06 describes the formation of two $\pi$ bonds. The situation where in each atom one electron occupies the z and the other either the x OAO or the y OAO is described by LC05. Singlet coupling  gives then either a $\sigma_p$ and a $\pi_x$ bond or a $\sigma_p$ and a $\pi_y$ bond. The positive linear combination of these two LCs has $\Sigma_g^+$ symmetry and is represented by LC05; the negative linear combination has  $\Delta_g$ symmetry and is represented by LC01.  Both $2s$ OAOs are always doubly occupied. At very long C-C distances, the two carbons atoms are completely interaction-free, the three ways of distributing two spins in three p orbitals are equivalent, the weight of LC05 is 2/3, and that of LC06 is 1/3, but when the atoms approach each other, the interaction along the molecule axis becomes more favorable so that the weight of LC05 increases. At a C-C distance of 3.5\,\AA, the weights of LC05 and LC06 are indeed 0.74 and 0.21, respectively; at  3.0\,\AA\, the weights are 0.78 and 0.14, respectively. At these distances, the missing LCs that describe either polarization in direction of the molecular axis or superposition of the $2p_z$ orbitals are not represented by large but only by significant LCs.

\begin{figure}[ht]
\caption{ \label{fig:CAS88St1largeLCs}Left: The distribution of the weights of the large LCs for the first $A_1$ singlet state. Right: The sum of the large weights and the complement.}
\includegraphics[width=0.48\textwidth]{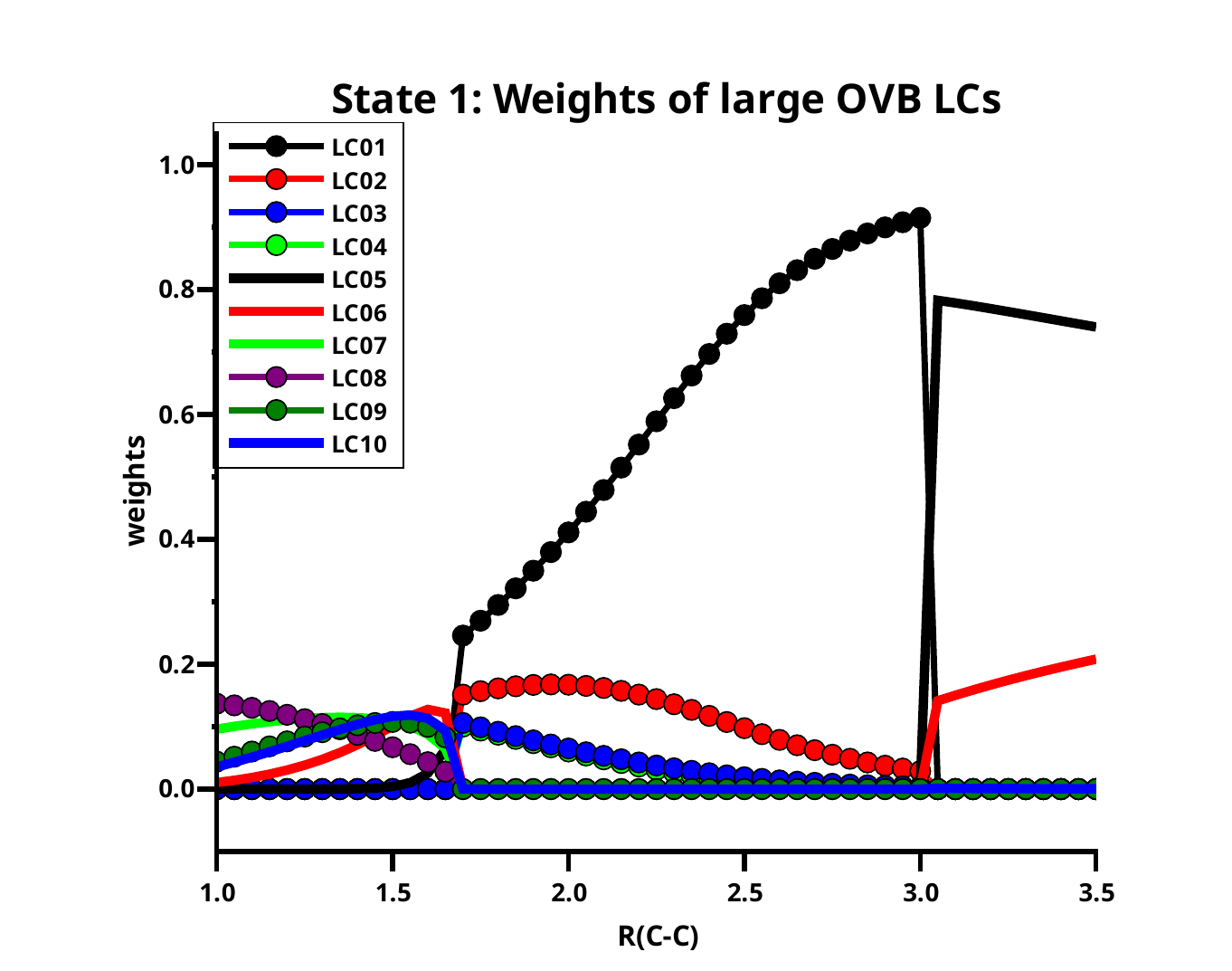}
\includegraphics[width=0.48\textwidth]{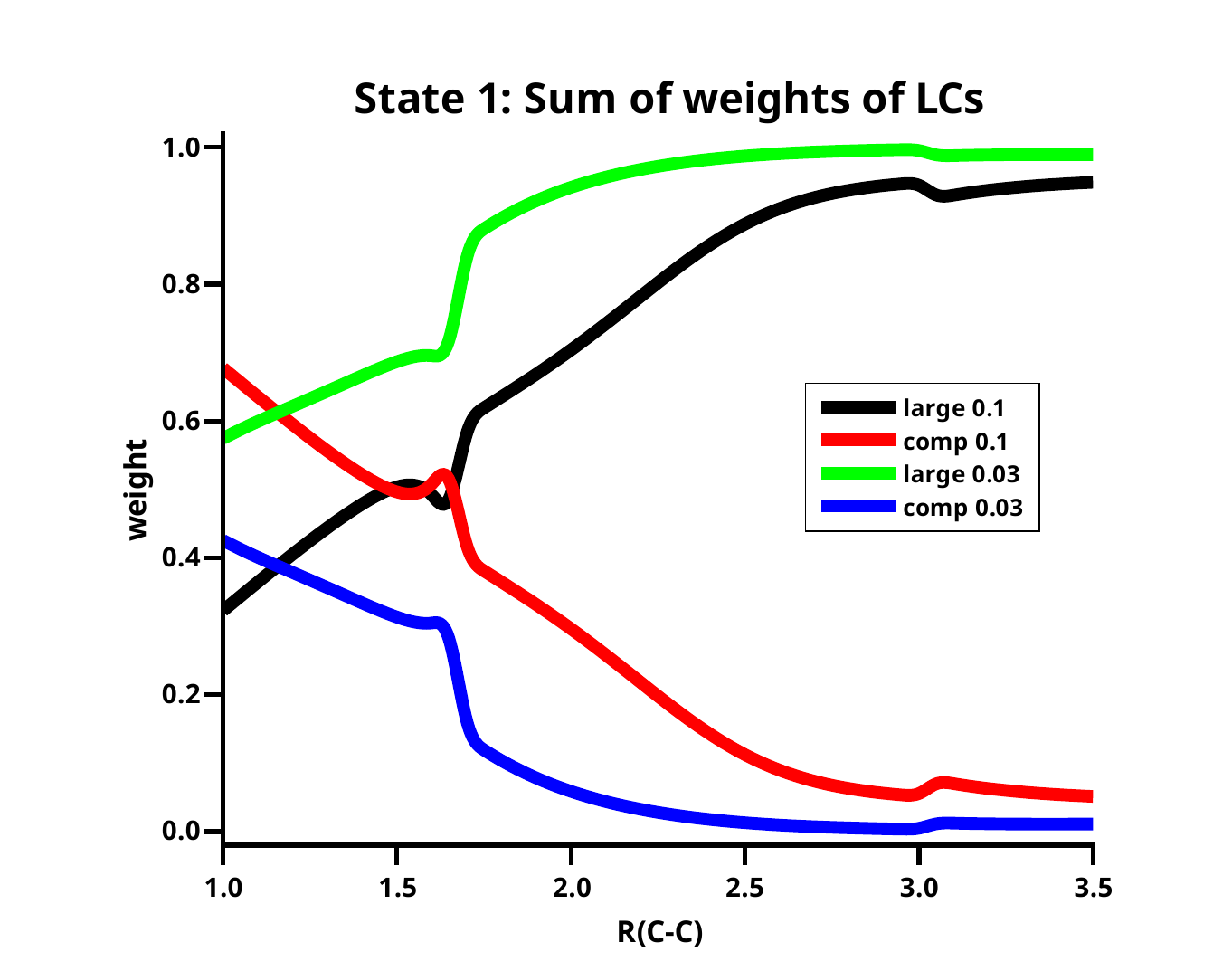}
\end{figure}

At C-C distances less than 3.0\,\AA, linear combinations LC01 to LC04 with $\Delta$ character  dominate the ground state. LC01 and LC03 describe neutral atomic charge distributions, LC02 and LC04 describe cation/anion pairs. LC01 describes  a $\sigma_p$ and a $\pi$  bond, it is the counterpart of LC05 with  different phase. LC01 describes the neutral Heitler-London component of the C-C $\sigma$ bond and $\pi$ bond, LC02 is the ionic component of the  $\sigma$ bond. The ionic LC04 describes deformations of the interacting C atoms caused by polarization in the $\sigma$ bond due to s-p hybridization. Polarization in the $\sigma$ bond is also described by the neutral LC03. For C-C distances less than 1.7\,\AA, the wave function is again dominated by LCs with \S\, character, the large LCs that are important at long distances contribute very little, LC05 is essentially vanished and LC06 goes to zero rapidly; the wave function is dominantly a superposition of small LCs that represent the deformation of the electron distribution of the C atoms due to polarization, interference, angular correlation and so on. Around the equilibrium geometry, LC07 gains weight, this neutral LC with \S\, character represents both C atoms in quintet high spin states coupled to a singlet. The quintet state is the result of the excitation of an electron from the doubly occupied 2s AO into the 2p subshell together with a spin flip. This LC has between 1.65\,\AA\, and 1.0\,\AA\, a rather constant weight of about 0.1. Formally, one could say LC07 represents a quadruple bond which becomes important around the C-C equilibrium distance. LC08 is an ionic LC, it describes polarisation of all four formal bonds as described by LC07; LC09 is a neutral LC that can be best described as angular correlation in the two $\pi$ bonds as described by LC06, whereas LC10 describes the polarization in the $\sigma$ bond. The sum of weights of the large LCs decreases dramatically with the C-C distance approaches the equilibrium value, at the same time the contribution of the small LCs becomes large. This is shown in Figure \ref{fig:CAS88St1largeLCs}, right. Only if the criterion for ``being large'' is reduced to 0.03, the ``large'' LCs contribute more than 50 percent along the whole reaction coordinate, and it is for short C-C distances where these LCs contribute most.

\begin{figure}[ht]
\caption{ \label{fig:CAS88St1sigLCs}Left: The weights of the significant $\Sigma_g$ LCs for the first $A_1$ singlet state. Right: The weights of the significant $\Delta_g$ LCs for the first $A_1$ singlet state.}
\includegraphics[width=0.48\textwidth]{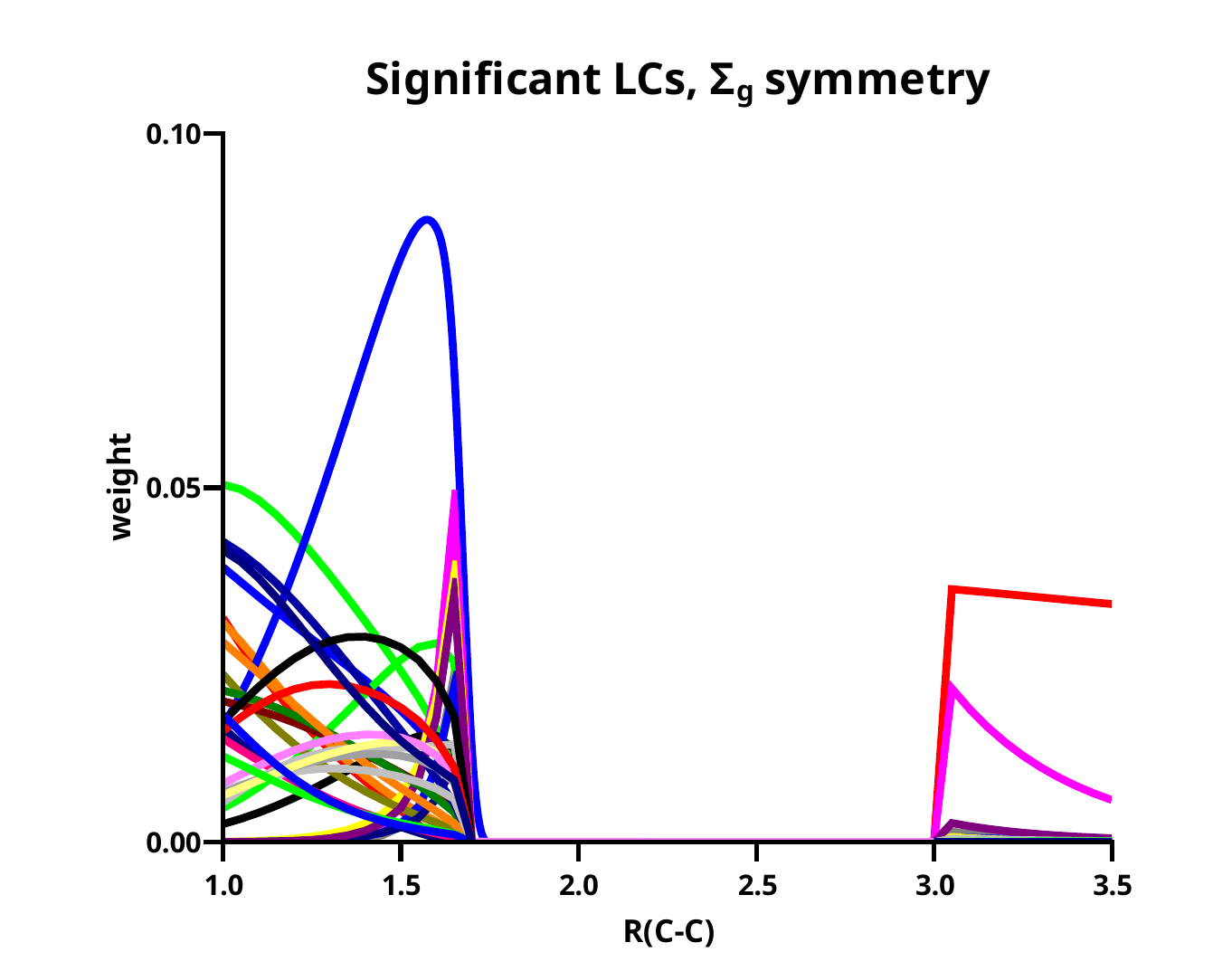}
\includegraphics[width=0.48\textwidth]{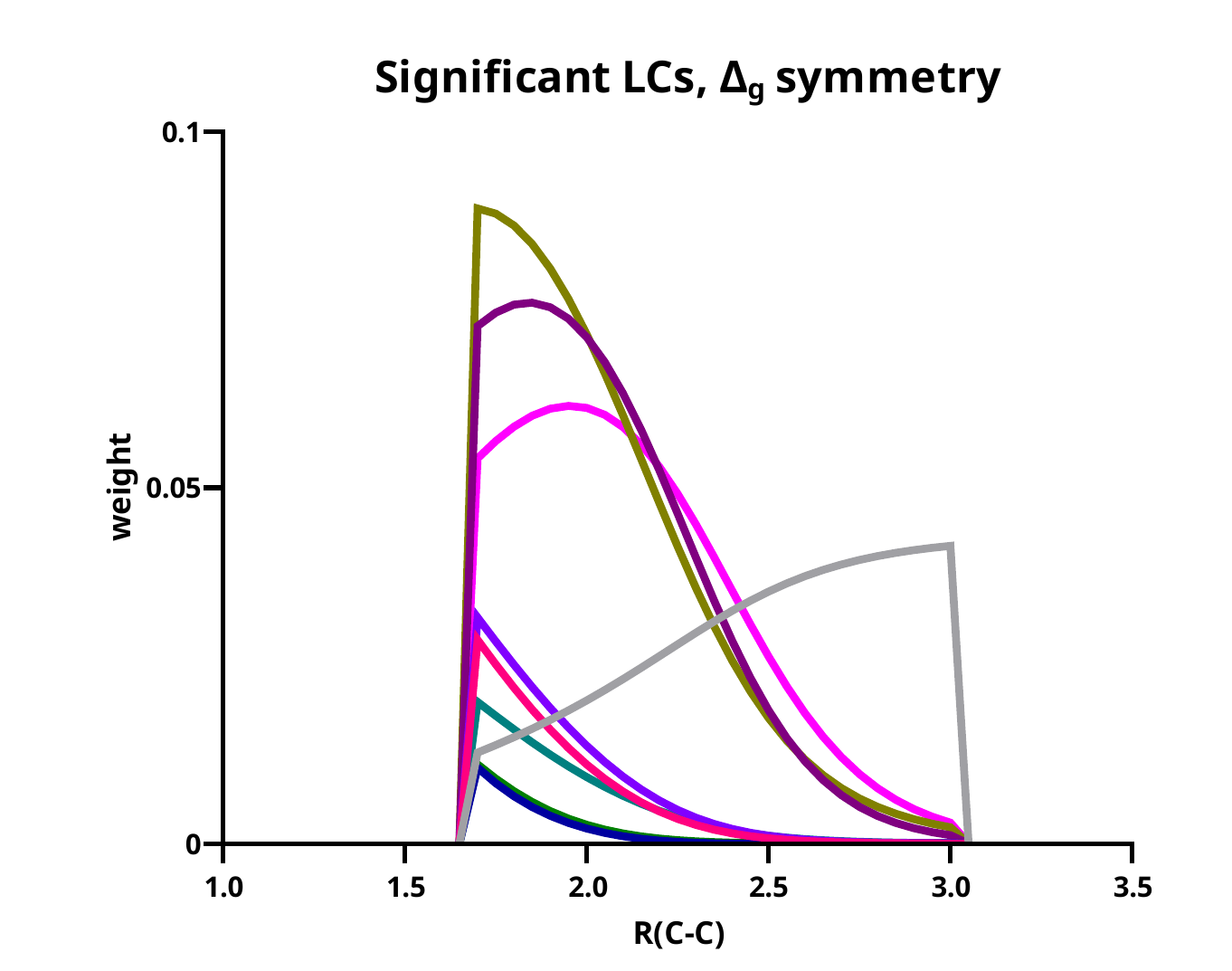}
\end{figure}

Figure  \ref{fig:CAS88St1sigLCs} shows the increase of the contributions of small LCs with decreasing C-C distance. Because there are so many of them the curves are not labelled.

\begin{table}
\caption{\label{tbl:table1}Description of the large LCs in the first $A_1$ singlet state. Col 2: IRREPs in $D_{\infty h}$. Col 3: Neutral (n) or singly ionic (s) LC. Col 4: AO symbols.}
\begin{tabular}{lllll}
LC01 &\D &n &  $\sAs\sBs\zA\zBb(\xA\xBb - \yA\yBb)$ \\
LC02 &\D &s &  $\sAs\sBs(\zAs+ \zBs)(\xA\xBb - \yA\yBb) $ \\
LC03 &\D &n &  $(\sAs\zA\sB\zBs - \zAs\sBs\sA\zBb)(\xA\xBb - \yA\yBb) $ \\
LC04 &\D &s &  $(\sAs\zAs\sBb\zB  - \sBs\zBs\sA\zAb) (\xA\xBb - \yA\yBb)$ \\
LC05 &\S &n &  $\sAs\sBs\zA\zBb(\xA\xBb + \yA\yBb)$ \\
LC06 &\S &n &  $\sAs\sBs\xA\xBb\yA\yBb$ \\
LC07 &\S &n &  $\sA\sBb\zA\zBb\xA\xBb\yA\yBb$ \\
LC08 &\D &s &  $((\yAs+\yBs)\xA\xBb - (\xAs+\xBs)\yA\yBb)\sA\sBb\zA\zBb$\\
LC09 &\D &n &  $\sAs\zA\sBb(\xBs\yA\yBb - \yBs\xA\xBb) + \sBs\sA\zBb(\yAs\xA\xBb - \xAs\yA\yBb)$ \\
LC10 &\S &s &  $(\sAs\zA\sBb- \sBs\sA\zBb)\xA\xBb\yA\yBb$
\end{tabular}
\end{table}

The second $A_1$ state changes three times its character: starting from short C-C distances, it changes at 1.2\,\AA\, from \S\, to \D\, character, at 1.7\,\AA\, from \D\, to \S\,, and at 3.0\,\AA\, again to \D\, character. This characterization is due to the eight large LC contributions that dominate the second $A_1$ singlet state, six of them contribute also to the first $A_1$ state. The dissociated molecule is solely described by LC01, the molecule with two $\pi$ bonds has always $\Sigma$ symmetry. In the region between 1.7\,\AA\, and 3.0\,\AA\, LC05 and LC06 dominate the wave function but, as mentioned above, because of the interaction between the atoms, one $\sigma$ and one $\pi$ bond are more stable than two $\pi$ bonds and therefore LC05 has much larger weight than LC06. The two  singly ionic LCs LC11 and LC12 describe in this region polarization in the $\sigma$ bond. The weight of the LC describing the singlet coupled quintet states is well below 0.1  in the second $A_1$ singlet state.

\begin{table}
\caption{\label{tbl:table2}Description of the large LCs in the second $A_1$ singlet stat. Col 2: IRREPs in $D_{\infty h}$. Col 3: Neutral (n) or singly ionic (s) LC. Col 4: AO symbols.}
\begin{tabular}{lllll}
LC01 &\D &n &  $\sAs\sBs\zA\zBb(\xA\xBb - \yA\yBb)$ \\
LC02 &\D &s &  $\sAs\sBs(\zAs+ \zBs)(\xA\xBb - \yA\yBb) $ \\
LC03 &\D &n &  $(\sAs\zA\sB\zBs - \zAs\sBs\sA\zBb)(\xA\xBb - \yA\yBb) $ \\
LC04 &\D &s &  $(\sAs\zAs\sBb\zB  - \sBs\zBs\sA\zAb) (\xA\xBb - \yA\yBb)$ \\
LC05 &\S &n &  $\sAs\sBs\zA\zBb(\xA\xBb + \yA\yBb)$ \\
LC06 &\S &n &  $\sAs\sBs\xA\xBb\yA\yBb$ \\
LC11 &\S &s &  $(\sAs\zAs\sBb\zB - \sBs\zBs\sA\zAb)(\xA\xBb+ \yA\yBb)$\\
LC12 &\S &n &  $\sAs\sBs(\zAs+\zBs) (\xA\xBb+ \yA\yBb)$
\end{tabular}
\end{table}

\begin{figure}[ht]
\caption{ \label{fig:CAS88St2largeLCs}Left: The distribution of the weights of the large LCs for the second $A_1$ singlet state. Right: The sum of the large weights and the complement.}
\includegraphics[width=0.48\textwidth]{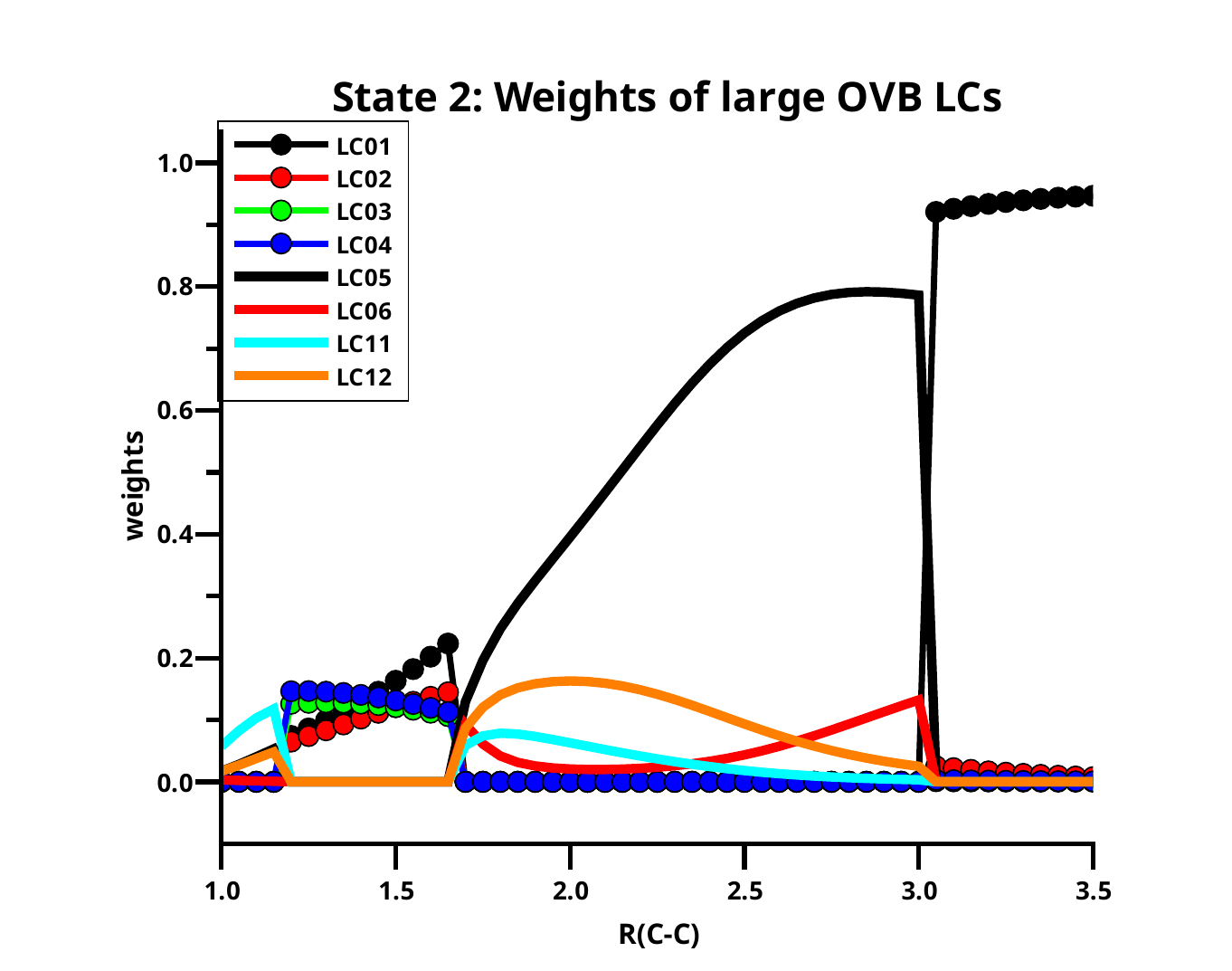}
\includegraphics[width=0.48\textwidth]{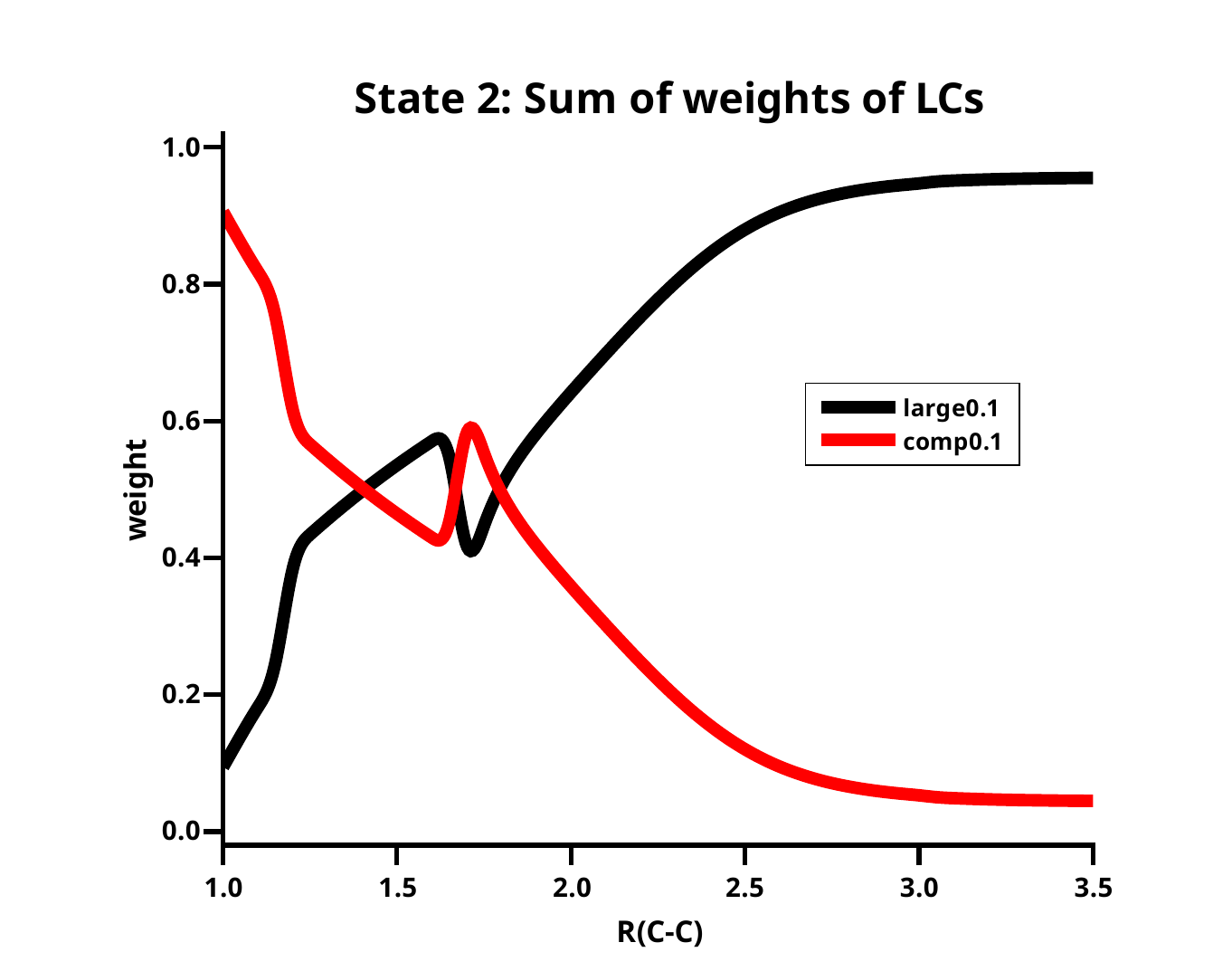}
\end{figure}

The third $A_1$ singlet state has \S\, character along the whole reaction coordinate, the dissociated molecule is, like the first state, described by LC05 and LC06 but now the ratio of the weights is 1:2. When the atoms approach the character of the state changes smoothly, the weights of LCs LC05 and LC06 decrease at C-C distance shorter then 2.5\,\AA\, where LC15 becomes more important. At 1.7\,\AA\, there is again a change of the dominant LCs and again at 1.2\,\AA. The sum of the large weights becomes nevertheless very small and the small LCs make the largest contribution.
\begin{table}
\caption{\label{tbl:table3}Description of the large LCs in the third $A_1$ singlet state. Col 2: IRREPs in $D_{\infty h}$. Col 3: Neutral (n) or singly ionic (s) LC. Col 4: AO symbols.}
\begin{tabular}{llll}
LC05 &\S &n &$\sAs\sBs\zA\zBb(\xA\xBb + \yA\yBb)$ \\
LC06 &\S &n &$\sAs\sBs\xA\xBb\yA\yBb$ \\
LC13 &\S &n &$\sAs\zAs\sBb\zB + \sA\zAb\sBs\zBs) (\xA\xBb+ \yA\yBb)$ \\
LC14 &\S &s &$\sAs\zBs\zA\sBb + \sA\zBb\zAs\sBs)(\xA\xBb + \yA\yBb) $ \\
LC15 &\S &n &$(\sAs\zA\sB +\sBs\sA\zBb)\xA\yA\xBb\yBb $ \\
LC16 &\S &s &$(\sAs\zAs\sBb\zB  - \sBs\zBs\sA\zAb) (\xA\xBb + \yA\yBb)$ \\
LC17 &\S &n &$\sAs\zBs\zA\sBb- \sA\zBb\zAs\sBs) (\xA\xBb + \yA\yBb)$ \\
LC18 &\S &s &$\sAs\sBs(\zAs+\zBs)(\xA\xBb+ \yA\yBb)$
\end{tabular}
\end{table}

\begin{figure}[ht]
\caption{ \label{fig:CAS88St3largeLCs}Left: The distribution of the weights of the large LCs for the third $A_1$ singlet state. Right: The sum of the large weights and the complement.}
\includegraphics[width=0.48\textwidth]{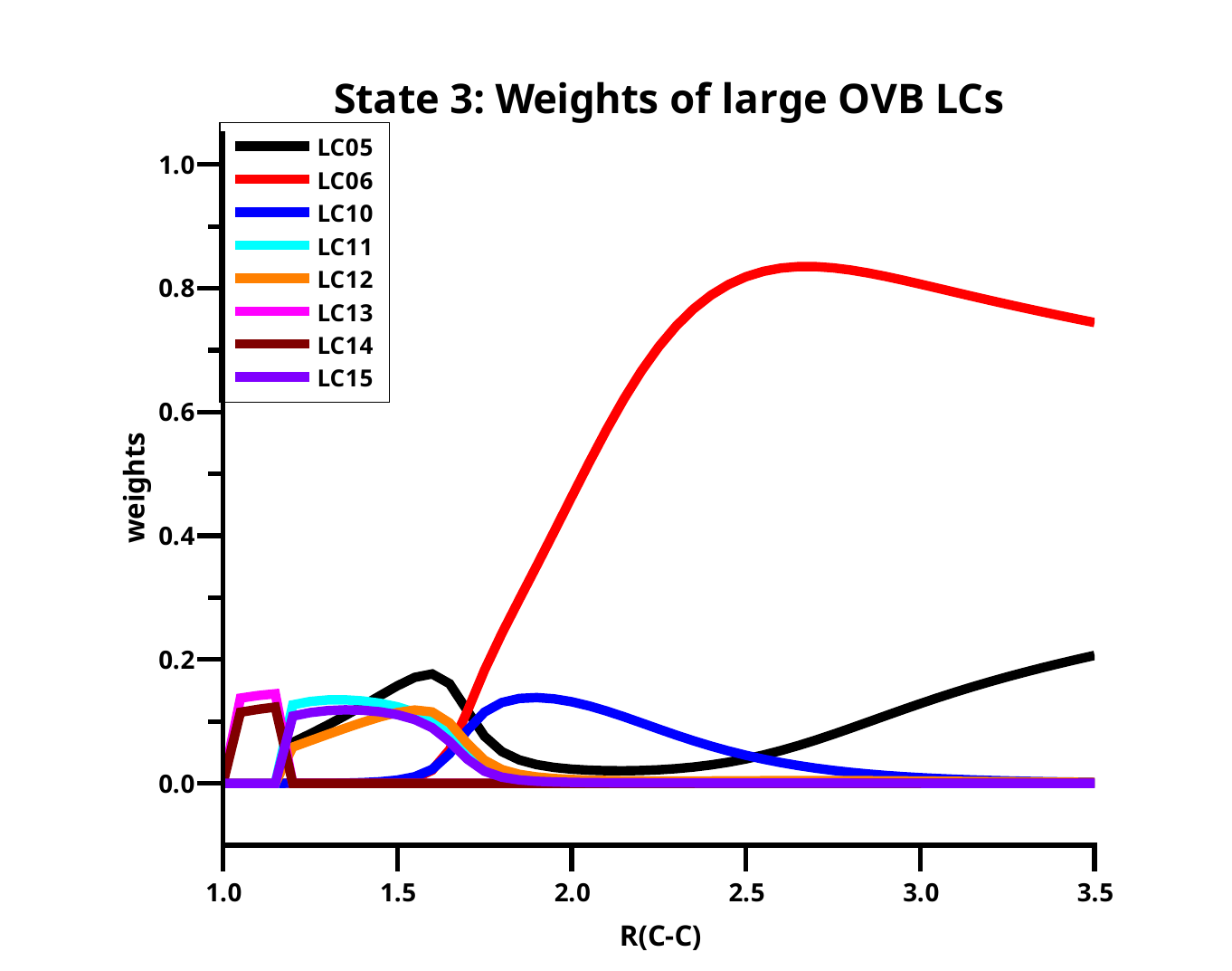}
\includegraphics[width=0.48\textwidth]{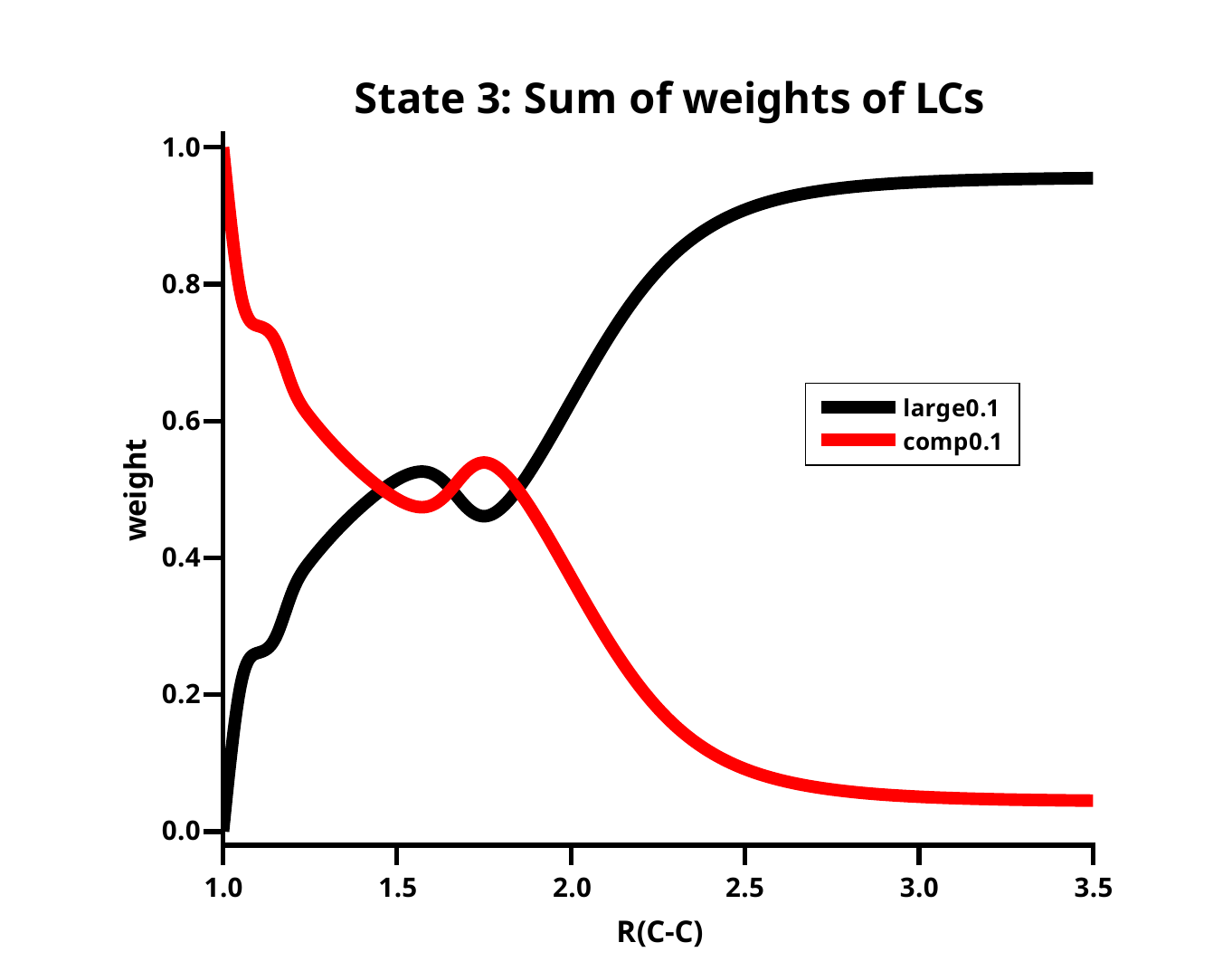}
\end{figure}

\section{Discussion}
Chemical bonding between molecular fragments is caused by a reduction of the total electronic energy when the bonded atoms are near to each other. Thereby the spatial region is enlarged in which bonding electrons can reside. Covalent chemical bonding is of purely quantum theoretical origin, it is the result of constructive interference when states of the interacting fragments are superimposed, meaning that the probability for finding the shared bonding electrons between the interacting fragments is higher than the sum of the probabilities calculated with the wave functions of the non-interacting fragments. This can also be interpreted as a charge shift, which  causes a deformation of the fragment's charge distributions together with classical interactions  like Coulomb attraction and repulsion of electrons and nuclei. This was shown to be responsible for the stabilization of one electron systems like H$_2^+$, which means that the energetic stabilization is a 1-electron effect but not a 2-electron effect as suggested by the important role of the Lewis electron pair. In many-electron systems, the fermionic character of the electrons becomes of utmost importance for the deformation of charge distributions due to the tendency of identical fermions to avoid coming spatially close,  as expressed by the Pauli exclusion principle (PEP). The PEP was expressed by L\'evy-Leblond and Balibar in the following way: \emph{A system of fermions can never occupy a configuration of individual states in which two individual states are identical.}\cite{Levy1990} This tendency is so important because it is independent of physical properties like the electric charge, after all, it holds also for protons and neutrons in nuclei where nuclear forces act. In chemistry, individual states of electrons are called  spinorbitals, they can be localized AOs as well as delocalized MOs, but that two identical electrons can never be found in the same place (Fermi correlation) becomes clear only if eigenstates of the position or localization operator are considered as individual states. However, this holds only if the electrons have identical properties, including the spin projection. Mathematically, the antisymmetry of the state function of a many-fermion system expresses this tendency, and it \emph{.. plays the role of a fictitious, although highly effective, mutual repulsion being exerted within the system, irrespective of any other actual forces or interactions [...] that might be present.}\cite{Levy1990} Using a loose language, one speaks of Pauli repulsion, which keeps identical electrons apart, thereby reducing the Coulomb repulsion. This tendency is not restricted to electrons in an atom or in a molecule but it is operative also between atoms or molecules when they come close, for example, in condensed matter or during chemical reactions. The PEP explains the shell structure of many-electron atoms, but also the origin of certain bond angles in molecules for which mainly the valence electrons are responsible. Before continuing with the role of the PEP, there is an important caveat: Electrons in atoms or molecules cannot be individualized, one says they are indistinguishable, and this means that it is not possible to attribute a certain individual state to each of them; one can only speak of a configuration of individual states and  say that all electrons together occupy these individual states. Nevertheless, using a sloppy language, one says that a certain electron is in a certain state or a certain electron has certain properties. In the following, I will also use this simple way to speak about a complex issue.   The valence electrons in an atom occupy a spherical shell with a characteristic radius and thickness, the radius of the spherical shell is approximately equal to the maximum of the radial density of the valence AOs. For atoms in the second row of the periodic table, the radial densities of the 2s and the 2p AOs are nearly identical, this is not true for all higher rows. So, 2s electrons and 2p electrons reside in the same spatial area irrespective of the different orbital energies, and, according to the PEP, the electrons with identical spin will prefer relative positions with maximum distance to all others. Two identical electrons will prefer to be on different sides of the nucleus, this means an electron-nucleus-electron angle of 180 degrees, three identical electrons will prefer a trigonal arrangement with three angles of about 120 degrees, and four identical electrons will prefer a tetrahedral arrangement.\cite{Lennard1954} In a noble gas, the valence shell is occupied by eight electrons four of which are identical $\alpha$ electrons and four are identical $\beta$ electrons, and therefore for both  groups of identical electrons the  probability for tetrahedral spatial arrangements will be highest of all possible. Coulomb repulsion maximizes the distance between $\alpha$ and $\beta$ electrons (Coulomb correlation), giving two interpenetrating tetrahedra inscribed into a cube. This was called a ``cubical atom'' by Lewis\cite{Lewis1916}; that such arrangements can be found in many-electron atoms was shown by Scemema \etal\cite{Scemama2007} using correlated electron structure methods. As soon as the free atom is disturbed, as it is in a chemical reaction, the electron distribution changes.  Starting from the noble gas electron configuration in, say, the fluoride anion F$^-$, the creation of an F-H single bond by the interaction with a proton can be seen as the rearrangement of the two tetrahedra when a proton approaches the F$^-$ and attracts electrons in the valence shell. One can assume that the electron at the corner of one tetrahedron, say of the $\alpha$ electrons,   will be attracted and the tetrahedron will rearrange so that the corner points towards the proton. But the proton can attract another electron, but this must be a $\beta$ electron, the Coulomb repulsion of the two electrons close to the proton is much smaller than the reluctance of two $\alpha$ electrons coming close. This causes a reorientation of the two tetrahedra bringing two corners in approximate coincidence, the two electrons are the bonding electron pair. The other six electrons can be thought of forming a regular hexagon with alternating $\alpha$ and $\beta$ electrons at the corners. Starting from the O$^{2-}$ dianion one can add stepwise two protons by which eventually all four corners of the two tetrahedra are brought into approximate coincidence giving two bonding and two lone pairs.\cite{Popelier2001} But, as Scemama \etal showed, maximum probability domains of electron pairs that are naively assumed to be typically placed in the midbond region can only be found with uncorrelated Hartree-Fock wave functions, as soon as correlated wave functions are used, \emph{...the bonding pairs separate along the bonds, 'pre-dissociate.'}\cite{Scemama2007}

In addition to rearrangements due to the PEP, energetic aspects must also be considered. The orbital energy of the 2s AO in the carbon atom is about 9 eV lower than that of the 2p AOs and therefore the 2s AO is always filled before any 2p AO is occupied; from boron to fluor the 2s AO is doubly occupied by one $\alpha$ and one $\beta$ electron. In carbon, the remaining two valence electrons occupy the triply degenerate 2p AOs, in accordance with Hund's first rule, with identical spins giving a $^3P_g$ high-spin ground state. In nitrogen, the three remaining electrons occupy the 2p AOs with identical spins resulting in a $^4S_u$ high-spin state. Any further electron must occupy an already singly occupied AO, this is only possible if it has different spin projection, giving a singlet coupled electron pair. This is what happens in the oxygen atom, but also in nitrogen when an electron is excited from the doubly occupied 2s AO. In the carbon atom, however, an electron can be excited from the doubly occupied 2s AO into the 2p subshell without and with spin flip. In the first case, the resulting multiplicity is still a triplet, but in the second case all four electrons have identical spin, this gives a quintet high-spin state and the electrons prefer a tetrahedral arrangement. It is noteworthy that the energy of the $^5S_u$ state is only 4.2 eV higher than the energy of the $^3P_g$ state,\cite{Strasburger2019} this is roughly half of the difference of the orbital energies. Although excitation energy must be provided, the repulsion energy in the high-spin state is considerably reduced, first, because the Coulomb repulsion of the electrons in the 2s AO, which are not Fermi correlated, is reduced, and moreover the Coulomb repulsion of four tetrahedrally arranged electrons is minimal in the spherical shell. Another consequence of the Fermi correlation is a contraction of the orbitals and thus an increase of the  attraction of the electron by the nucleus. All these effects are important when molecules come close and the Pauli repulsion between them increases. Increase of the inter-molecular distance reduces it, but if this is not possible, changes from local low-spin to local high-spin arrangements in the interacting molecules can reduce the Pauli repulsion. In any case, energy is needed for the excitation, and, moreover, something must trigger the spin flip.

In the ground state of the dissociated C$_2$ system, both carbon atoms are in their $^3P_g$ ground states.  Coupling of the atoms gives 18 molecular terms, 6 singlets, 6 triplets, and 6 quintets. Among the singlets are two $\Sigma_g^+$ states and one $\Delta_g$ state. Only these states are responsible for the stabilization of the system when the atoms approach, the large weight of LC05 suggests that the bonding situation is dominated by a $\sigma$ and a $\pi$ bond, but the weight of LC06, which represents two $\pi$ bonds without a $\sigma$ bonds shows, that even at long distances the number and kind of bonds is not definite. At short C-C distances, the weight of LC07 becomes large, this LC describes two singlet coupled atomic quintet states. The weight of LC07 is larger than those of LC05 and LC06, but  ionic LCs or LCs describing intra-atomic charge shifts contribute together much more to the ground state wave function than LC07. The attempt to claim that C$_2$ has around the equilibrium distance a quadruple bond  ignores the fact, that the occurrence of LC07 at the equilibrium does not mean that the carbon atom is there in a local quintet state. After all, no interacting subsystem of a system is in a pure state but only in a mixed state, which allows only to say with which probability a certain pure state can be expected. To get this information, one must get the reduced density matrix for the subsystem considered. But then, the answer that can be given is definitively different from what those scientists expect who want describe the bonding situation using concepts like bond order that are not compatible with electron structures that must be described by multi-configurational wave functions.

\section{Method}
All calculations were made with CAS(n,n) wave functions, where $n$ active electrons are distributed among the same number of active orbitals, the wave functions are linear combination of configuration state functions (CSF) generated with the GUGA technique. All calculations were done with a local version of GAMESS.\cite{Gamess}
For all systems but ethane, the cc-pVTZ basis set was used, the ethane system was calculated with the cc-pVDZ basis. The single bond in ethane is represented by a CAS(2,2) wave functions, the double bond in ethene by a CAS(4,4) wave function, the triple bond in N$_2$ by a CAS(6,6) wave function. The electron distribution in C$_2$ is described by a CAS(8,8) wave function. In all systems, the two lowest MOs (positive and negative linear combination of 1s AOs) are kept frozen. For the calculation of the dissociation reactions the reaction coordinates, that is the C-C and N-N bond lengths, respectively, were incremented in steps of 0.1\,\AA; the geometries of ethene and ethane were optimized for each frozen C-C distance. The fragments of the four systems are the C and the N atom for C$_2$ and N$_2$, the methyl radical for ethane and carbene for ethene. The fragment wave functions were calculated for high spin states using low level methods, e.g., UHF; the methyl and carbene geometries were taken from the optimized molecular geometries. For each bond length, the optimized CASSCF MOs are localized on the respective fragments, using an orthogonal Procrustes transformation.\cite{Sax2012} Doubly occupied non-active MOs are transformed into doubly occupied fragment MOs (FMO), which are delocalized in case of methyl and carbene; active MOs are transformed into FMOs that resemble AOs or hybrid AOs. The CSFs constructed with these FMOs are dubbed OVB CSFs (orthogonal valence bond). Finally, the CI matrix is set up with the OVB CSFs and diagonalized. This gives the energies and weights for all OVB CSFs.

\begin{acknowledgement}
The author acknowledges helpful comments by W.H.E. Schwarz.
\end{acknowledgement}
\newpage
\bibliography{C2arX}

%%%%%%%%%% Merge with supplemental materials %%%%%%%%%%
\pagebreak
%\widetext
%\begin{center}
%\textbf{\large Supporting Information: Chemical Bonding in the C$_2$ Molecule}
%\end{center}
%%%%%%%%%% Merge with supplemental materials %%%%%%%%%%
%%%%%%%%%% Prefix a "S" to all equations, figures, tables and reset the counter %%%%%%%%%%
\setcounter{equation}{0}
\setcounter{figure}{0}
\setcounter{table}{0}
\setcounter{page}{1}
\makeatletter
\renewcommand{\theequation}{S\arabic{equation}}
\renewcommand{\thefigure}{S\arabic{figure}}
\renewcommand{\thetable}{S\arabic{table}}
\renewcommand{\bibnumfmt}[1]{[S#1]}
\renewcommand{\citenumfont}[1]{S#1}
%%%%%%%%%% Prefix a "S" to all equations, figures, tables and reset the counter %%%%%%%%%%

\section{Supporting Information}
All molecular states of C$_2$ belong to IRREPs of $D_{\infty h}$ symmetry,  $\Sigma$ IRREPs are one dimensional, all others are two dimensional.
Since actual calculations are done in Abelian subgroups of $D_{\infty h}$, that is either $D_{2h}$ or $C_{2v}$, very few  CSFs  have the correct $D_{\infty h}$ symmetry, in general, only linear combinations of CSFs do. Symmetry adapted CSFs, that is linear combinations of CSFs, will be labelled as LC together with a running index. Calculations with delocalized MOs are done in $D_{2h}$, linear combinations of two CSFS with $\pi$ MOs may be necessary to describe $\Sigma$ or $\Delta$ states; OVB CSFs, that is CSFs constructed with localized frgamnet MOs, are calculated in $C_{2v}$, therefore LCs of two or four OVB CSFs may be needed to represent both parity and rotational symmetry.

\subsection{MO description of the three lowest C$_2$ singlet states}
The number of CSFs that can be constructed with $N$ electrons and $n$ MOs for spin  state with spin quantum number $S$ is
$\displaystyle \frac{2S+1}{n+1} \binom{n+1}{\frac{N}{2} - S}\binom{n+1}{\frac{N}{2}+S+1}$; therefore, with eight electrons and eight MOs 1764 singlet CSFs can be made. This is the number of CSFs in the single IRREP of symmetry group $C_1$,\; for higher symmetry groups with more than one IRREP the number of CSFs in each IRREP is consequently smaller. If actual calculations cannot be done in a high point group symmetry but must be done in a subgroup of lower symmetry, states belonging to different IRREPs in  high symmetry may be in the same IRREP in low symmetry and are allowed to mix; this leads to avoided crossings.

The ground state of C$_2$ is a singlet state of $\Sigma_g^+$ symmetry in $D_{\infty h}$ and of $A_g$ symmetry in $D_{2h}$. Also one component of the $\Delta_g$ state becomes $A_g$ in $D_{2h}$ so these states will mix and avoided crossings will occur. Three singlet $A_g$ states were calculated in this study. The lowest $^1A_g$ state has avoided crossings at 1.70\,\AA, and at 3.00\,\AA; the second $^1A_g$ state has avoided crossings at  1.15\,\AA, at 1.65\,\AA\, and at 3.00\,\AA, and the third $^1A_g$ state has one (obvious) avoided crossings at  1.15\,\AA. The left branch of the ground state has $\Sigma_g^+$ character and is dominated by the LC $\sigma(2s)^2 \sigma^*(2s)^2 \pi_x^2 \pi_y^2$;  in the middle branch $\sigma(2s)^2 \sigma^*(2s)^2 \sigma(2p)^2 (\pi_x^2 - \pi_y^2)$ dominates, and this has $\Delta_g$ character. The right branch has again $\Sigma_g^+$ character with the dominating LC $\sigma(2s)^2 \sigma^*(2s)^2 \sigma(2p)^2(\pi_x^2 + \pi_y^2)$. In the first excited state, this LC dominates up to 1.15\,\AA; from 1.20\,\AA\, up to 1.65\,\AA\, $\sigma(2s)^2 \sigma^*(2s)^2 \sigma(2p)^2 (\pi_x^2 - \pi_y^2)$ dominates the $\Delta_g$ character; from 1.70\,\AA\, to 3.00\,\AA\, the $\Sigma_g^+$ character is due to the LC of LCs $\sigma(2s)^2 \sigma^*(2s)^2 \sigma(2p)^2(\pi_x^2 + \pi_y^2)$, and then the state has again $\Delta_g$ character. The second excited state has up to 1.15\,\AA\, $\Delta_g$ character, then $\Sigma_g^+$ character with dominance of $\sigma(2s)^2 \sigma^*(2s)^2 \sigma(2p)^2(\pi_x^2 + \pi_y^2)$; from 1.70\,\AA\, on the $\Sigma_g^+$ character is dominated by $\sigma(2s)^2 \sigma^*(2s)^2 \pi_x^2 \pi_y^2$, and from 3.05\,\AA\, on
$\sigma(2s)^2 \sigma^*(2s)^2(\pi_x^2 \pi_y^{*2}+ \pi_y^2\pi_x^{*2})$ dominates.

A change in the character of a state can also be seen by looking at the number of MO LCs that contribute to its description. In GAMESS, non-zero CSFs must have a weight larger than $10^{-6}$, many of the listed CSFs have very small weights; the weights of the LCs are of the same magnitude. In the following, LCs with weight larger than 0.01 will be called significant; LCs having weight larger than 0.1 will be called large.

\begin{figure}[ht]
\caption{ \label{fig:CAS88NCSFs}Top left: Number of non-zero MO LCs for the three lowest $^1A_g$  states.
Top right: Number of significant LCs for the three lowest $^1A_g$  states. Bottom: Number of large MO LCs for the three lowest $^1A_g$  states.}
\includegraphics[width=0.48\textwidth]{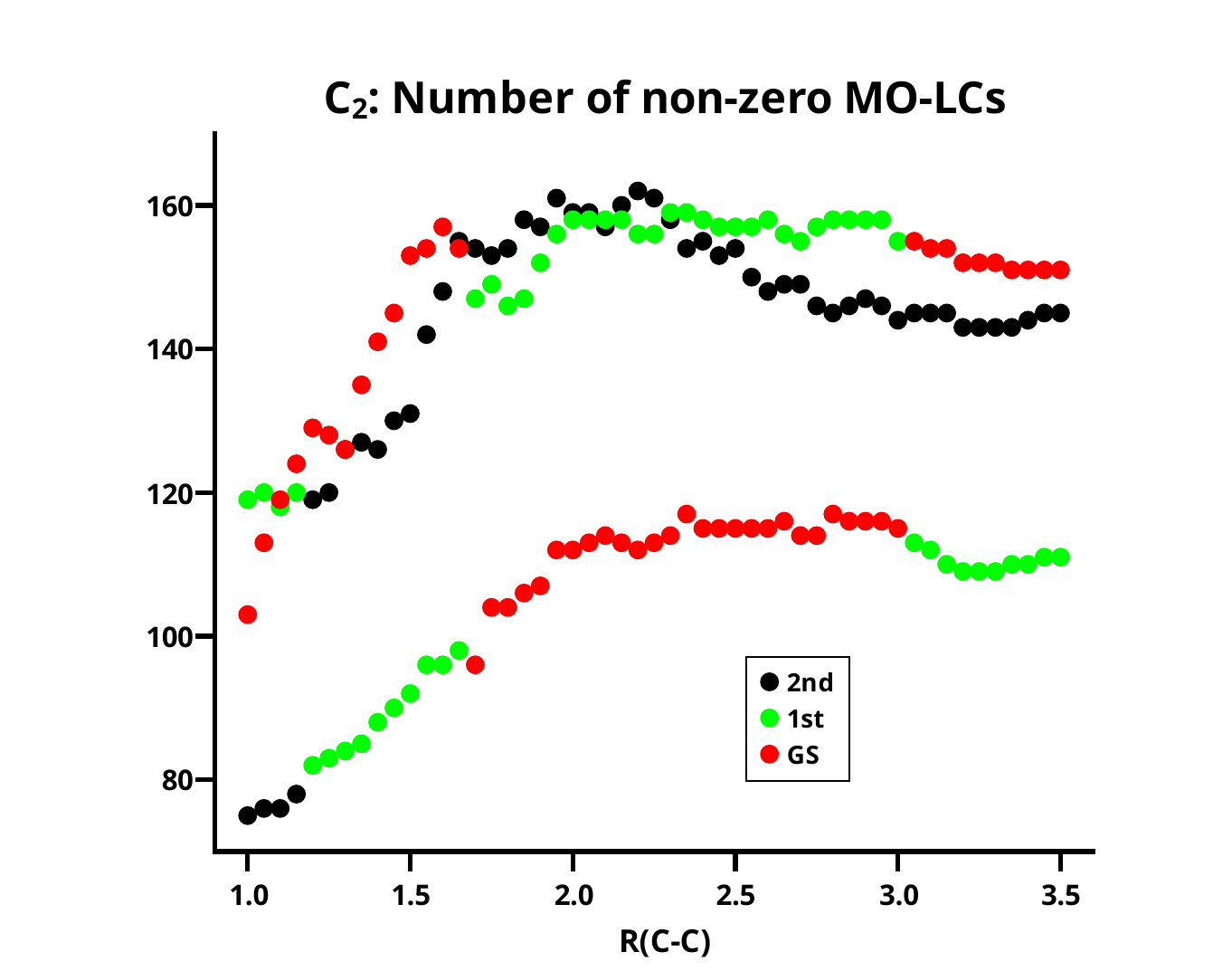}
\includegraphics[width=0.48\textwidth]{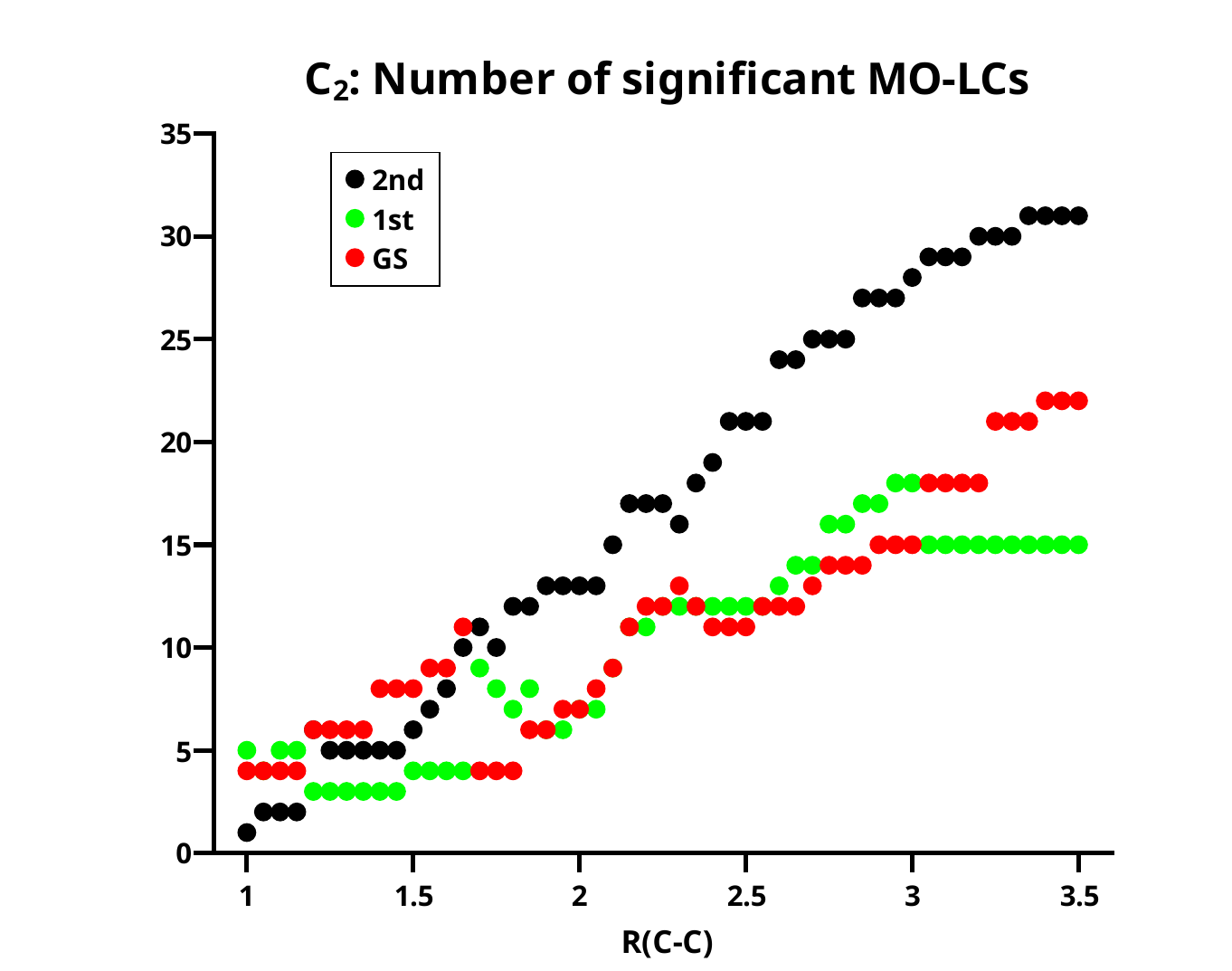}
\includegraphics[width=0.48\textwidth]{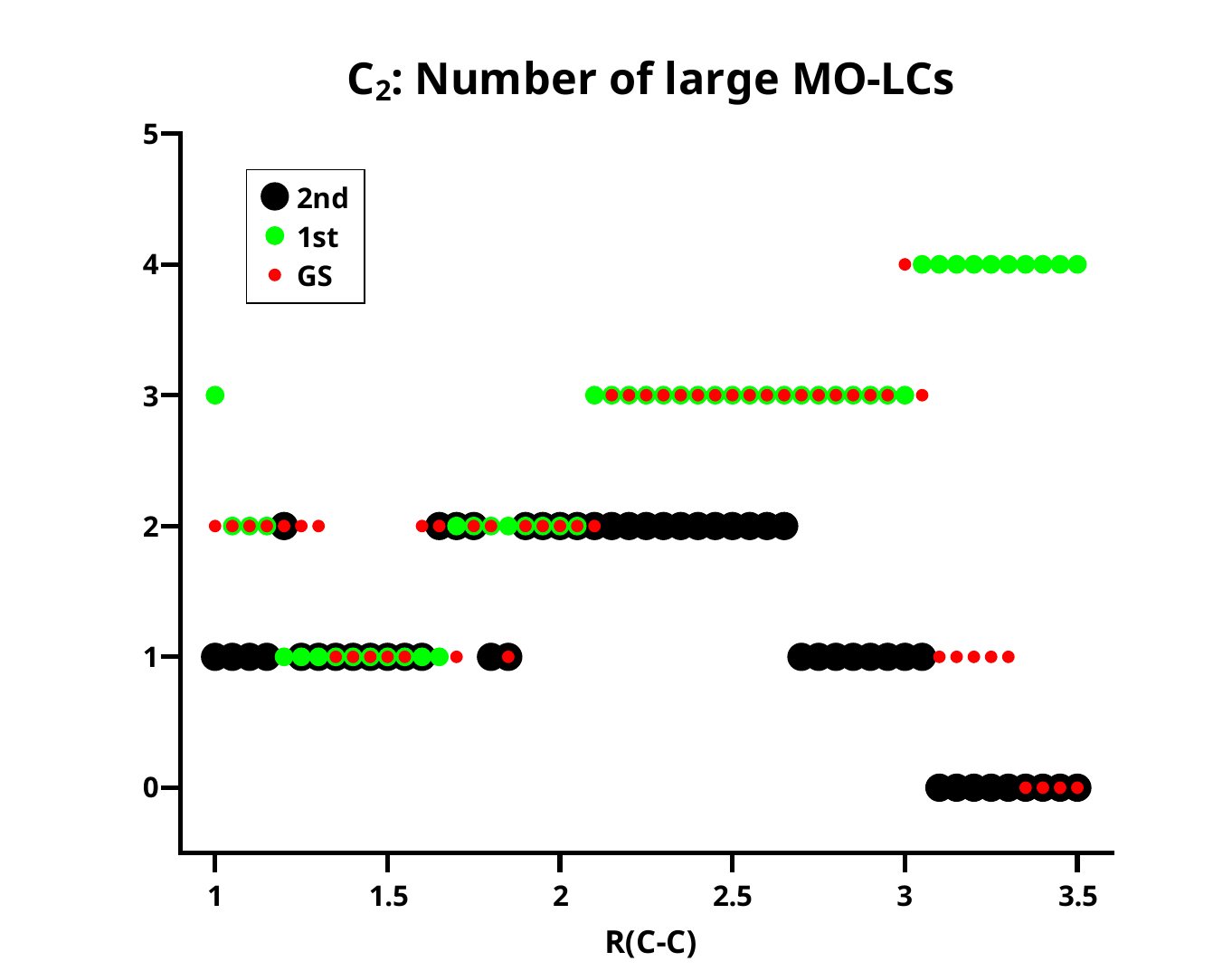}
\end{figure}

Figure \ref{fig:CAS88NCSFs} shows the change in the number of non-zero, significant, and large LCs. The left branch of the ground state with $\Sigma_g^+$ character starts with about 100 non-zero LCs and goes up to about 160 LCs, then, in the middle branch, jumps down to about 100 LCs and rises in to about 110 LCs, the right branch finally  jumps up to 150 LCs. This corresponds to the changes from $\Sigma_g^+$ to $\Delta_g$ and back to $\Sigma_g^+$ character. For the second $A_g$ state, the left branch ($\Sigma_g^+$ character) has about 120 LCFs, the next branch ($\Delta_g$ character) jumps to about 80 LCs and rises to about 100 LCs, the next branch has again $\Sigma_g^+$ character and is described by 140 to 160 LCs, the last branch has again $\Delta_g$ character and is described by about 110 LCs. The third $A_g$ state has for short C-C distances  $\Delta_g$ character and is described by less than 80 LCs, then it has $\Sigma_g^+$ character and is described by 120 to 160 LCs.

The number of significant LCs is by a factor of ten smaller. At short C-C distances, the number of significant LCs is small and increases with increasing C-C increasing C-C distance. This is found for all states and just shows that few MOs are needed to describe stable molecules around the equilibrium geometry; many MOs and, therefore, many LCs are needed to describe dissociated structures. This differs from state to state, the third $A_g$ state needs twice as many LCs as the second state. But it is clear that the weight of the significant LCs must decrease when the number of LCs increases, and, eventually all LCs may have a weight smaller than 0.01.  Both the first and the third state have for long distances $\Sigma_g^+$ character and are represented by a large number of significant but by no large LCs; the second state at long C-C distances, where the state has $\Delta_g$ character,  is represented by four large LCs and by a relatively few significant LCs.

In the following, all significant LCs are listed. CSFs are represented by strings describing the occupation of the active orbitals, orbitals can be doubly occupied (occupation symbol 2), not occupied (occupation symbol 0), or singly occupied (occupations symbols a or b, indicating occupation by an $\alpha$ or by a $\beta$ electron. The order of the active MOs is: $\gs$, $\px$, $\py$,  $\us$,  $\gp$, $\ux$, $\uy$,  $\up$. String ababbaba means $\gs\pxb\py\usb\gpb\ux\uyb\up$.

\begin{longtable}{*{4}{l}}%
  \caption{Description of the significant MO LCs as linear combinations of MO CSFs. Col 2: IRREP in $D_{\infty h}$. Col 3: Description by occupation strings. Col 4: Description by MO symbols.}
 \label{tbl:largeMOCSFs} \\
  \toprule\addlinespace[-2pt]
  LC & IRREP & Occupation strings &MO symbols \\   % \thead{Occupation \\ strings.}
  \midrule\addlinespace[-2pt]
  \endfirsthead
  \multicolumn{4}{c}{\tablename~\thetable~ (continued)} \\
  \addlinespace
  \toprule\addlinespace[-2pt]
  LC & IRREP & \thead{Occupation \\ strings.} &MO symbols \\
  \midrule\addlinespace[-1pt]
  \endhead
%  \midrule
%  \addlinespace[-8pt]
%  \multicolumn{4}{r@{}}{\footnotesize To be continued}
%  \endfoot
  \bottomrule
  \endlastfoot

LC01 &\S & 22220000             & $\gss\pxs\pys\uss $ \\
LC02 &\S & 20220200 +  22020020 & $\gss\uss\left(\pys\uxs+\pxs\uys\right) $ \\
LC03 &\D & 2b2aba00 -- 22bab0a0 & $\gss{}\,^1(\us\gpb) \left( \pys{}\,^1(\pxb\ux)- \pxs{}\,^1(\pyb\uy)\right)$ \\
LC04 &\S & 22202000             & $\gss\pxs\pys\gps $ \\
LC05 &\D & 2a2abb00 -- 22aab0b0 & $\gss{}\,^1(\us\gpb)\left(\pys{}\,^1(\px\uxb) -\pxs{}\,^1(\py\uyb)\right) $ \\
LC06 &\S & 20222000 +  22022000 & $\gss\uss\gps\left(\pys + \pxs\right)$ \\
LC07 &\D & 20222000 -- 22022000 & $\gss\uss\gps \left(\pys-\pxs\right)$ \\
LC08 &\D & 20022200 -- 20022020 & $\gss\uss\gps\left(\uxs - \uys\right) $ \\
LC09 &\S & 20022200 +  20022020 & $\gss\uss\gps\left(\uxs + \uys\right) $ \\
LC10 &\S & 2ba20ab0             & $\gss\uss{}\,^1(\pxb\py) {}\,^1(\ux\uyb) $ \\
LC11 &\S & 20202200 +  22002020 & $\gss\gps\left(\pys\uxs + \pxs\uys\right)$ \\
LC12 &\D & 202a200b -- 220a200b & $\gss\gps\left(\pys-\pxs\right){}\,^1(\us\upb)$ \\
LC13 &\S & 202a200b +  220a200b & $\gss\gps\left(\pys+\pxs\right){}\,^1(\us\upb)$ \\
LC14 &\S & 222a000b             & $\gss\pxs\pys{}\,^1(\us\upb) $ \\
LC15 &\D & 2a02bb0a -- 20a2b0ba & $\gss\uss{}\,^1(\gpb\up)\left(^1(\px\uxb) - {}^1(\py\uyb)\right) $ \\
LC16 &\S & a222b000             & $\pxs\pys\gps{}\,^1(\gs\gpb) $ \\
LC17 &\D & aa2b2b00 -- a2ab20b0 & $^1(\gs\usb)\pys \left(^1(\px\uxb)-{} ^1(\py\uyb)\right) $ \\
LC18 &\S & 2a02bb0a +  20a2b0ba & $\gss\uss{}\,^1(\gpb\up)\left(^1(\px\uxb) + {}^1(\py\uyb)\right) $ \\
LC19 &\S & 20020220             & $\gss\uss\uxs\uys $ \\
LC20 &\S & 202a020b +  220a002b & $\gss{}\,^1(\us\upb)\left(\pys\uxs + \pxs\uys\right) $ \\
LC21 &\D & 20202200 -- 22002020 & $\gss \gps\left(\pys\uxs - \pxs\uys\right)$ \\
LC22 &\D & 200a220b -- 200a202b & $\gss\gps{}\,^1(\us\upb)\left(\uxs - \uys\right) $ \\
LC23 &\S & 200a220b +  200a202b & $\gss\gps{}\,^1(\us\upb)\left(\uxs + \uys\right) $ \\
LC24 &\S & a22ab00b             & $\pxs\pys{}\,^3(\gs\us)^3(\gpb\upb) $ \\
LC25 &\S & a22bb00a             & $\pxs\pys{}\,^1(\gs\usb)^1(\gpb\up) $ \\
LC26 &\D & 20220002 -- 22020002 & $\gss\ups\uss\left(\pys - \pxs\right) $ \\
LC27 &\S & a022b200 +  a202b020 & $^1(\gs\gpb)\us2\left(\pys\uxs+\pxs\uys\right) $ \\
LC28 &\S & 20220002 +  22020002 & $\gss\ups\uss\left(\pys + \pxs\right) $ \\
LC29 &\D & ab022a0b -- a0b220ab & $^1(\gs\upb)\uss\gps\left(^1(\pxb\ux) - {} ^1(\pyb\uy)\right) $ \\
LC30 &\D & 20020202 -- 20020022 & $\gss\uss\ups\left(\uxs - \uys\right) $\\
LC31 &\S & 20020202 +  20020022 & $\gss\uss\ups\left(\uxs + \uys\right)$\\
LC32 &\S  & ab022a0b +  a0b220ab & $^1(\gs\gpb)\uss\gps\left(^1(\pxb\ux) + {} ^1(\pyb\uy)\right)$ \\
LC33 &\S  & 200a022b             & $\gss\uxs\uys{}\,^1(\us\upb) $ \\
LC34 &\S  & a02ab20b +  a20ab02b & $^3(\gs\us)^3(\gpb\upb)\left(\pys\uxs + \pxs\uys\right) $ \\
LC35 &\D  & 2a02ba0b -- 20a2b0ab & $\gss\uss{}\, ^3(\gpb\upb)\left(^3(\px\ux) - {}^3(\py\uy)\right) $ \\
LC36 &\D  & a022b002 -- a202b002 & $\uss\ups{}\,^1(\gs\gpb)(\pys - \pxs) $ \\
LC37 &\S  & 2a02ba0b +  20a2b0ab & $\gss\uss{}\, ^3(\gpb\upb)\left(^3(\px\ux) + {}^3(\py\uy)\right) $ \\
LC38 &\S  & a022b002 +  a202b002 & $\uss\ups{}\,^1(\gs\gpb)(\pys + \pxs) $ \\
LC39 &\S  & 2baa0abb             & $\gss{}^1(\us\upb)\,^1(\pxb\py)^1(\ux\uyb) $ \\
LC40 &\S  & a02bb20a +  a20bb02a & $^1(\gs\usb)^1(\gpb\up)\left(\pys + \pxs\right) $ \\
LC41 &\D  & a002b202 -- a002b022 & $\uss\ups{}\,^1(\gs\gpb)\left(\uxs-\uys\right) $ \\
LC42 &\S  & a002b202 +  a002b022 & $\uss\ups{}\,^1(\gs\gpb)\left(\uxs+\uys\right) $ \\
LC43 &\S  & 2aa20bb0             & $\gss\uss{}\,^3(\px\py)^3(\uxb\uyb) $ \\
LC44 &\D  & 20202002 -- 22002002 & $\gss\gps\ups\left(\pys-\pxs\right) $ \\
LC45 &\S  & 20202002 +  22002002 & $\gss\gps\ups\left(\pys+\pxs\right) $ \\
LC46 &\S  & a002b220             & $^1(\gs\gpb)\uss\uxs\uys $ \\
LC47 &\S  & ababbaba             & $^1(\gs\pxb)^1(\py\usb)^1(\gsb\px)^1(\pyb\us) $ \\
LC48 &\S  & aba2bab0             & $^1(\gs\gpb)^1(\pxb\py)^1(\ux\uyb)\uss $ \\
LC49 &\S  & 2aba0abb             & $\gss{}\,^1(\px\pyb)^1(\ux\uyb)^1(\us\upb) $ \\
LC50 &\S  & 2baa0bab             & $\gss{}\,^1(\pxb\py)^1(\uxb\uy)^1(\us\upb) $ \\
LC51 &\D  & 2b0aba02 -- 20bab0a2 & $\gss\ups{}\,^1(\us\gpb)\left(^1(\pxb\ux) - {}^1(\pyb\uy) \right) $ \\
LC52 &\S  & 2b0aba02 +  20bab0a2 & $\gss\ups{}\,^1(\us\gpb)\left(^1(\pxb\ux) + {}^1(\pyb\uy) \right) $ \\
LC53 &\S  & a00ab22b             & $^3(\gs\us)^3(\gpb\upb)\uxs\uys $ \\
LC54 &\S  & a00bb22a             & $^1(\gs\usb)^1(\gpb\up)\uxs\uys $ \\
LC55 &\D  & 20002202 -- 20002022 & $\gss\gps\ups\left(\uxs-\uys\right) $ \\
LC56 &\D  & ab0a2b02 -- a0ba20b2 & $\gps\ups{}\,^3(\gs\us)\left(^3(\pxb\uxb)-{}^3(\pyb\uyb)\right) $ \\
LC57 &\S  & 20002202 +  20002022 & $\gss\gps\ups\left(\uxs+\uys\right) $ \\
LC58 &\S  & ab0a2b02 +  a0ba20b2 & $\gps\ups{}\,^3(\gs\us)\left(^3(\pxb\uxb)+{}^3(\pyb\uyb)\right) $ \\
LC59 &\D  & 2a0abb02 -- 20aab0b2 & $\gss\ups{}\,^1(\us\gpb)\left( ^1(\px\uyb) -{} ^1(\py\uyb)\right) $ \\
LC60 &\S  & aabababb +  abaabbab & $^3(\gs\us)^3(\gpb\upb)\left(^1(\px\pyb)^1(\ux\uyb)+ {}^1(\pxb\py)^1(\uxb\uy)\right) $ \\
LC61 &\S  & 2a0abb02 +  20aab0b2 & $\gss\ups{}\,^1(\us\gpb)\left( ^1(\px\uyb) +{} ^1(\py\uyb)\right) $
\end{longtable}
21 of the 61 significant  LCs have $\Delta_g$ character.

\begin{figure}[ht]
\caption{ \label{fig:CAS88State123w}Top left: Weights of large MO LCs for the ground state. Top left: Weights of large MO LCs for the 1st excited state. Bottom: Weights of large LCs for the  2nd excited state. }
\includegraphics[width=0.48\textwidth]{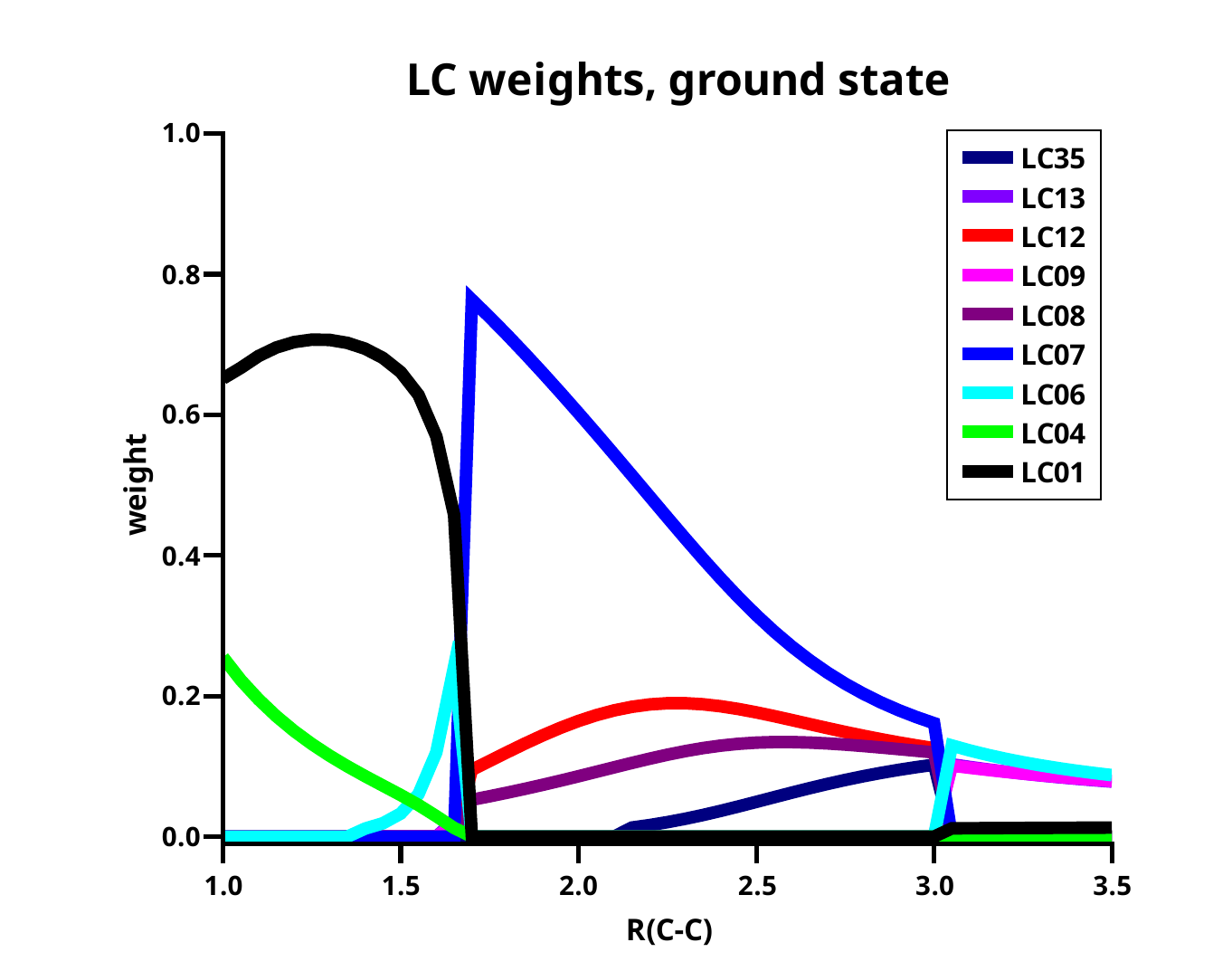}
\includegraphics[width=0.48\textwidth]{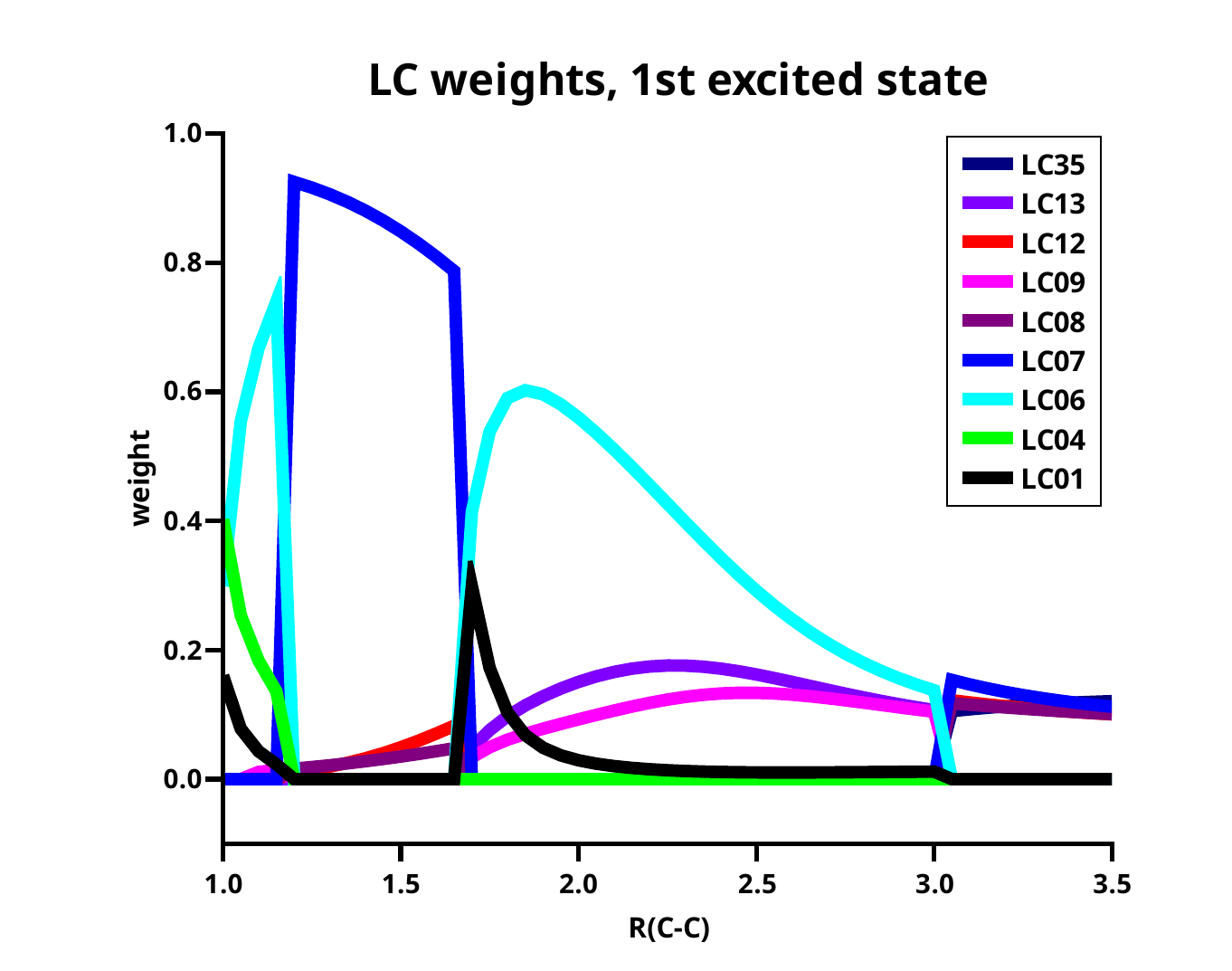}
\includegraphics[width=0.48\textwidth]{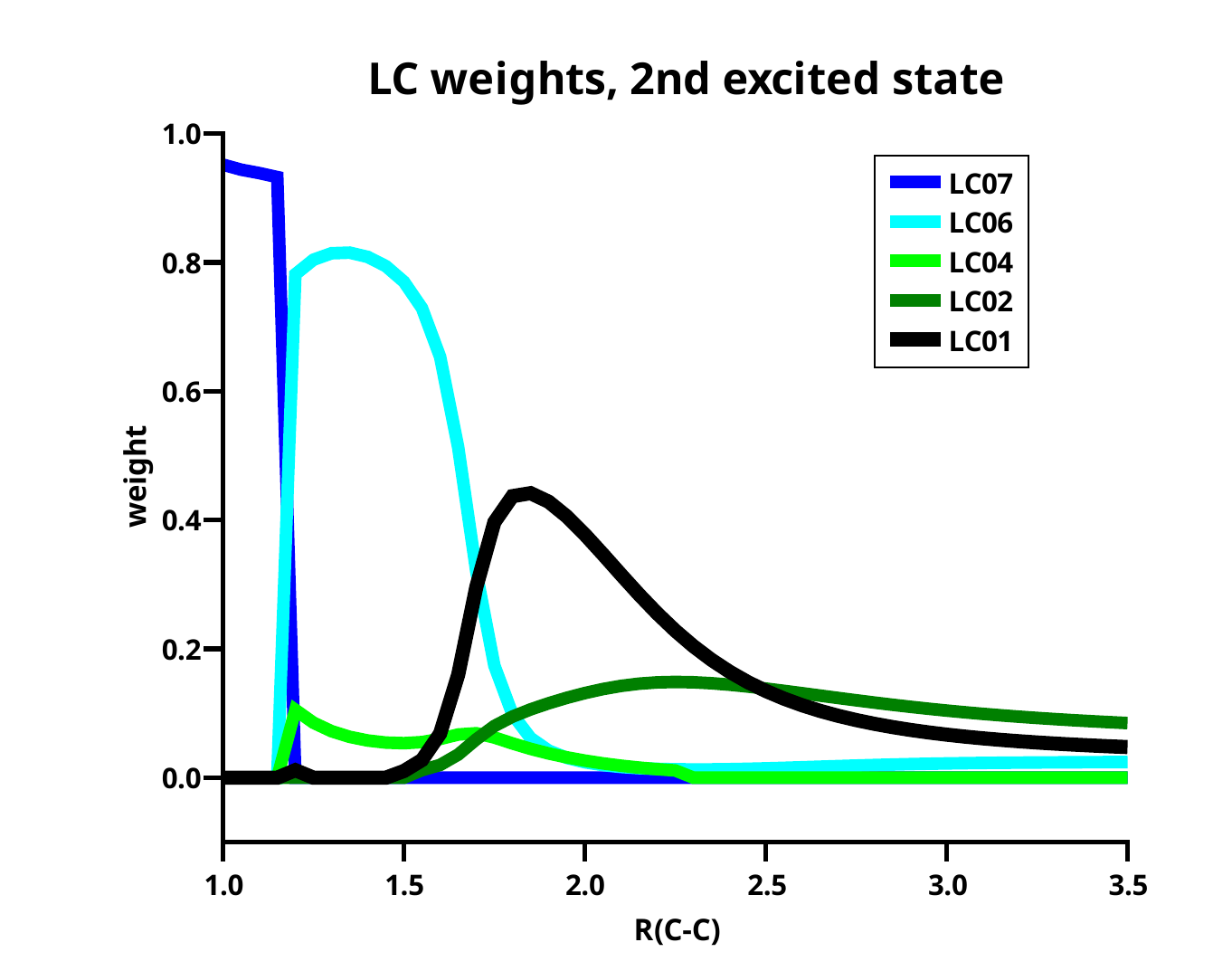}
\end{figure}

In Figures \ref{fig:CAS88State123w}  the weights of those LCs are shown that have large weight somewhere along the reaction coordinate. The avoided crossings can be clearly identified by the discontinuities in the LC curves.

\subsection{OVB description of the three lowest C$_2$ singlet states}
The point group is $C_{2v}$, the rotation axis is the z-axis; the active orbitals are the OAOs, the order in the actual calculations is $s_A, z_A, x_A, y_A, s_B, z_B, x_B, y_B$, A and B indicate the two  C atoms. An occstring like aaa2bbb0 represents the OAO configuration  $\displaystyle s_A z_A x_A y_A^2 \overline{s_B}\, \overline{z_B}\, \overline{x_B}$. In general, only LC have $D_{2h}$ symmetry.

The number of non-zero LCs for each state is large, it varies with the C-C distance; the number of significant LCs is by a factor of 5 smaller but still rather large. In total, 128 significant LCs occur in the wave functions describing the three $^1A_1$ states. See
Table \ref{tbl:sigOVBCSFs}.
The number of large LCs is by a factor on 10 smaller,  only 15 LCs are large in any of the three lowest $^1A_g$ states.

Comparison of Tables  \ref{fig:CAS88NCSFs} and \ref{tbl:sigOVBCSFs} shows the antagonistic development of the number of contributing LCs for wave functions constructed with canonical MOs and with localized FMOs.

\begin{figure}[ht]
\caption{ \label{fig:CAS88NLCs}Top left: Number of non-zero LCs for the three lowest $A_g$ singlet states.
Top right: Number of significant LCs for the three lowest $A_g$ singlet states. Bottom: Number of large LCs for the three lowest $A_g$ singlet states.}
\includegraphics[width=0.48\textwidth]{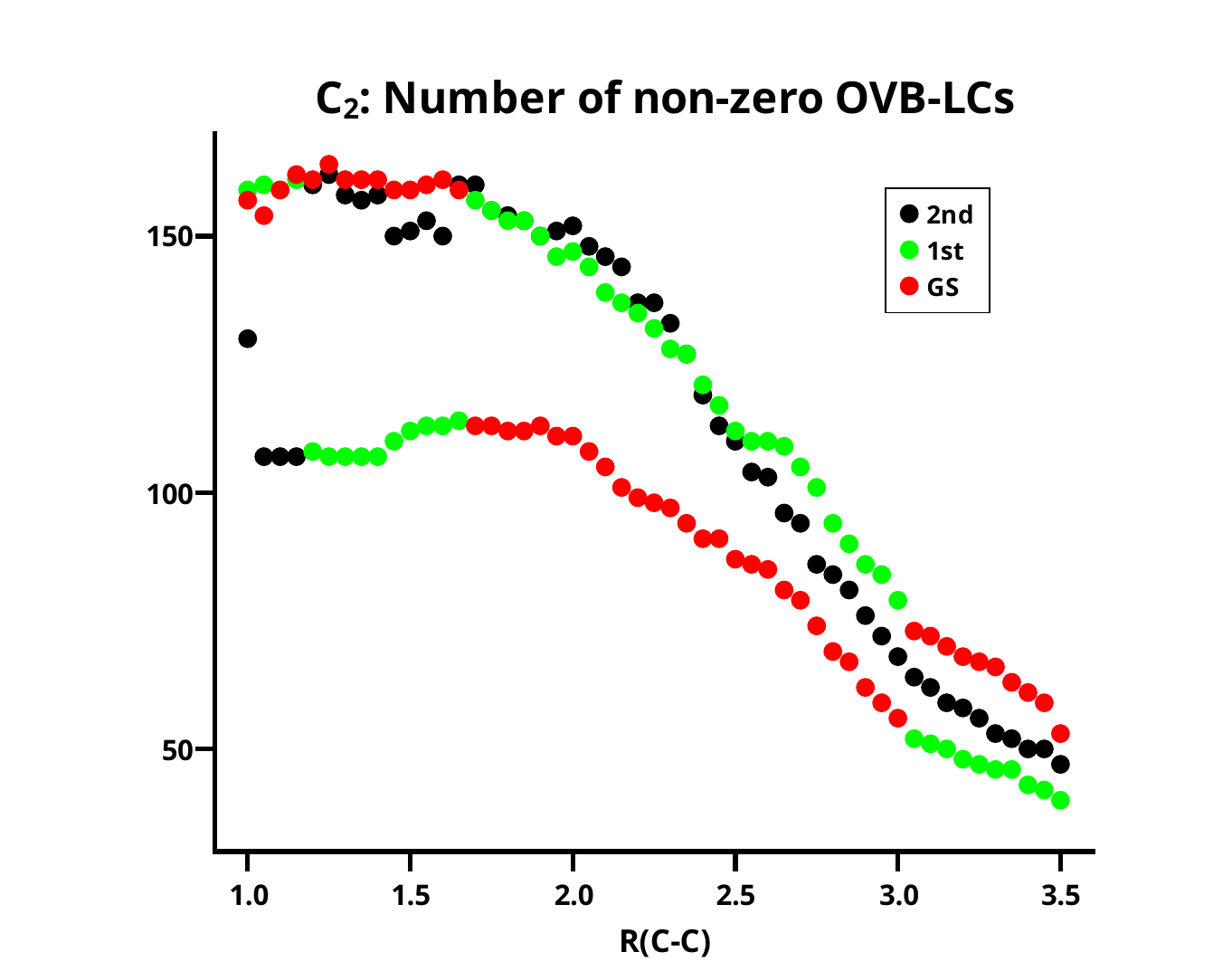}
\includegraphics[width=0.48\textwidth]{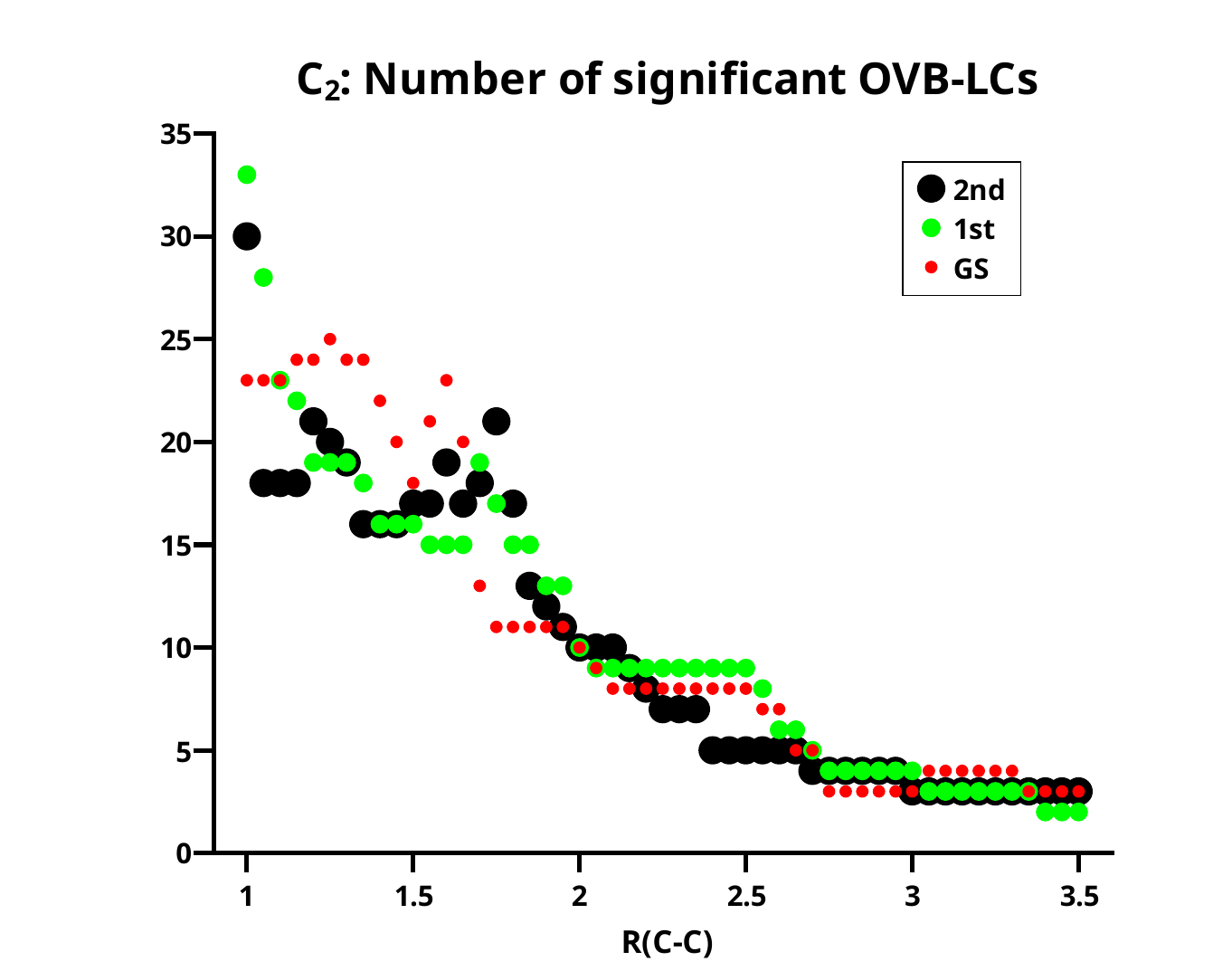}
\includegraphics[width=0.48\textwidth]{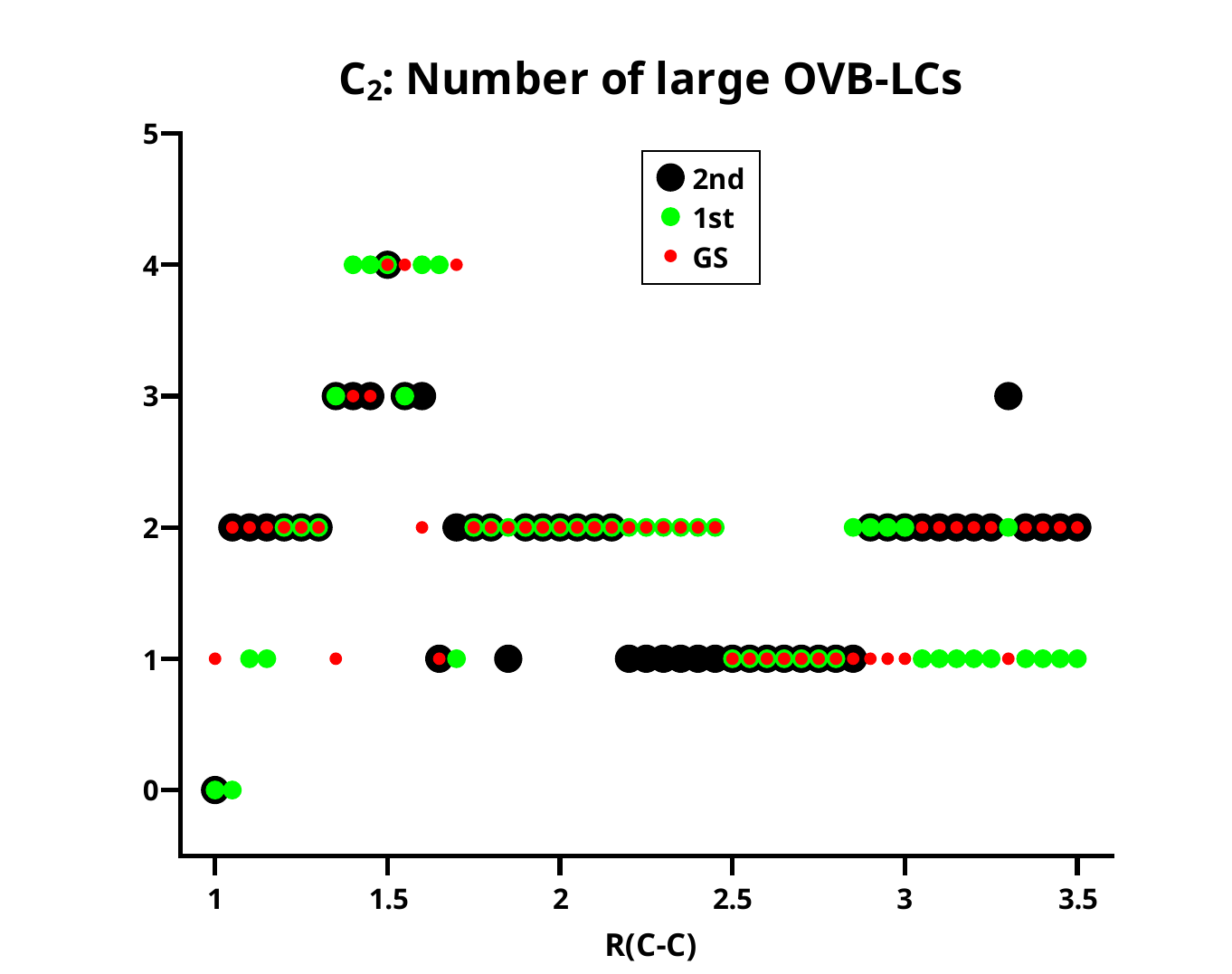}
\end{figure}

\begin{figure}[ht]
\caption{ \label{fig:CAS88largeLCs}Top left: The distribution of the weights of the large LCs for the first $A_1$ singlet state. Top right: The distribution of the weights of the large LCs for the second $A_1$ singlet state. Bottom: The distribution of the weights of the large LCs for the third $A_1$ singlet state. }
\includegraphics[width=0.48\textwidth]{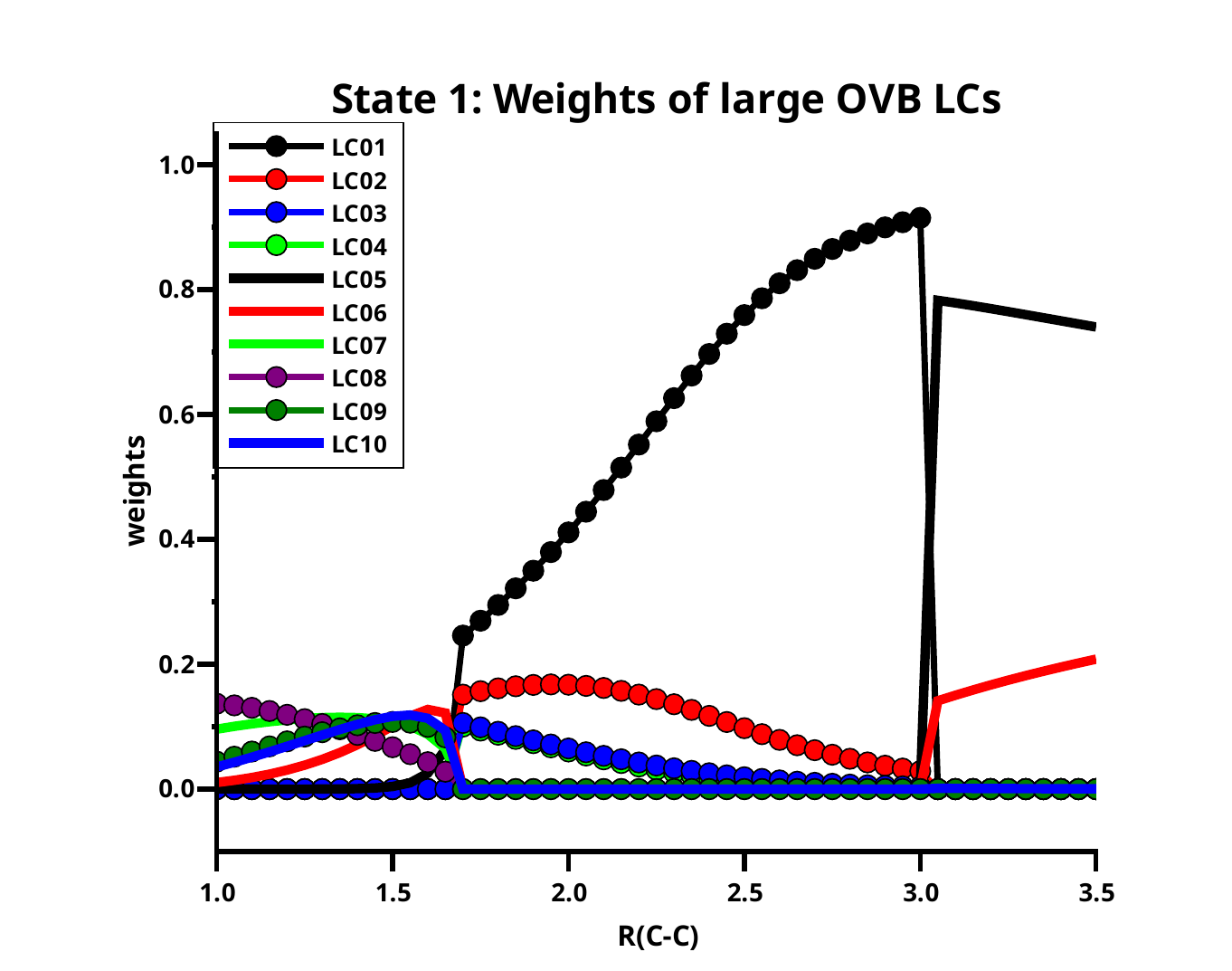}
\includegraphics[width=0.48\textwidth]{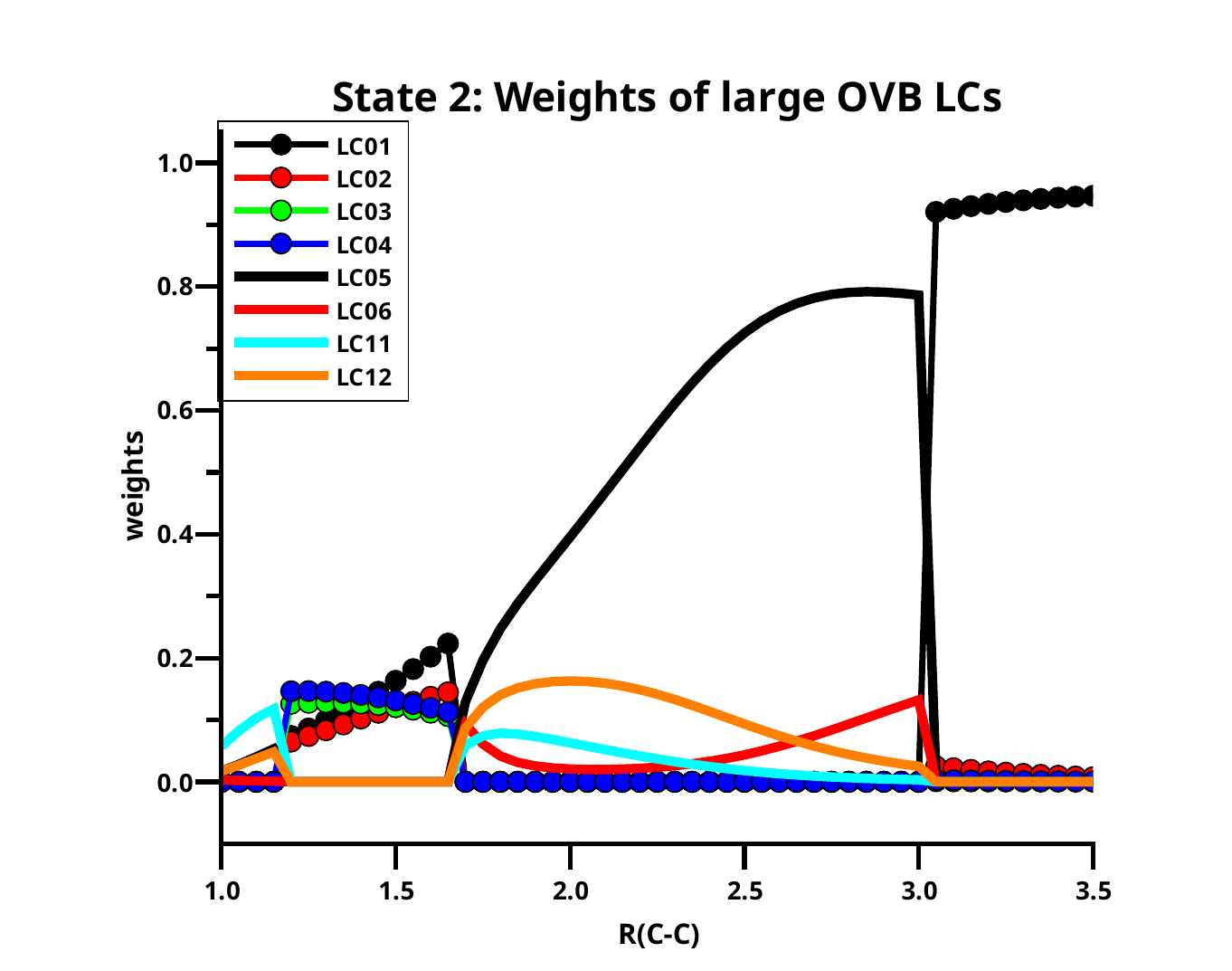}
\includegraphics[width=0.48\textwidth]{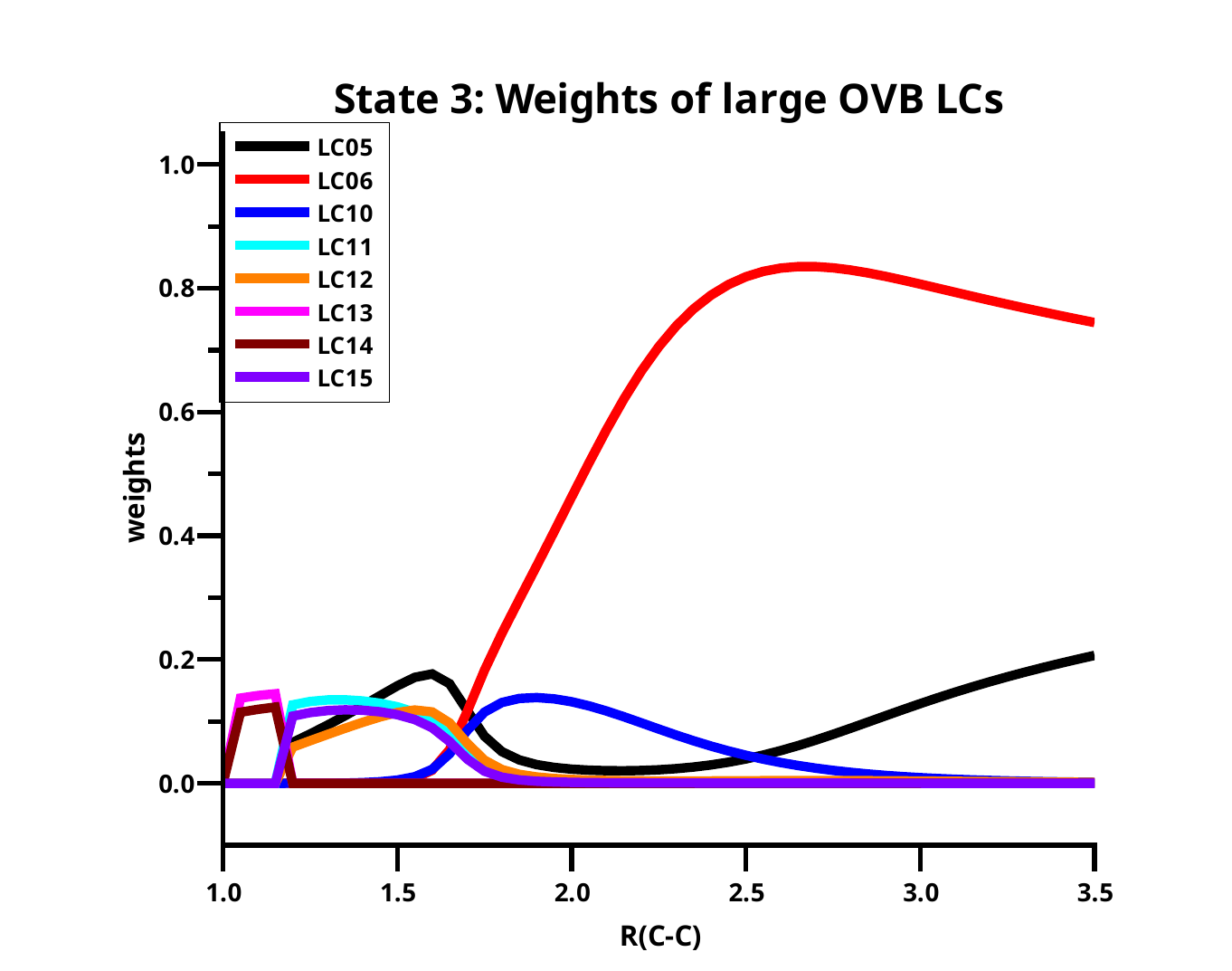}
\end{figure}

Figure \ref{fig:CAS88NLCs} shows for all three singlet states how the number of large LCs increases when going from large C-C distances towards short C-C distances; Figure \ref{fig:CAS88largeLCs} shows this for the weights of the large LCs and one can see that the increase of the number of large LCs is accompanied by a decrease of the weights, this holds not only for the individual LCs but also for the sum of the weights. The reason is the strong increase of the number of the significant but not large LCs; for the ground state, Figure \ref{fig:CAS88St1sigLCs} shows the development of the weights separately for LCs with \S\, and with \D\, character.

\newpage
\begin{landscape}
\begin{longtable}{*{7}{l}}
\caption{Description of all significant LCs. Col 2: IRREP in $D_{\infty h}$. Col 3, 4, 5: State where LC occurs. Col 6: Description by occupation strings. Col 7: Description by MO symbols.}

\label{tbl:sigOVBCSFs}\\

  \toprule\addlinespace[-2pt]

%  LC & IRREP &  \thead{State \\ 1} & \thead{State \\ 2}  &  \thead{State \\ 3}  &  \thead{Occupation \\ strings.} &Occupation strings &MO %symbols \\
  LC & IRREP &  \multicolumn{3}{c}{State} & Occupation  strings. & MO symbols \\

  \midrule\addlinespace[-2pt]
  \endfirsthead

  \multicolumn{7}{c}{\tablename~\thetable~ (continued)} \\
  \addlinespace
  \toprule\addlinespace[-2pt]
%  LC & IRREP &  \thead{State \\ 1} & \thead{State \\ 2}  &  \thead{State \\ 3}  &  \thead{Occupation \\ strings.} &Occupation strings &MO %symbols \\
%  LC & IRREP &  \multicolumn{3}{c}{\thead{State \\ 1\quad 2\quad 3}} & Occupation  strings. & MO symbols \\
  LC & IRREP &  \multicolumn{3}{c}{State} & Occupation  strings. & MO symbols \\
  \midrule\addlinespace[-1pt]
  \endhead
%  \midrule
%  \addlinespace[-8pt]
%  \multicolumn{7}{r@{}}%{\footnotesize To be continued}
%  \endfoot
  \bottomrule
  \endlastfoot

LC001 &\D &1&2&3&$\rm 2aa02bb0-2a0a2b0b                       $&$\sAs\zA\sBs\zBb(\xA\xBb-\yA\yBb)                                        $\\
LC002 &\D &1&2& &$\rm 22a020b0-220a200b+20a022b0-200a220b$&$\sAs\zAs\sBs(\xA\xBb-\yA\yBb)+\sAs\sBs\zBs(\xA\xBb-\yA\yBb)                  $\\
LC003 &\D &1&2& &$\rm 2aa0b2b0-2a0ab20b-a2a02bb0+a20a2b0b$&$\sAs\zA\sBb\zBs(\xA\xBb-\yA\yBb)+\sA\zAs\sBs\zBb(\xA\xBb-\yA\yBb)            $\\
LC004 &\D &1&2& &$\rm 22a0bab0-220aba0b-aba022b0+ab0a220b$&$\sAs\zAs\sBb\zB(\xA\xBb-\yA\yBb)+\sA\zAb\sBs\zBs(\xA\xBb-\yA\yBb)            $\\
LC005 &\S &1&2&3&$\rm 2aa02bb0+2a0a2b0b                       $ &$\sAs\zA\sBs\zBb(\xA\xBb+\yA\yBb)                                       $\\
LC006 &\S &1&2&3&$\rm 20aa20bb                                  $ &$\sAs\xA\yA\sBs\xBb\yBb                                               $\\
LC007 &\S &1&2&3&$\rm aaaabbbb                                  $ &$\sA\zA\xA\yA\sBb\zBb\xBb\yBb                                         $\\
LC008 &\D &1&2& &$\rm aaa2bbb0-aa2abb0b-aa0abb2b+aaa0bbb2$&$\sA\zA\sBb\zBb(\xA\yAs\xBb-\xAs\yA\yBb)+\sA\zA\sBb\zBb(\yA\xBs\yBb-xA\xBb\yBs)$\\
LC009 &\D &1&2& &$\rm 2a0ab02b-2aa0b0b2+a0a22bb0-a02a2b0b$&$\sAs\zA\sBb(\yA\xBs\yBb-\xA\xBb\yBs)+\sA\sBs\zBb(\xA\yAs\xBb-\xAs\yA\yBb)    $\\
LC010 &\D &1&2&3&$\rm 2aaab0bb-a0aa2bbb                       $&$\xA\yA\xBb\yBb(\sAs\zA\sBb-\sA\sBs\zBb)                                 $\\
LC011 &\S &1&2&3&$\rm 22a0bab0+220aba0b-aba022b0-ab0a220b$&$\sAs\zAs\sBb\zB(\xA\xBb+\yA\yBb)+\sA\zAb\sBs\zBs(\xA\xBb+\yA\yBb)            $\\
LC012 &\S &1&2&3&$\rm 22a020b0+220a200b+20a022b0+200a220b $ &$\sAs\zAs\sBs(\xA\xBb+\yA\yBb)+\sAs\sBs\zBs(\xA\xBb+\yA\yBb)                $\\
LC013 &\S & & &3&$\rm 22a0bab0+220aba0b+aba022b0+ab0a220b $ &$\sAs\zAs\sBb\zB(\xA\xBb+\yA\yBb)+\sA\zAb\sBs\zBs(\xA\xBb+\yA\yBb)          $\\
LC014 &\S & & &3&$\rm 2aa0b2b0+2a0ab20b+a2a02bb0+a20a2b0b $ &$\sAs\zA\sBb\zBs(\xA\xBb+\yA\yBb)+\sA\zAs\sBs\zBb(\xA\xBb+\yA\yBb)          $\\
LC015 &\S & & &3&$\rm 2aa0b2b0+2a0ab20b-a2a02bb0-a20a2b0b$&$\sAs\zA\sBb\zBs(\xA\xBb+\yA\yBb)+\sA\zAs\sBs\zBb(\xA\xBb+\yA\yBb)            $\\
LC016 &\D &1&2& &$\rm 220a002b-22a000b2-00a222b0+002a220b$&$\sAs\zAs(\yA\xBs\yBb-\xA\xBb\yBs)+\sBs\zBs(\xA\yAs\xBb-\xAs\yA\yBb)          $\\
LC017 &\D &1& & &$\rm 2aa20bb0-2a2a0b0b-0a0a2b2b+0aa02bb2$&$\sAs\zA\zBb(\xA\yAs\xBb-\xAs\yA\yBb)+\zA\sBs\zBb(\yA\xBs\yBb-\xA\xBb\yBs)    $\\
LC018 &\D &1& & &$\rm a2a20bb0-a22a0b0b+0a0ab22b-0aa0b2b2$&$\sA\zAs\zBb(\xA\yAs\xBb-\xAs\yA\yBb)+\zA\sBb\zBs(\yA\xBs\yBb-\xA\xBb\yBs)    $\\
LC019 &\S &1& & &$\rm 2a020b20+2a200b02+0a022b20+0a202b02 $ &$\sAs\zA\zBb(\yAs\xBs+\xAs\yBs)+\zA\sBs\zBb(\yAs\xBs+\xAs\yBs)              $\\
LC020 &\S &1& & &$\rm a2020b20+a2200b02-0a02b220-0a20b202$&$\sA\zAs\zBb(\yAs\xBs+\xAs\yBs)+\zA\sBb\zBs(\yAs\xBs+\xAs\yBs)                $\\
LC021 &\S &1& &3&$\rm 2aaa0bbb+0aaa2bbb                       $ &$\zA\xA\yA\zBb\xBb\yBb(\sAs+\sBs)                                       $\\
LC022 &\D &1& &3&$\rm a2aa0bbb-0aaab2bb                       $&$\xA\yA\xBb\yBb(\sA\zAs\zBb-\zA\sBb\zBs)                                 $\\
LC023 &\D &1&2& &$\rm 2a0a0b2b-2aa00bb2-0aa22bb0+0a2a2b0b$&$\sAs\zA\zBb(\yA\xBs\yBb-\xA\xBb\yBs)+\zA\sBs\zBb(\xA\yAs\xBb-\xAs\yA\yBb)    $\\
LC024 &\D &1&2& &$\rm a20a0b2b-a2a00bb2+0aa2b2b0-0a2ab20b$&$\sA\zAs\zBb(\yA\xBs\yBb-\xA\xBb\yBs)+\zA\sBb\zBs(\xA\yAs\xBb-\xAs\yA\yBb)    $\\
LC025 &\D &1& & &$\rm 02a202b0-022a020b-020a022b+02a002b2$&$\zAs\zBs(\xA\yAs\xBb-\xAs\yA\yBb)+\zAs\zBs(\yA\xBs\yBb-\xA\xBb\yBs)          $\\
LC026 &\S &1& & &$\rm 02aa02bb                                  $ &$\zAs\xA\yA\zBs\xBb\yBb                                               $\\
LC027 &\D &1& & &$\rm 2aa2b0b0-2a2ab00b+a00a2b2b-a0a02bb2$&$\sAs\zA\sBb(\xA\yAs\xBb-\xAs\yA\yBb)+\sA\sBs\zBb(\yA\xBs\yBb-\xA\xBb\yBs)    $\\
LC028 &\S &1&2&3&$\rm 2a02b020+2a20b002-a0022b20-a0202b02$&$\sAs\zA\sBb(\yAs\xBs+\xAs\yBs)+\sA\sBs\zBb(\yAs\xBs+\xAs\yBs)                $\\
LC029 &\S &1&2&3&$\rm 2a02b200+2a20b200-a2002b20-a2002b02$&$\sAs\zA\sBb\zBs(\yAs+\xAs)+\sA\zAs\sBs\zBb(\xBs+\yBs)                        $\\
LC030 &\S &1& &3&$\rm 2abab0ab-a0ba2bab                       $&$\xAb\yA\xB\yBb(\sAs\zA\sBb-\sA\sBs\zBb)                                 $\\
LC031 &\S &1& &3&$\rm 2aabb0ba-a0ab2bba                       $&$\xA\yAb\xBb\yB(\sAs\zA\sBb-\sA\sBs\zBb)                                 $\\
LC032 &\D &1& & &$\rm 2a0bb02a-2ab0b0a2+a0b22ba0-a02b2b0a$&$\sAs\zA\sBb(\yAb\xBs\yB-\xAb\xB\yBs)+\sA\sBs\zBb(\xAb\yAs\xB-\xAs\yAb\yB)    $\\
LC033 &\D &1& & &$\rm 2a00b022-a0222b00                       $&$(\sAs\zA\sBb\xBs\yBs-\sA\xAs\yAs\sBs\zBb)                               $\\
LC034 &\S &1& & &$\rm aa22bb00+aa00bb22                       $ &$\sA\zA\sBb\zBb(\xAs\yAs+\xBs\yBs)                                      $\\
LC035 &\D &1& & &$\rm aab2bba0-aa2bbb0a-aa0bbb2a+aab0bba2$&$\sA\zA\sBb\zBb(\xAb\yAs\xB-\xAs\yAb\yB)+\sA\zA\sBb\zBb(\yAb\xBs\yB-xAb\xB\yBs)$\\
LC036 &\S &1& & &$\rm aa02bb20+aa20bb02                       $ &$\sA\zA\sBb\zBb(\yAs\xBs+\xAs\yBs)                                      $\\
LC037 &\S &1& & &$\rm aababbab                                  $ &$\sA\zA\xAb\yA\sBb\zBb\xB\yBb                                         $\\
LC038 &\S &1& & &$\rm aaabbbba                                  $ &$\sA\zA\xA\yAb\sBb\zBb\xBb\yB                                         $\\
LC039 &\D &1&2& &$\rm 20a220b0-202a200b-200a202b+20a020b2$&$\sAs\sBs(\xA\yAs\xBb-\xAs\yA\yBb)+\sAs\sBs(\yA\xBs\yBb-\xA\xBb\yBs)          $\\
LC040 &\S &1&2&3&$\rm 22002020+22002002+20022200+20202200 $ &$\sAs\zAs\sBs(\xBs+\yBs)+\sAs\sBs\zBs(\yAs+\xAs)                            $\\
LC041 &\S &1&2&3&$\rm 20022020+20202002                       $ &$\sAs\sBs(\yAs\xBs+\xAs\yBs)                                            $\\
LC042 &\S &1& &3&$\rm 20ba20ab                                  $ &$\sAs\xAb\yA\sBs\xB\yBb                                               $\\
LC043 &\S &1&2&3&$\rm 2a022b00+2a202b00+2a002b20+2a002b02 $ &$\sAs\zA\sBs\zBb(\yAs+\xAs)+\sAs\zA\sBs\zBb(\xBs+\yBs)                      $\\
LC044 &\S &1&2&3&$\rm 2ab02ba0+2a0b2b0a                       $ &$\sAs\zA\sBs\zBb(\xAb\xB+\yAb\yB)                                       $\\
LC045 &\S &1&2& &$\rm 2a0a0b2b+2aa00bb2+0aa22bb0+0a2a2b0b $ &$\sAs\zA\zBb(\yA\xBs\yBb+\xA\xBb\yBs)+\zA\sBs\zBb(\xA\yAs\xBb+\xAs\yA\yBb)  $\\
LC046 &\D &1&2& &$\rm 2200ba20-2200ba02+ab022200-ab202200$&$\sAs\zAs\sBb\zB(\xBs-\yBs)+\sA\zAb\sBs\zBs(\yAs-\xAs)                        $\\
LC047 &\D &1&2& &$\rm 2a02b200-2a20b200+a2002b20-a2002b02$&$\sAs\zA\sBb\zBs(\yAs-\xAs)+\sA\zAs\sBs\zBb(\xBs-\yBs)                        $\\
LC048 &\D &1&2&3&$\rm a2a0b2b0-a20ab20b                       $&$\sA\zAs\sBb\zBs(\xA\xBb-\yA\yBb)                                        $\\
LC049 &\D &1&2& &$\rm 2ab0b2a0-2a0bb20a-a2b02ba0+a20b2b0a$&$\sAs\zA\sBb\zBs(\xAb\xB-\yAb\yB)+\sA\zAs\sBs\zBb(\xAb\xB-\yAb\yB)            $\\
LC050 &\D &1&2&3&$\rm 2a00b220-2a00b202+a2022b00-a2202b00$&$\sAs\zA\sBb\zBs(\xBs-\yBs)+\sA\zAs\sBs\zBb(\yAs-\xAs)                        $\\
LC051 &\D &1&2& &$\rm 22002020-22002002-20022200+20202200$&$\sAs\zAs\sBs(\xBs-\yBs)+\sAs\sBs\zBs(\yAs-\xAs)                              $\\
LC052 &\D &1&2& &$\rm 2a022b00-2a202b00-2a002b20+2a002b02$&$\sAs\zA\sBs\zBb(\yAs-\xAs)+\sAs\zA\sBs\zBb(\xBs-\yBs)                        $\\
LC053 &\D &1&2&3&$\rm 2ab02ba0-2a0b2b0a                       $&$\sAs\zA\sBs\zBb(\xAb\xB-\yAb\yB)                                        $\\
LC054 &\S & &2& &$\rm 22020020+22200002+00022220+00202202 $ &$\sAs\zAs(\yAs\xBs+\xAs\yBs)+\sBs\zBs(\yAs\xBs+\xAs\yBs)                    $\\
LC055 &\S & &2& &$\rm 22aa00bb+00aa22bb                       $ &$\xA\yA\xBb\yBb(\sAs\zAs+\sBs\zBs)                                      $\\
LC056 &\S & &2& &$\rm 22000022+00222200                       $ &$(\sAs\zAs\xBs\yBs+\xAs\yAs\sBs\zBs)                                    $\\
LC057 &\D & &2& &$\rm a2000b22-0a22b200                       $&$(\sA\zAs\zBb\xBs\yBs-\zA\xAs\yAs\sBb\zBs)                               $\\
LC058 &\S & &2&3&$\rm 22020200+22200200+02002220+02002202 $ &$\sAs\zAs\zBs(\yAs+\xAs)+\zAs\sBs\zBs(\xBs+\yBs)                            $\\
LC059 &\S & &2&3&$\rm 22a002b0+220a020b+02a022b0+020a220b $ &$\sAs\zAs\zBs(\xA\xBb+\yA\yBb)+\zAs\sBs\zBs(\xA\xBb+\yA\yBb)                $\\
LC060 &\D & &2& &$\rm 20a202b0-202a020b-020a202b+02a020b2$&$\sAs\zBs(\xA\yAs\xBb-\xAs\yA\yBb)+\zAs\sBs(\yA\xBs\yBb-\xA\xBb\yBs)          $\\
LC061 &\S & &2&3&$\rm 22000220+22000202+02022200+02202200 $ &$\sAs\zAs\zBs(\xBs+\yBs)+\zAs\sBs\zBs(\yAs+\xAs)                            $\\
LC062 &\S & &2&3&$\rm 20aa02bb+02aa20bb                       $ &$\xA\yA\xBb\yBb(\sAs\zBs+\zAs\sBs)                                      $\\
LC063 &\D & &2& &$\rm 200a022b-20a002b2-02a220b0+022a200b$&$\sAs\zBs(\yA\xBs\yBb-\xA\xBb\yBs)+\zAs\sBs(\xA\yAs\xBb-\xAs\yA\yBb)          $\\
LC064 &\D & &2& &$\rm a2a2b0b0-a22ab00b-a00ab22b+a0a0b2b2$&$\sA\zAs\sBb(\xA\yAs\xBb-\xAs\yA\yBb)+\sA\sBb\zBs(\yA\xBs\yBb-\xA\xBb\yBs)    $\\
LC065 &\S & &2& &$\rm a202b020+a220b002+a002b220+a020b202 $ &$\sA\zAs\sBb(\yAs\xBs+\xAs\yBs)+\sA\sBb\zBs(\yAs\xBs+\xAs\yBs)              $\\
LC066 &\S & &2& &$\rm a2aab0bb+a0aab2bb                       $ &$\sA\xA\yA\sBb\xBb\yBb(\zAs+\zBs)                                       $\\
LC067 &\D & &2& &$\rm a20ab02b-a2a0b0b2-a0a2b2b0+a02ab20b$&$\sA\zAs\sBb(\yA\xBs\yBb-\xA\xBb\yBs)+\sA\sBb\zBs(\xA\yAs\xBb-\xAs\yA\yBb)    $\\
LC068 &\D & &2& &$\rm a20bb02a-a2b0b0a2-a0b2b2a0+a02bb20a$&$\sA\zAs\sBb(\yAb\xBs\yB-\xAb\xB\yBs)+\sA\sBb\zBs(\xAb\yAs\xB-\xAs\yAb\yB)    $\\
LC069 &\S & &2& &$\rm a200b022+a022b200                       $ &$\sA\sBb(\zAs\xBs\yBs+\xAs\yAs\zBs)                                     $\\
LC070 &\S & &2&3&$\rm 2202ba00+2220ba00-ab002220-ab002202$&$\sAs\zAs\sBb\zB(\yAs+\xAs)+\sA\zAb\sBs\zBs(\xBs+\yBs)                        $\\
LC071 &\D & &2& &$\rm aba2bab0-ab2aba0b-ab0aba2b+aba0bab2$&$\sA\zAb\sBb\zB(\xA\yAs\xBb-\xAs\yA\yBb)+\sA\zAb\sBb\zB(\yA\xBs\yBb-xA\xBb\yBs)$\\
LC072 &\S & &2&3&$\rm 2200ba20+2200ba02-ab022200-ab202200$&$\sAs\zAs\sBb\zB(\xBs+\yBs)+\sA\zAb\sBs\zBs(\yAs+\xAs)                        $\\
LC073 &\S & &2&3&$\rm a202b200+a220b200+a200b220+a200b202 $ &$\sA\zAs\sBb\zBs(\yAs+\xAs)+\sA\zAs\sBb\zBs(\xBs+\yBs)                      $\\
LC074 &\S & &2&3&$\rm a2a0b2b0+a20ab20b                       $ &$\sA\zAs\sBb\zBs(\xA\xBb+\yA\yBb)                                       $\\
LC075 &\S & &2&3&$\rm 2ab0b2a0+2a0bb20a-a2b02ba0-a20b2b0a$&$\sAs\zA\sBb\zBs(\xAb\xB+\yAb\yB)+\sA\zAs\sBs\zBb(\xAb\xB+\yAb\yB)            $\\
LC076 &\S & &2&3&$\rm a2b0b2a0+a20bb20a                       $ &$\sA\zAs\sBb\zBs(\xAb\xB+\yAb\yB)                                       $\\
LC077 &\S & &2&3&$\rm 2a00b220+2a00b202-a2022b00-a2202b00$&$\sAs\zA\sBb\zBs(\xBs+\yBs)+\sA\zAs\sBs\zBb(\yAs+\xAs)                        $\\
LC078 &\D & &2& &$\rm 22020200-22200200-02002220+02002202$&$\sAs\zAs\zBs(\yAs-\xAs)+\zAs\sBs\zBs(\xBs-\yBs)                              $\\
LC079 &\D & &2& &$\rm 22a002b0-220a020b+02a022b0-020a220b$&$\sAs\zAs\zBs(\xA\xBb-\yA\yBb)+\zAs\sBs\zBs(\xA\xBb-\yA\yBb)                 $\\
LC080 &\D & &2& &$\rm 22000220-22000202-02022200+02202200$&$\sAs\zAs\zBs(\xBs-\yBs)+\zAs\sBs\zBs(\yAs-\xAs)                             $\\
LC081 &\D & &2& &$\rm 2202ba00-2220ba00+ab002220-ab002202$&$\sAs\zAs\sBb\zB(\yAs-\xAs)+\sA\zAb\sBs\zBs(\xBs-\yBs)                       $\\
LC082 &\D & &2& &$\rm a202b200-a220b200-a200b220+a200b202$&$\sA\zAs\sBb\zBs(\yAs-\xAs)+\sA\zAs\sBb\zBs(\xBs-\yBs)                       $\\
LC083 &\D & &2&3&$\rm a2b0b2a0-a20bb20a                       $&$\sA\zAs\sBb\zBs(\xAb\xB-\yAb\yB)                                       $\\
LC084 &\D & &2& &$\rm 22022000-22202000-20002220+20002202$&$\sAs\zAs\sBs(\yAs-\xAs)+\sAs\sBs\zBs(\xBs-\yBs)                             $\\
LC085 &\D & & &3&$\rm 22a200b0-222a000b+000a222b-00a022b2$&$\sAs\zAs(\xA\yAs\xBb-\xAs\yA\yBb)+\sBs\zBs(\yA\xBs\yBb-\xA\xBb\yBs)         $\\
LC086 &\S & & &3&$\rm 22020020+22200002-00022220-00202202$&$\sAs\zAs(\yAs\xBs+\xAs\yBs)+\sBs\zBs(\yAs\xBs+\xAs\yBs)                     $\\
LC087 &\D & & &3&$\rm 22aa00bb-00aa22bb                       $&$\xA\yA\xBb\yBb(\sAs\zAs-\sBs\zBs)                                      $\\
LC088 &\S & & &3&$\rm 22ba00ab-00ba22ab                       $&$\xAb\yA\xB\yBb(\sAs\zAs-\sBs\zBs)                                      $\\
LC089 &\D & & &3&$\rm 220a002b-22a000b2+00a222b0-002a220b$&$\sAs\zAs(\yA\xBs\yBb-\xA\xBb\yBs)+\sBs\zBs(\xA\yAs\xBb-\xAs\yA\yBb)         $\\
LC090 &\D & & &3&$\rm 22000022-00222200                       $&$(\sAs\zAs\xBs\yBs-\xAs\yAs\sBs\zBs)                                    $\\
LC091 &\D & & &3&$\rm 2a0a0b2b-2aa00bb2+0aa22bb0-0a2a2b0b$&$\sAs\zA\zBb(\yA\xBs\yBb-\xA\xBb\yBs)+\zA\sBs\zBb(\xA\yAs\xBb-\xAs\yA\yBb)   $\\
LC092 &\D & & &3&$\rm a20a0b2b-a2a00bb2-0aa2b2b0+0a2ab20b$&$\sA\zAs\zBb(\yA\xBs\yBb-\xA\xBb\yBs)+\zA\sBb\zBs(\xA\yAs\xBb-\xAs\yA\yBb)   $\\
LC093 &\S & & &3&$\rm a2000b22+0a22b200                       $ &$(\sA\zAs\zBb\xBs\yBs+\zA\xAs\yAs\sBb\zBs)                             $\\
LC094 &\S & & &3&$\rm 22020200+22200200-02002220-02002202$&$\sAs\zAs\zBs(\yAs+\xAs)+\zAs\sBs\zBs(\xBs+\yBs)                             $\\
LC095 &\S & & &3&$\rm 22a002b0+220a020b-02a022b0-020a220b$&$\sAs\zAs\zBs(\xA\xBb+\yA\yBb)+\zAs\sBs\zBs(\xA\xBb+\yA\yBb)                 $\\
LC096 &\D & & &3&$\rm 20a202b0-202a020b+020a202b-02a020b2$&$\sAs\zBs(\xA\yAs\xBb-\xAs\yA\yBb)+\zAs\sBs(\yA\xBs\yBb-\xA\xBb\yBs)         $\\
LC097 &\D & & &3&$\rm aba202b0-ab2a020b-020aba2b+02a0bab2$&$\sA\zAb\zBs(\xA\yAs\xBb-\xAs\yA\yBb)+\zAs\sBb\zB(\yA\xBs\yBb-\xA\xBb\yBs)   $\\
LC098 &\S & & &3&$\rm 22000220+22000202-02022200-02202200$&$\sAs\zAs\zBs(\xBs+\yBs)+\zAs\sBs\zBs(\yAs+\xAs)                             $\\
LC099 &\D & & &3&$\rm 20aa02bb-02aa20bb                       $&$\xA\yA\xBb\yBb(\sAs\zBs-\zAs\sBs)                                      $\\
LC100 &\S & & &3&$\rm abaa02bb+02aababb                       $ &$\xA\yA\xBb\yBb(\sA\zAb\zBs+\zAs\sBb\zB)                               $\\
LC101 &\D & & &3&$\rm 2aa2b0b0-2a2ab00b-a00a2b2b+a0a02bb2$&$\sAs\zA\sBb(\xA\yAs\xBb-\xAs\yA\yBb)+\sA\sBs\zBb(\yA\xBs\yBb-\xA\xBb\yBs)   $\\
LC102 &\D & & &3&$\rm a2a2b0b0-a22ab00b+a00ab22b-a0a0b2b2$&$\sA\zAs\sBb(\xA\yAs\xBb-\xAs\yA\yBb)+\sA\sBb\zBs(\yA\xBs\yBb-\xA\xBb\yBs)   $\\
LC103 &\S & & &3&$\rm 2a02b020+2a20b002+a0022b20+a0202b02 $ &$\sAs\zA\sBb(\yAs\xBs+\xAs\yBs)+\sA\sBs\zBb(\yAs\xBs+\xAs\yBs)             $\\
LC104 &\S & & &3&$\rm a202b020+a220b002-a002b220-a020b202$&$\sA\zAs\sBb(\yAs\xBs+\xAs\yBs)+\sA\sBb\zBs(\yAs\xBs+\xAs\yBs)               $\\
LC105 &\S & & &3&$\rm 2aaab0bb+a0aa2bbb                       $ &$\xA\yA\xBb\yBb(\sAs\zA\sBb+\sA\sBs\zBb)                               $\\
LC106 &\D & & &3&$\rm a2aab0bb-a0aab2bb                       $&$\sA\xA\yA\sBb\xBb\yBb(\zAs-\zBs)                                       $\\
LC107 &\S & & &3&$\rm a2bab0ab-a0bab2ab                       $&$\sA\xAb\yA\sBb\xB\yBb(\zAs-\zBs)                                       $\\
LC108 &\D & & &3&$\rm 2a0ab02b-2aa0b0b2-a0a22bb0+a02a2b0b$&$\sAs\zA\sBb(\yA\xBs\yBb-\xA\xBb\yBs)+\sA\sBs\zBb(\xA\yAs\xBb-\xAs\yA\yBb)   $\\
LC109 &\D & & &3&$\rm a20ab02b-a2a0b0b2+a0a2b2b0-a02ab20b$&$\sA\zAs\sBb(\yA\xBs\yBb-\xA\xBb\yBs)+\sA\sBb\zBs(\xA\yAs\xBb-\xAs\yA\yBb)   $\\
LC110 &\S & & &3&$\rm a2abb0ba-a0abb2ba                       $&$\sA\xA\yAb\sBb\xBb\yB(\zAs-\zBs)                                       $\\
LC111 &\D & & &3&$\rm 2a0bb02a-2ab0b0a2-a0b22ba0+a02b2b0a$&$\sAs\zA\sBb(\yAb\xBs\yB-\xAb\xB\yBs)+\sA\sBs\zBb(\xAb\yAs\xB-\xAs\yAb\yB)   $\\
LC112 &\D & & &3&$\rm a20bb02a-a2b0b0a2+a0b2b2a0-a02bb20a$&$\sA\zAs\sBb(\yAb\xBs\yB-\xAb\xB\yBs)+\sA\sBb\zBs(\xAb\yAs\xB-\xAs\yAb\yB)   $\\
LC113 &\S & & &3&$\rm 2a00b022+a0222b00                       $ &$(\sAs\zA\sBb\xBs\yBs+\sA\xAs\yAs\sBs\zBb)                             $\\
LC114 &\D & & &3&$\rm a200b022-a022b200                       $&$\sA\sBb(\zAs\xBs\yBs-\xAs\yAs\zBs)                                     $\\
LC115 &\D & & &3&$\rm aaa2bbb0-aa2abb0b+aa0abb2b-aaa0bbb2$&$\sA\zA\sBb\zBb(\xA\yAs\xBb-\xAs\yA\yBb)+\sA\zA\sBb\zBb(\yA\xBs\yBb-xA\xBb\yBs)$\\
LC116 &\S & & &3&$\rm 2202ba00+2220ba00+ab002220+ab002202 $& $\sAs\zAs\sBb\zB(\yAs+\xAs)+\sA\zAb\sBs\zBs(\xBs+\yBs)                     $\\
LC117 &\D & & &3&$\rm 20a2bab0-202aba0b-ab0a202b+aba020b2$&$\sAs\sBb\zB(\xA\yAs\xBb-\xAs\yA\yBb)+\sA\zAb\sBs(\yA\xBs\yBb-\xA\xBb\yBs)   $\\
LC118 &\S & & &3&$\rm 2200ba20+2200ba02+ab022200+ab202200 $& $\sAs\zAs\sBb\zB(\xBs+\yBs)+\sA\zAb\sBs\zBs(\yAs+\xAs)                     $\\
LC119 &\S & & &3&$\rm 2002ba20+2020ba02+ab022020+ab202002 $& $\sAs\sBb\zB(\yAs\xBs+\xAs\yBs)+\sA\zAb\sBs(\yAs\xBs+\xAs\yBs)              $\\
LC120 &\S & & &3&$\rm 20aababb+abaa20bb                       $& $\xA\yA\xBb\yBb(\sAs\sBb\zB+\sA\zAb\sBs)                                $\\
LC121 &\D & & &3&$\rm 200aba2b- 20a0bab2-aba220b0+ab2a200b$  & $\sAs\sBb\zB(\yA\xBs\yBb-\xA\xBb\yBs)+\sA\zAb\sBs(\xA\yAs\xBb-\xAs\yA\yBb)$\\
LC122 &\S & & &3&$\rm 2a02b200+2a20b200+a2002b20+a2002b02 $& $\sAs\zA\sBb\zBs(\yAs+\xAs)+\sA\zAs\sBs\zBb(\xBs+\yBs)                      $\\
LC123 &\S & & &3&$\rm a202b200+a220b200-a200b220-a200b202$&$\sA\zAs\sBb\zBs(\yAs+\xAs)+\sA\zAs\sBb\zBs(\xBs+\yBs)                        $\\
LC124 &\S & & &3&$\rm 2ab0b2a0+2a0bb20a+a2b02ba0+a20b2b0a $& $\sAs\zA\sBb\zBs(\xAb\xB+\yAb\yB)+\sA\zAs\sBs\zBb(\xAb\xB+\yAb\yB)          $\\
LC125 &\S & & &3&$\rm 22a020b0+220a200b-20a022b0-200a220b$&$\sAs\zAs\sBs(\xA\xBb+\yA\yBb)+\sAs\sBs\zBs(\xA\xBb+\yA\yBb)                  $\\
LC126 &\D & & &3&$\rm 20a220b0-202a200b+200a202b-20a020b2$&$\sAs\sBs(\xA\yAs\xBb-\xAs\yA\yBb)+\sAs\sBs(\yA\xBs\yBb-\xA\xBb\yBs)          $\\
LC127 &\S & & &3&$\rm 22002020+22002002-20022200-20202200$&$\sAs\zAs\sBs(\xBs+\yBs)+\sAs\sBs\zBs(\yAs+\xAs)                              $\\
LC128 &\S & & &3&$\rm 2a022b00+2a202b00-2a002b20-2a002b02$&$\sAs\zA\sBs\zBb(\yAs+\xAs)+\sAs\zA\sBs\zBb(\xBs+\yBs)                        $
\end{longtable}
\end{landscape}

\begin{landscape}
\begin{longtable}{*{8}{l}}
\caption{Description of all large LCs. Col 2: IRREPs in $D_{\infty h}$. Col 3: Neutral (n) or singly ionic (s) LC. Col 4: LC of CSFs.  Col 5: MO symbols.}
\label{tbl:largeOVBCSFs}\\

  \toprule\addlinespace[-2pt]

%  LC & IRREP &  \thead{State \\ 1} & \thead{State \\ 2}  &  \thead{State \\ 3}  &  \thead{Occupation \\ strings.} &Occupation strings &MO %symbols \\
  LC & IRREP & charge & \multicolumn{3}{c}{State} & Occupation  strings. & MO symbols \\

  \midrule\addlinespace[-2pt]
  \endfirsthead

  \multicolumn{8}{c}{\tablename~\thetable~ (continued)} \\
  \addlinespace
  \toprule\addlinespace[-2pt]
%  LC & IRREP &  \thead{State \\ 1} & \thead{State \\ 2}  &  \thead{State \\ 3}  &  \thead{Occupation \\ strings.} &Occupation strings &MO %symbols \\
%  LC & IRREP &  \multicolumn{3}{c}{\thead{State \\ 1\quad 2\quad 3}} & Occupation  strings. & MO symbols \\
  LC & IRREP &  charge & \multicolumn{3}{c}{State} & Occupation  strings. & MO symbols \\
  \midrule\addlinespace[-1pt]
  \endhead
%  \midrule
%  \addlinespace[-8pt]
%  \multicolumn{7}{r@{}}%{\footnotesize To be continued}
%  \endfoot
  \bottomrule
  \endlastfoot

LC01 &\D &n &1&2& & 2aa02bb0 - 2a0a2b0b                       &$\sAs\sBs\zA\zBb(\xA\xBb - \yA\yBb)$ \\
LC02 &\D &s &1&2& & 22a020b0 - 220a200b + 20a022b0 - 200a220b &$\sAs\sBs(\zAs+ \zBs)(\xA\xBb - \yA\yBb) $ \\
LC03 &\D &n &1&2& & 2aa0b2b0 - 2a0ab20b - a2a02bb0 + a20a2b0b &$(\sAs\zA\sB\zBs - \zAs\sBs\sA\zBb)(\xA\xBb - \yA\yBb) $ \\
LC04 &\D &s &1&2& & 22a0bab0 - 220aba0b - aba022b0 + ab0a220b &$(\sAs\zAs\sBb\zB  - \sBs\zBs\sA\zAb) (\xA\xBb - \yA\yBb)$ \\
LC05 &\S &n &1&2&3& 2aa02bb0 + 2a0a2b0b                       &$\sAs\sBs\zA\zBb(\xA\xBb + \yA\yBb)$ \\
LC06 &\S &n &1&2&3& 20aa20bb                                  &$\sAs\sBs\xA\xBb\yA\yBb$ \\
LC07 &\S &n &1& & & aaaabbbb                                  &$\sA\sBb\zA\zBb\xA\xBb\yA\yBb$ \\
LC08 &\D &s &1& & & aaa2bbb0 - aa2abb0b - aa0abb2b + aaa0bbb2 &$(\yAs\xA\xBb - \xAs\yA\yBb - \xBs\yA\yBb + \yBs\xA\xBb)\sA\sBb\zA\zBb$\\
LC09 &\D &n &1& & & 2a0ab02b - 2aa0b0b2 + a0a22bb0 - a02a2b0b &$\sAs\zA\sBb(\xBs\yA\yBb - \yBs\xA\xBb) + \sBs\sA\zBb(\yAs\xA\xBb - \xAs\yA\yBb)$\\
LC10 &\S &s &1& &3& 2aaab0bb - a0aa2bbb                       &$(\sAs\zA\sBb- \sBs\sA\zBb)\xA\xBb\yA\yBb$\\
LC11 &\S &s & &2&3& 22a0bab0 + 220aba0b - aba022b0 - ab0a220b &$(\sAs\zAs\sBb\zB - \sBs\zBs\sA\zAb)(\xA\xBb+ \yA\yBb)$\\
LC12 &\S &n & &2&3& 22a020b0 + 220a200b + 20a022b0 + 200a220b &$\sAs\sBs(\zAs+\zBs) (\xA\xBb+ \yA\yBb)$\\
LC13 &\S &n & & &3& 22a0bab0 + 220aba0b + aba022b0 + ab0a220b &$\sAs\zAs\sBb\zB + \sA\zAb\sBs\zBs) (\xA\xBb+ \yA\yBb)$ \\
LC14 &\S &s & & &3& 2aa0b2b0 + 2a0ab20b + a2a02bb0 + a20a2b0b &$\sAs\zBs\zA\sBb + \sA\zBb\zAs\sBs)(\xA\xBb + \yA\yBb) $ \\
LC15 &\S &n & & &3& 2aa0b2b0 + 2a0ab20b - a2a02bb0 - a20a2b0b &$\sAs\zBs\zA\sBb- \sA\zBb\zAs\sBs) (\xA\xBb + \yA\yBb)$
\end{longtable}
\end{landscape}

\end{document}